\documentclass{aa}
\usepackage{hyperref}
\hypersetup{colorlinks, allcolors=blue}
\usepackage{graphicx}    
\usepackage{color}  
\usepackage{txfonts}
\usepackage{inputenc}
\usepackage{multicol}
\usepackage{multirow}
\usepackage{longtable}
\usepackage{subcaption}
\usepackage{siunitx}
\DeclareSIUnit{\year}{yr}
\DeclareSIUnit{\arcsec}{arcsec}
\usepackage{amsmath}
\usepackage{xcolor}
\usepackage{booktabs}
\usepackage{color}

\usepackage{algorithm} 
\usepackage{algpseudocode}
\usepackage{placeins}

\begin{document} 

    \title{HURACAN: A comparison of stellar and interstellar proper motions inside L1688}
    \titlerunning{HURACAN: A comparison of stellar and interstellar proper motions inside L1688}
    
    \author{M. Piecka\inst{1}
            \and
            J. Alves\inst{1}
            \and
            A. Rottensteiner\inst{1}
            \and
            S. Hutschenreuter\inst{1}
            \and
            S. Meingast\inst{1}
            \and
            L. Posch\inst{1}
            }

    \institute{University of Vienna, Department of Astrophysics,
               T\"urkenschanzstrasse 17, 1180 Vienna, Austria\\
               \email{martin.piecka@univie.ac.at}}

    \date{Received X XX, XXXX; accepted X XX, XXXX}

\abstract{The assumption that molecular clouds are kinematically coupled to the young stellar populations forming within them is observationally still poorly constrained. With the transverse component of the interstellar medium (ISM) motion being largely unknown, the three-dimensional kinematic state of even the closest clouds remains incomplete. We provide the first direct measurement of the transverse motion of the ISM inside the Ophiuchus cloud L1688 and compare it with the kinematics of the surrounding young stellar populations. The measurement was obtained using intensity-based image registration applied to archival data, allowing us to generate a proper motion field of the studied regions and determine the cloud's typical local motion, $(\mu_{\alpha^*},\mu_\delta)^{\textrm{ISM}}_{\textrm{median}}=(-3.0, -24.1)$ mas\,yr$^{-1}$. This motion differs from the kinematics of the youngest stellar population in Upper~Scorpius by $3\pm1$~km\,s$^{-1}$, highlighting the possibility of a bias when using stars as a proxy for the ISM motion near star-forming regions. By comparing the cloud's motion with that of the youngest stellar population in the region and the B-star Oph~S1 located within the cloud, we find that stellar feedback likely dynamically shaped the motion of the ISM over the last few million years. Furthermore, by studying a shell-like ISM structure appearing in L1688, we find evidence suggesting that this object is likely neither a typical Herbig-Haro object nor a supernova remnant, with its origin remaining a mystery. Our results demonstrate that archival and future high-resolution near-infrared observations can and will continue to enable measurements of ISM transverse motions in nearby clouds, opening a new window for studying ISM dynamics that is not accessible when using radial velocities alone.}


\maketitle
\nolinenumbers

\section{Introduction}\label{section:1}

L1688 is one of the closest star-forming clouds \citep[$\approx 140$~pc,][]{Zucker2020} and is a part of the Ophiuchus molecular complex located at the border of the Ophiuchus and the Scorpius constellations. This complex is associated with multiple dark clouds and surrounded by several bright B-type stars that belong to the Upper~Scorpius (USco) stellar population \citep[see the review by][]{Wilking2008}. Owing to its proximity to Earth and the presence of young massive stars, the Ophiuchus complex is considered as one of the most important benchmarks for testing and refining theories of star formation \citep[e.g.][]{Montmerle1983,deGeus1992,Tachihara2000,Simpson2008,Lomax2014,Williams2019,Testi2022,Ratzenbock2023,Posch2025}. Observational evidence suggests that the region was (and possibly still is) heavily influenced by stellar feedback \citep[e.g.][]{deGeus1992,Robitaille2018,Krause2018,Alves2025}. Furthermore, it is widely recognised that the star-formation history of the Scorpius-Centaurus stellar association, which includes USco, is associated with over a dozen supernovae that occurred over the last 20~Myr, transporting momentum into the surrounding interstellar medium (ISM) and forming the large superbubble known as the Local~Bubble \citep{MaizApellaniz2001,Fuchs2006,Zucker2022}. A by-product of this history is the presence of low-density ($\lesssim0.1$~cm$^{-3}$) ISM flows within the Local~Bubble \citep{Frisch1995,Frisch2011,Piecka2024,Zucker2025}, which may influence the environments of planetary systems, including our Solar~System \citep{Ellis1995,Benitez2002,Breitschwerdt2016,Schulreich2017}.

L1688 is the most investigated cloud in Ophiuchus. It is by far the most massive part of the complex \citep{Loren1989} and it hosts multiple active star-forming sites \citep{Wilking1983,Bontemps2001}. Recent works based on {\it Gaia} astrometry \citep{GaiaDR2,GaiaDR3} and infrared surveys have provided a detailed picture of the kinematic and spatial substructure of stellar populations associated with L1688 \citep[e.g.][]{Grasser2021,MiretRoig2022,SigMA,Hutschenreuter2026}. Among the members of these populations, particular attention has been given to the B-type stars $\rho$~Oph~S1 (GSS~35) and HD~147889, which are located near the northern tip of L1688 and represent the most important radiation sources in this section of the cloud \citep{Howard2021,Ordonez2025}.

Turning our attention to the ISM, the region of L1688 encompasses various ISM features, ranging from jets to dense cores with sizes of up to 0.1~pc \citep[e.g.][]{Gomez2003,Chen2019,Nakamura2025}. Unlike in the case of the stars, a precise 3D kinematic picture of the gas in L1688 is still missing, primarily due to the difficulty of obtaining the transverse components ($v_{\textrm{t}}$) of the ISM motion. Although several kinematic studies of L1688 exist \citep{Andre2007,Rigliaco2016,Yun2021}, they are either based on radial velocities obtained directly for the gas or 3D~kinematics obtained for the stars.

Young stellar objects (YSOs) embedded within their parental clouds are commonly assumed to be kinematically coupled with the surrounding gas \citep{Ducourant2017,Grossschedl2021,Megeath2022}. While masers also represent a useful tracer of the ISM, they are far less abundant and arguably more poorly understood when compared with YSOs \citep[e.g.][]{Wilking1987,Bontemps2001}. This makes YSOs a promising proxy for ISM studies. However, the coupling between YSOs and ISM was so far only tested using radial velocities \citep[e.g.][]{Covey2006,Hacar2016,Grossschedl2021,Rottensteiner2026}, with studies suggesting a tight but not perfect agreement between the gas and stellar kinematics ($\Delta v_{\mathrm{r}} \lesssim 1$~km\,s$^{-1}$), which requires further investigation. A direct ISM proper motion measurement would be preferred for a more accurate understanding of the dynamical state of the gas and to verify the assumption of kinematical coupling between YSOs and the ISM. This is especially emphasised by the recent findings presented by \citet{Le2024} and \citet{Alves2025} that highlight the roles of magnetic fields and stellar feedback, respectively, in dynamically shaping the molecular cloud.

Recently, \citet{CrA_motion} published a historically first direct measurement of a molecular cloud proper motion. The authors investigated the multi-epoch Corona~Australis data from the VISIONS survey \citep{VISIONS1,VISIONS2} and applied an image registration procedure to align images of ISM structures, yielding a proper motion measurement. Their results suggest a possible discrepancy between the proper motion of YSOs and the ISM. Verifying this finding will require additional data due to the relatively short time baseline \citep[4.5~yr,][]{CrA_motion} that resulted in an observed ISM position offset corresponding to only about one-third of a pixel in the images.

In this paper, we present the proper motion analysis of the first target of our High-resolution survey Utilising Registration-based image Alignments to obtain proper motions of Clouds Across the solar Neighbourhood (HURACAN). We focus on the ISM in L1688 and compare its kinematic properties with those of the YSO population in this region. In Sections~\ref{section:2} and~\ref{section:3}, we discuss the archival images available for such a study and outline the calibration of the astrometric solution. In Section~\ref{section:4}, we report the available information concerning YSOs in L1688 and describe the procedures used to measure the ISM proper motion. Section~\ref{section:5} focuses on our analysis of the presented measurements, resulting in a comparison between ISM and YSO kinematics. Section~\ref{section:6} summarises the results of this work.

\section{Archival images}\label{section:2}

The near-infrared imaging capabilities of the Hubble Space Telescope (HST) and the James Webb Space Telescope (JWST) offer good opportunities to study proper motions of interstellar clouds. Firstly, the filters used by HST/WFC3 and JWST/NIRCam (F160W, and F140M, F150W, or F200W, respectively) probe roughly the same ISM optical depths and ensure that the imaged ISM morphology does not significantly vary between the two instruments. Furthermore, the throughput, the apertures, and the location outside Earth's atmosphere allow achieving good signal-to-noise (S/N) levels for the ISM with reasonable exposure times ($\lesssim1$~h). Last but not least, the high resolution of HST and JWST should theoretically permit a more precise proper motion measurement when compared with lower-resolution instruments.

The available HST and JWST fields in L1688 overlap only in the regions surrounding Oph~S1. The oldest archival HST F160W-band image was taken on September 11, 2016, with a total exposure time of 1596.9~s (\texttt{hst\_14181\_0a\_wfc3\_ir\_f160w}, program ID:~14181). In the case of JWST, the F200W-band images of the same region were taken between March~7 and April~6, 2023, as part of the JWST Cycle~1 outreach campaign (total exposure time of 2125.8~s, program ID: DD~2739). Therefore, the archival images presently provide a time baseline of 6.5~yr, which should result in an offset in the ISM positions of about 1.9~px, based on the pixel scale of 90~mas\,px$^{-1}$ in the HST/WFC3 image and a YSO-based proper motion estimate of 26.8~mas\,yr$^{-1}$ for L1688 \citep{Zhang2023}.

We note that some point sources and ISM structures seen in the JWST images show duplication artefacts (see Appendix~\ref{section:A}) that we identify to be a result of a problem in the tile stacking process. The affected tiles belong to the observation number 9, visits 5 and 6. We downloaded the calibrated level~2 tiles for this observation and stacked the tiles for each of the two visits separately using the JWST \texttt{Image3Pipeline}. The resulting images (visit~5, visit~6, remainder of the JWST Ophiuchus images) were co-added using \texttt{SWarp} \citep[][version 2.42.0]{SWarp} to form an image free of the above mentioned artefacts. We note that some parts of the co-added image still show artefacts related to uneven flux distribution, but these appear to be localised to the vicinity of GSS~37, are also present in the published version of the image, and do not affect imaging of ISM features targeted by this study.

\section{Calibration of astrometric solution}\label{section:3}

Providing an accurate astrometric solution for images is vital for all astrometric studies. While HST and JWST level~3 data products are treated for distortions, rotations, and pointing offsets \citep[e.g.][]{Bellini2009,Griggio2023}, significant offsets ($\lesssim100$~mas) from standard coordinate systems, such as that of {\it Gaia}, can be identified in images from different epochs. Optimally, the astrometric solution of an image should be corrected by using information from {\it Gaia} \citep{Griggio2023}. However, not all observed fields contain enough {\it Gaia} sources to allow for such a correction, especially within high-extinction regions such as L1688.

In this study, we circumvent this issue by making use of relative astrometry. Several galaxies appear in the HST and JWST images of the Oph~core region, which makes them the perfect candidates for reference targets due to their expected small proper motion. We measured the positions of these galaxies using \texttt{DAOStarFinder} from \texttt{photutils} and present them in Table~\ref{table:A} (see Appendix~\ref{section:A} for further information). The median offset ($\Delta_{\textrm{RA}}=-52.4$~mas, $\Delta_{\textrm{DEC}}=+109.6$~mas) was used to align images in a common frame of reference. Finally, since the JWST image is of the higher resolution, we used \texttt{reproject} to match its pixel scale and image shape with HST. These two procedures represent the initial pre-processing of the image data.

\begin{figure*}
 \centering
 \includegraphics[width=\textwidth]{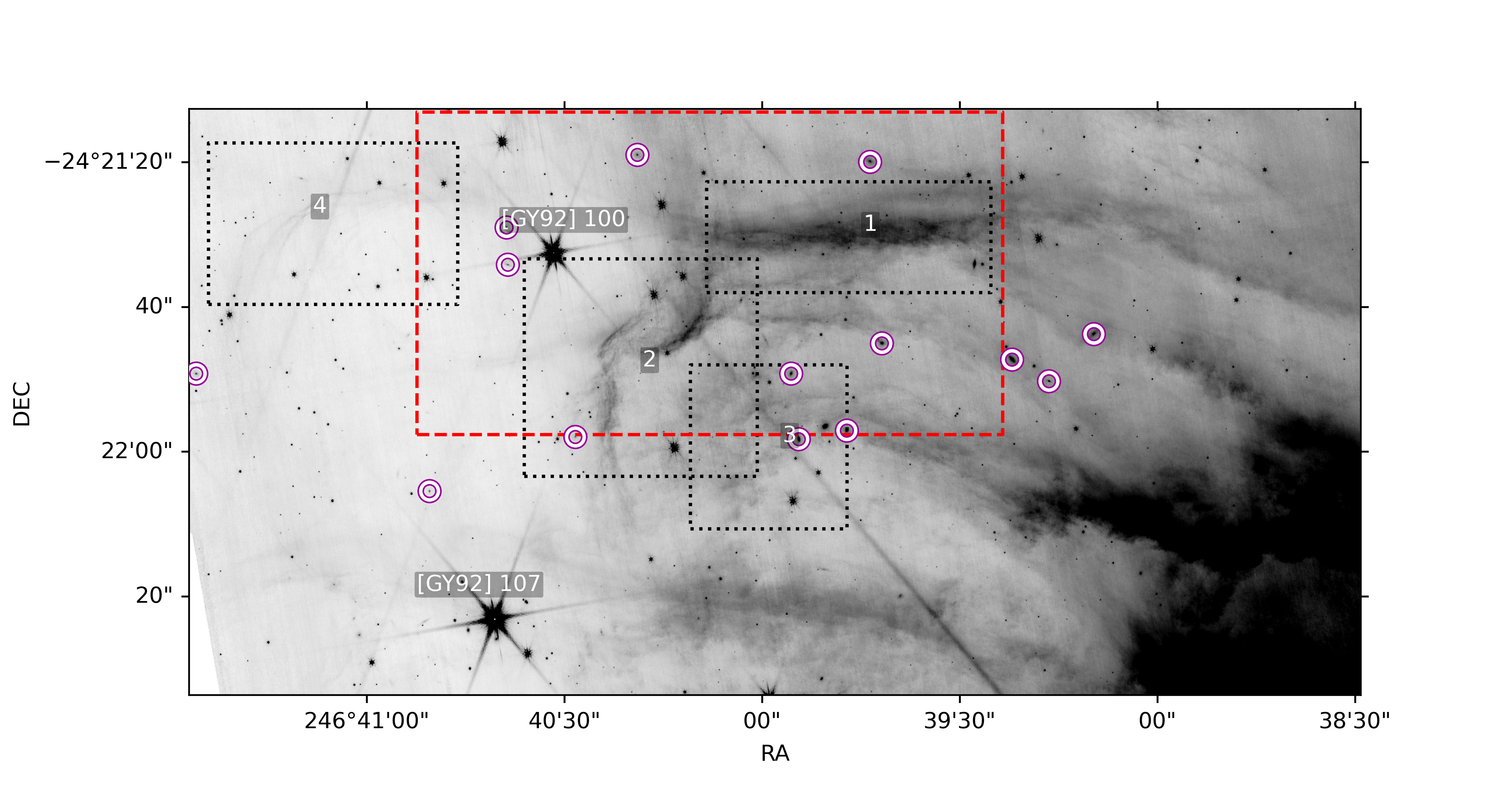}
 \caption{JWST (F200W, negative) map of the investigated region in L1688, between GSS~37 (northern bright diffraction spike) and Oph~S1 (southern bright diffraction spike). We aim to measure the ISM proper motion within the region contained by the dashed red rectangle. The individual sub-fields used for navigating the region are labelled and highlighted by dotted rectangles. Positions of the galaxies used for matching the HST and JWST reference frames are emphasised using the white circles with magenta outlines (four additional galaxies located to the north). The two brightest stars within the image are labelled.}
 \label{fig:infomap}
\end{figure*}


The standard deviations, or residuals, of the offsets ($\sigma_{\textrm{RA}}=14.1$~mas, $\sigma_{\textrm{DEC}}=19.4$~mas) highlight measurement errors and possible residual distortions. In practice, the distribution will likely be affected by a combination of both effects, and the resulting noise-based error is expected to be $\lesssim3$~mas\,yr$^{-1}$ ($\lesssim2$~km\,s$^{-1}$) when considering the standard deviations and the baseline of 6.5~yr. We also find weak indications of spatially correlated residuals: marginally significant correlations are present between $\Delta_{\textrm{RA}}$ and $\Delta_{\textrm{DEC}}$ ($+0.519$), and between RA and $\Delta_{\textrm{DEC}}$ ($-0.618$), when compared with the two-tailed test at $\alpha=0.05$ significance level (0.468). Furthermore, while the distribution of $\Delta_{\textrm{RA}}$ is approximately Gaussian, the distribution of $\Delta_{\textrm{DEC}}$ is asymmetric and possibly bimodal (Fig.~\ref{fig:corrs}). These features may indicate small residual systematic effects in at least one of the astrometric solutions, which we do not attempt to correct for given the limited number of reference galaxies. Their possible contribution to the overall systematic uncertainty is discussed in Sect.~\ref{section:5.2} and Appendix~\ref{section:A}.

\section{Proper motion measurements}\label{section:4}

\begin{table}
\caption{List of image sub-fields used for navigating the investigated region in L1688.}\label{table:1}
    \begin{center}
    \begin{tabular}{r|rr}
    \hline
    \hline
    ID & RA [deg] & DEC [deg] \\
    \hline
    Oph~ISM~1 & 246.6624 & $-24.3582$ \\
    Oph~ISM~2 & 246.6717 & $-24.3634$ \\
    Oph~ISM~3 & 246.6658 & $-24.3663$ \\
    Oph~ISM~4 & 246.6856 & $-24.3575$ \\
    \hline
    \end{tabular}
    \end{center}
\end{table}

\subsection{Stellar objects}\label{section:4.1}

One of the primary aims of this work is to compare stellar and ISM proper motions, which requires a compilation of stellar populations in the studied region. \citet{Grasser2021} proposed the existence of two distinct stellar populations in Ophiuchus, Pop~1 and Pop~2, with the latter being up to 10~Myr older. Very similar results were obtained by \citet{Ratzenbock2023,SigMA} who confirmed the existence of these populations and associated them with $\rho$~Oph/L1688 ($\textrm{age}\approx4$~Myr) and Antares ($\textrm{age}\approx10$~Myr) populations, respectively. We make use of the list of members from \citet{Grasser2021} and focus on the stars for which the authors note signs of infrared excess. These stars should be embedded in the gas and represent the best proxy among all cluster members for the ISM kinematics. We also include astrometric information for Oph~S1 and HD~147889 \citep[{\it Gaia}~DR3,][]{GaiaDR3} since they represent the primary source of radiation in the region, and both were identified as members of the L1688 population by \citet{SigMA}. We are also interested in the proper motions of several massive stars in the region, which are provided by {\it Hipparcos} \citep[][$\sigma$~Sco and $\rho$~Oph]{Hipparcos2007} and {\it Gaia} ($i$~Sco).

\citet{Hutschenreuter2026} presented the first 3D velocity field model of Sco-Cen. In their analysis, the authors note that the USco population cannot be described by the same velocity field as the rest of Sco-Cen. Instead, they include a second field-component in their fit, independently identifying the same two stellar populations that were originally presented by \citet{Grasser2021}. In our analysis, these two fields are used to represent the general stellar motion in Ophiuchus that is not necessarily tied to embedded YSOs.

\begin{figure*}
 \centering
 \includegraphics[width=\textwidth]{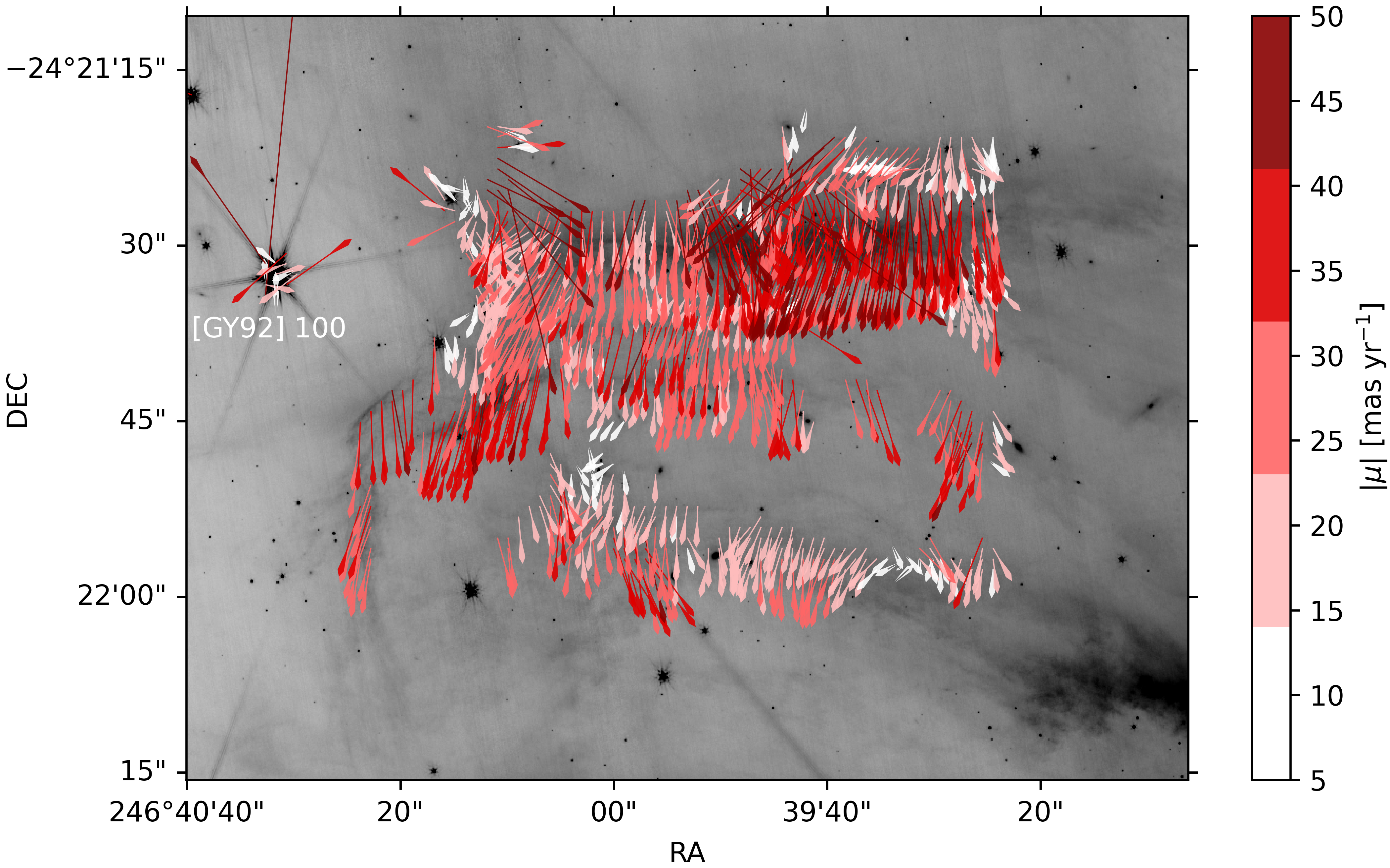}
 \caption{Filtered proper motion map of an ISM field in L1688, between GSS~37 (north-east) and Oph~S1 (south-west). The background gray-scale map represents the JWST (F200W, negative) image of this region. Both the arrow lengths and their colours highlight the magnitude of the motion. The displayed map roughly coincides with the red rectangle in Fig.~\ref{fig:infomap}, with only a minor cut in the eastern section where no proper motion measurements remain after applying our filter.}
 \label{fig:velmap}
\end{figure*}

Furthermore, we make use of the HST and JWST images to extract positions of detectable stellar sources using point-spread-function fitting in IRAF \citep[for further description, see][]{Rottensteiner2026}. Stellar proper motions (Appendix~\ref{section:B}) were estimated using relative position offsets. Although these calculations are based on only two epochs, the distribution of the measurements should provide sufficient information about the kinematics of these stars. We note that none of the stars presented in Table~\ref{table:B} were included in the {\it Gaia}~DR3 catalogue, and that two of our measured stars are kinematically associated with the young USco stellar population. While three {\it Gaia} stars ([GY92]~100 and the two stars of the binary system GSS~37) are present in the images, they are too bright to offer reliable proper motion measurements with IRAF. Our registration-based estimate of the motion of these two stars (see Appendix~\ref{section:B}) agrees with that from {\it Gaia}, with calculated mean proper motion differences of only $(\Delta\mu_{\alpha^*},\Delta\mu_\delta) = (-1.1,-1.0)$~mas\,yr$^{-1}$. These differences are consistent with the proper motion error estimate based on the residuals of the galaxy positions (see Sect.~\ref{section:3}) but remain uncorrected for the parallax effect (discussed in Sect.~\ref{section:4.3}).

\subsection{ISM -- proper motion field map}\label{section:4.2}

Unlike in the case of stars, which can be described as point sources, morphological structures of the ISM tend to be complex and irregular, leading to difficulties in position measurements. This suggests that standard methods for proper motion measurements, which rely on fitting on-sky positions with an on-sky motion curve, will generally not be feasible. Image registration procedures do not require the assignment of a single position and provide offset measurements through image alignment by matching larger parts of images. In our study, we focus on intensity-based image registration, variations of which have also been applied to protostellar outflows \citep[e.g.,][]{Lopez2022,Hodapp2026}.

Our proper motion measurements utilise image registration tools from SimpleITK \citep{ITK,SimpleITK1,SimpleITK2,SimpleITK3}. Using the same setup as described by \citet{CrA_motion}, the moving (target) and the fixed (reference) images are aligned by optimising the correlation metric with a regular-step gradient descent (learning rate 1.0, minimum step size $10^{-5}$, 300 iterations). The transformation used to align the images is a simple 2D translation. This procedure ensures sub-pixel alignment, which is necessary for accurate proper motion measurements. Our exact use of these procedures is explained in Appendix~\ref{section:C1}. Similar to \citet{CrA_motion}, we mask the image regions outside the chosen measurement window. There is no ideal way of measuring only targeted ISM features and it is necessary to keep in mind that overlaps with unrelated features (emitting, reflecting, or absorbing) might affect our interpretation of the registration results.

In the images of L1688, we selected a primary sub-field of interest defined in the HST image using $X_{\textrm{min}}=550$~px, $X_{\textrm{max}}=1450$~px, $Y_{\textrm{min}}=700$~px, and $Y_{\textrm{max}}=1375$~px. This sub-field is highlighted in Fig.~\ref{fig:infomap} by the dashed red rectangle. To aid navigation across the region, we also defined several smaller sub-fields containing easily identifiable ISM features with different morphologies (central coordinates in Table~\ref{table:1}). We utilised a sliding window approach with a small window of size $60\times60$~px$^2$, scanning the primary sub-field on a grid with a step of $\Delta X = \Delta Y = 10$~px. For each of these scans, we obtained a proper motion estimate using SimpleITK, resulting in a grid of proper motion vectors akin to a velocity field map.

In order to estimate measurement uncertainties with SimpleITK, we sampled the analysed images. In addition to flux sampling from the image noise \citep{CrA_motion}, we make use of a jittering window approach to account for possible differences in the registration performance due to window shape and size. The positions of window edges are drawn from a uniform random distribution centred on the pre-determined window shape with offsets of up to $\pm3$~px. Considering a $60\times60$~px$^2$ window, this results in an area change of $\pm10$\% relative to the original window size. This should be sufficient to capture the effect of the choice of window. For the individual measurements provided by the sliding-window strategy, the numerical errors tend to be $\lesssim1$~mas\,yr$^{-1}$. Additional details regarding our sampling strategy can be found in Appendix~\ref{section:C2}.

Since we are primarily interested in the brighter regions of the ISM features, as their higher signal-to-noise ratios provide more precise measurements of the motion, we applied a local brightness filter that was designed to take all pixels within a 5~px radius of the sliding window centre, determine the median intensities, and include in our analysis only measurements with filter values above the 75th percentile of pixel intensities of the whole primary sub-field (Fig.~\ref{fig:infomap}, red dashed rectangle). The filtered proper motion map is presented in Fig.~\ref{fig:velmap}. Some parts of the unfiltered map show either extreme motion or small-scale motion variations. We argue that these originate from:
\begin{itemize}
    \item The presence of stellar halos due to imperfections in stellar masking
    \item The presence of artefacts, primarily diffraction spikes from outlying sources in the JWST image
    \item The absence of a strong ISM intensity gradient relative to the S/N, resulting in a lack of motion information
\end{itemize}
The above-described filter eliminates many of these effects, resulting in a more robust proper motion map. The remaining unphysical measurements should be statistically insignificant. See Appendix~\ref{section:D} for a discussion about the impact of the velocity-map filter.

The stars present in the images can severely affect image registration (Fig.~\ref{fig:mask}) since they represent sharp brightness gradients. In SimpleITK, a single mask image can be provided to offer users the ability to exclude certain pixels from being analysed. We used \texttt{DAOStarFinder} to detect sources and created flux-based circular masks with sizes large enough to capture the stellar proper motions. The individual masks were then combined into a single mask image that serves as an input for SimpleITK for each proper motion measurement. For a further description of the masking method, see Appendix~\ref{section:E}.

The results of our application of SimpleITK on the individual ISM structures are presented in Sect.~\ref{section:5.1}. We note that the individual measurement uncertainties do not account for potential systematic errors, but these can be picked out in the proper motion map. A graphical demonstration of the motion is presented in Appendix~\ref{section:C3} (Fig.~\ref{fig:demo}) and in our online repository\footnote{\href{https://github.com/mpiecka/ISM-Proper-Motions-Ophiuchus}{https://github.com/mpiecka/ISM-Proper-Motions-Ophiuchus}} (image alignment based on \texttt{image\_registration.fft\_tools.shift} and the derived offset from image registration).

\subsection{Systematics from parallaxes}\label{section:4.3}

Due to the relatively longer time baseline in Ophiuchus, the effect of parallax on the extracted proper motions should be slightly smaller than what was suggested in Corona~Australis \citep[$\lesssim1.5$~mas\,yr$^{-1}$,][]{CrA_motion}. While small, this effect introduces systematics to our proper motion measurements. These are impossible to treat for individual stellar motions based on only two epochs, but we can provide a correction term for the ISM proper motion by making use of the known distance of L1688.

\begin{figure}
 \centering
 \includegraphics[width=\columnwidth]{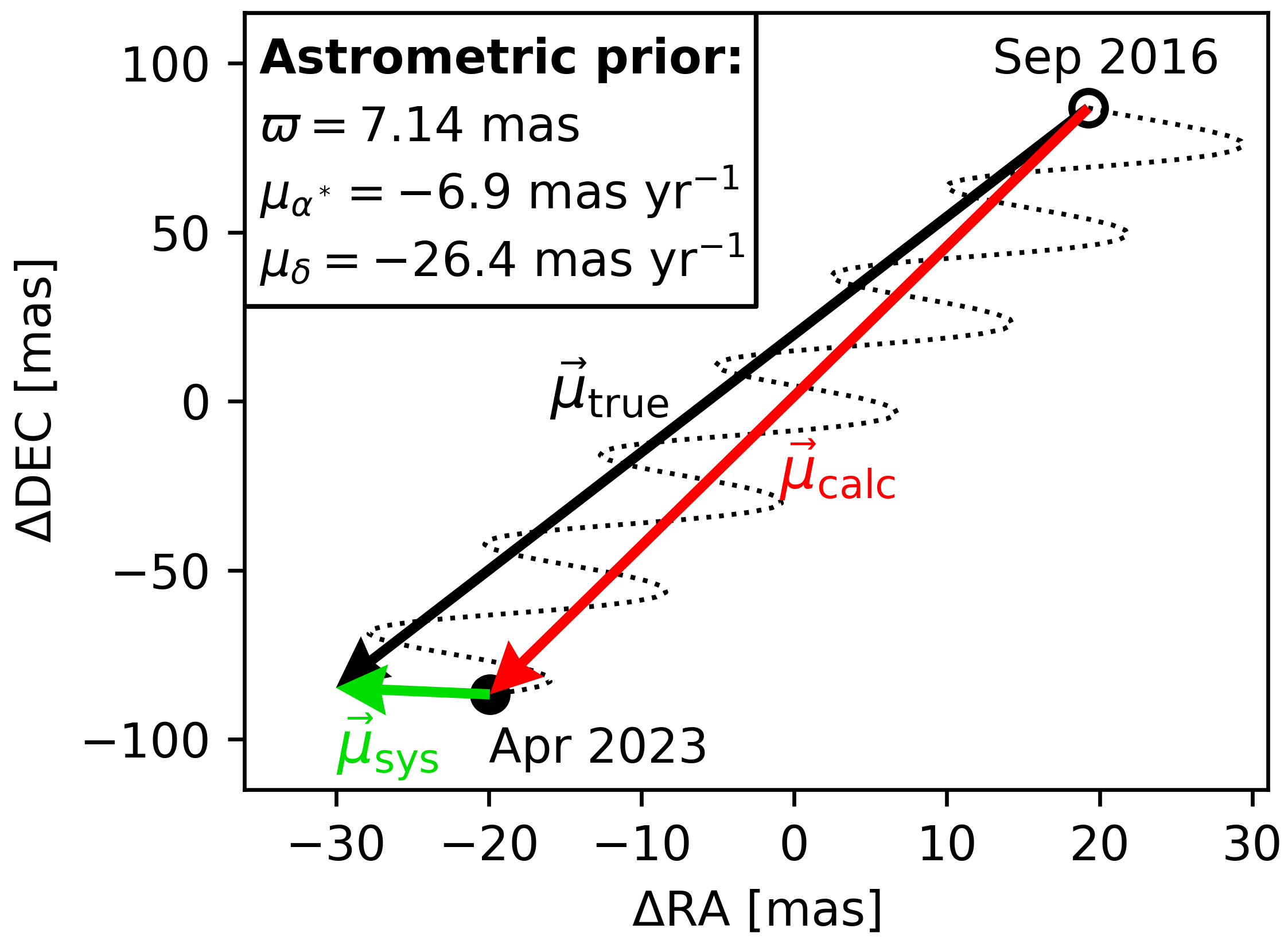}
 \caption{Theoretical model of the ISM on-sky motion in the vicinity of Oph~S1, with the distance and proper motion priors being used to evaluate the correction vector ($\vec{\mu}_{\textrm{sys}}$) resulting from the parallax effect.}
 \label{fig:plxeffect}
\end{figure}

Noting that the orbits of the spacecrafts have negligible effect on the measurements, we simulated the geocentric on-sky orbit of the cloud using a prior model consisting of the proper motion of the young USco stellar population and the distance of the cloud \citep[140~pc, e.g.][]{Zucker2020}. We make use of the epochs specified in Sect.~\ref{section:2} and calculate the barycentric on-sky motion using \texttt{astropy}'s \texttt{SkyCoord.apply\_space\_motion}. To evaluate the motion in the geocentric reference frame, we subtract Earth's barycentric motion utilising \texttt{astropy}'s \texttt{get\_body\_barycentric} and the Cartesian vectors
\begin{equation*}
\vec{r}_{\,\textrm{target}}^{\,\textrm{bary}} = \vec{r}_{\,\textrm{target}}^{\,\textrm{geo}} + \vec{r}_{\,\textrm{Earth}}^{\,\textrm{bary}} \,\,.
\end{equation*}
In the simulation we assume that the resulting orbit is a result of true astrometric parameters. On the other hand, our calculated proper motions ($\vec{\mu}_{\textrm{calc}}$) are based only on the offset between the initial and the final epoch in the simulation. To estimate the systematic error of such a procedure due to the parallax effect, we calculate
\begin{equation*}
\vec{\mu}_{\textrm{sys}} = \vec{\mu}_{\textrm{true}} - \vec{\mu}_{\textrm{calc}} = (-1.4, +0.3) \,\textrm{mas\,yr}^{-1} \,\,,
\end{equation*}
using the cloud's distance prior of 140~pc ($\varpi\approx7.14$~mas) and the median proper motion of the young USco stellar population, $(\mu_{\alpha^*},\mu_{\delta})^{\textrm{USco}}=(-6.9,-26.4)$~mas\,yr$^{-1}$. Unless stated otherwise, we used $\vec{\mu}_{\textrm{sys}}$ as a correction for the parallax effect in our measured ISM proper motions (see Fig.~\ref{fig:plxeffect} for visualisation).

The estimated correction vector is almost independent of $\vec{\mu}_{\textrm{true}}$. Given the small size of the analysed ISM proper motion field, the correction vector mainly relies on the assumed distance of the cloud. Considering a conservative distance Gaussian uncertainty of 10~pc and proper motion Gaussian uncertainties of 1~mas\,yr$^{-1}$, we sample the prior astrometric parameters using the Monte Carlo approach and find that the correction vector terms vary by $(\pm0.10, \pm0.02)$~mas\,yr$^{-1}$, based on the standard deviations of the posterior distribution. These uncertainties are smaller that those resulting from the astrometric solution, suggesting that errors in the correction vector are negligible.

\section{Results and discussion}\label{section:5}

\subsection{L1688 proper motion}\label{section:5.1}

The proper motion distribution inside the studied region should be more informative about the kinematics of the imaged ISM than the individual sliding-window measurements. We derive a proper motion estimate for the typical local motion of L1688 (i.e. the characteristic motion of the ISM within the investigated region near the head of the cloud) by making use of the median of the proper motion velocity map (Fig.~\ref{fig:velmap}) and accounting for the parallax effect. The median was selected as the measure of central tendency because it is robust to outliers and therefore less sensitive to extreme observations than, for example, the mean. The estimated motion of the cloud is found to be $(\mu_{\alpha^*},\mu_\delta)^{\textrm{ISM}}_{\textrm{median}}=(-3.0\pm1.1, -24.1\pm1.7)$ mas\,yr$^{-1}$, with the uncertainties based on the standard error on the median computed from the median absolute deviations $(\mu_{\alpha^*},\mu_\delta)^{\textrm{ISM}}_{\textrm{MAD}}=(5.0, 7.2)$ mas\,yr$^{-1}$ and assuming the ratio of the total number of sliding-window scans and the effective number of scans $\frac{N}{N_{\textrm{eff}}}\approx31.5$, where $N_{\textrm{eff}}=\frac{\textrm{image area}}{\textrm{scan area}}$. We note that the uncertainty estimate may also be affected by a turbulence term, which cannot presently be quantified because only two epochs are available. See Sect.~\ref{section:5.5} for a further discussion about the presented measurement uncertainties.

As can be clearly seen in the proper motion field map of L1688 (Fig.~\ref{fig:velmap}), the measured positional offsets between the two investigated epochs vary across the studied region. This results in proper motion gradients. In the case of the Oph~ISM~1 feature ($246^{\circ} 39\arcmin 25\arcsec \lesssim\textrm{RA}\lesssim 246^{\circ} 40\arcmin 10\arcsec$, $\textrm{DEC} \approx -24^{\circ} 21\arcmin 30\arcsec$), we notice a sharp proper motion gradient of $\sim10$~mas\,yr$^{-1}$ (along right ascension, changes mainly in $\mu_\delta$) over an angular distance of 40\arcsec, which can be converted to $\approx240$~km\,s$^{-1}$\,pc$^{-1}$, assuming Oph~ISM~1 is not significantly extended in the radial direction. There is presently no observational evidence that would support the existence of any source capable of driving such a strong gradient across a single ISM feature in L1688. By aligning the epoch-images using the median of the ISM motion, we find that the western, slightly brighter, section of Oph~ISM~1 shows residual motion along negative declination, as suggested by the measurement. This apparent residual motion seems to be the result of a shift in the intensity distribution but not in 2D geometry. Similar residuals can also be identified in Oph~ISM~3 (see our online repository). The most likely source of these systematics is the use of the different filters (F160W and F200W) in our analysis. In the future, this issue can be overcome by utilising identical or at least strongly overlapping filters. However, we do not rule out the possibility of features with on-sky overlapping located at different different distances that can display different kinematic behaviour.

\subsection{ISM vs stellar kinematics}\label{section:5.2}

\begin{figure*}
 \centering
 \includegraphics[width=\textwidth]{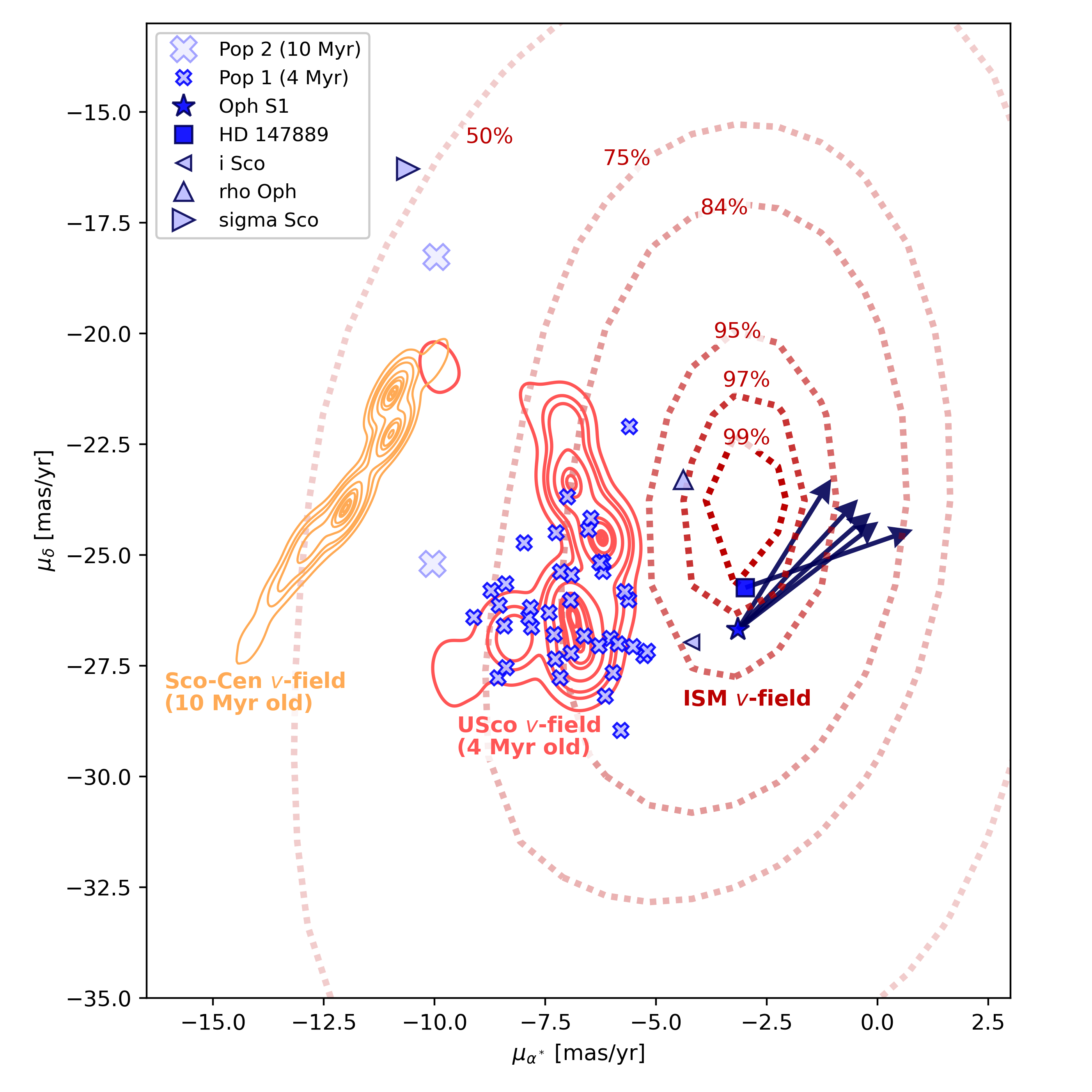}
 \caption{Comparison of YSO and ISM proper motions. The blue crosses correspond to YSOs for which \citet{Grasser2021} noted infrared excess and with proper motion component errors $<0.5$~mas\,yr$^{-1}$. The noted ages represent means of the isochrone-based values published by \citet{Ratzenbock2023}. The dark-blue star and square symbols highlight {\it Gaia} astrometry for the feedback-dominant stars in the region, with the arrows noting the expected direction of acceleration from Oph~S1 and HD~147889 to the ISM structures due to feedback. The solid line contours represent velocity fields from \citet{Hutschenreuter2026} that should coincide with the YSO kinematics. The dotted contours highlight 50th, 75th, 84th, 95th, 97th, and 99th percentile KDE density levels of the ISM proper motion map (corrected for the parallax effect) extracted from the brightest sections of the JWST image (Fig.~\ref{fig:velmap}).}
 \label{fig:main}
\end{figure*}

Our primary aim is to determine whether the classical assumption that the embedded stars represent a good kinematic proxy for the associated molecular cloud. We focus on the stellar populations in the vicinity of Oph~S1 (see Sect.~\ref{section:4.1}). Their proper motions are displayed in Fig.~\ref{fig:main}. There is significant motion of the ISM with respect to the young USco stars, resulting in a velocity difference of
\begin{equation*}
\Delta\mu_{\alpha^*}^{\textrm{ISM -- USco}} \approx+3.9 \,\, \textrm{mas}\,\textrm{yr}^{-1} \,\,,
\end{equation*}
\begin{equation*}
\Delta\mu_{\delta}^{\textrm{ISM -- USco}} \approx+2.3 \,\, \textrm{mas}\,\textrm{yr}^{-1} \,\,,
\end{equation*}
\begin{equation*}
\Delta v_{\textrm{t}}^{\textrm{ISM -- USco}} \approx 3.0 \pm 0.9 \,\, \textrm{km}\,\textrm{s}^{-1} \,\,,
\end{equation*}
with uncertainties being dominated by the presented measurements of the ISM kinematics (Sect.~\ref{section:5.1}). The motion of the ISM better resembles that of HD~147889 and Oph~S1, with a velocity difference
\begin{equation*}
\Delta\mu_{\alpha^*}^{\textrm{ISM} - \textrm{Oph S1}} \approx+0.2 \,\, \textrm{mas}\,\textrm{yr}^{-1} \,\,,
\end{equation*}
\begin{equation*}
\Delta\mu_{\delta}^{\textrm{ISM} - \textrm{Oph S1}} \approx+2.6 \,\, \textrm{mas}\,\textrm{yr}^{-1} \,\,,
\end{equation*}
\begin{equation*}
\Delta v_{\textrm{t}}^{\textrm{ISM} - \textrm{Oph S1}} \approx 1.7 \pm 1.1 \,\, \textrm{km}\,\textrm{s}^{-1} \,\,,
\end{equation*}
when compared to Oph~S1. Since the proper motion errors for the ISM and the stars are ranging between 0.1--1.7~mas\,yr$^{-1}$ (0.07--1.33~km\,s$^{-1}$), we note that the difference in the declination component is statistically significant but close to the measurement accuracy limits, and marginally significant if adopting MAD as the uncertainty estimator ($\Delta v_{\textrm{t}} \approx 3 \pm 4$~km\,s$^{-1}$). In the future, these results could be verified by including additional epochs using HST, JWST, or other high-resolution space telescopes.

The above mentioned differences do not account for all systematic effects that can affect both our and {\it Gaia} measurements. In our case, the systematics remain largely unknown due the the limitations of two-epoch observations, but they are unlikely to exceed an error of 3~mas\,yr$^{-1}$ (Sect.~\ref{section:3}). As for the {\it Gaia} measurements, the USco population of stars is large enough for the median of the distribution to constrain any significant systematic effects. On the other hand, Oph~S1 is a known binary system with a sub-solar-mass companion and a 1.7-yr orbital period \citep{Ordonez2025}. Given the 2.8-yr long time baseline of {\it Gaia}~DR3, it is possible that the provided astrometry is affected by systematics originating from the unmodelled binary nature of the star. By using the astrometric information from \citet{Ordonez2025}, we find
\begin{equation*}
\Delta\mu_{\alpha^*}^{\textrm{ISM} - \textrm{Oph S1, VLBA}} \approx-0.5 \,\, \textrm{mas}\,\textrm{yr}^{-1} \,\,,
\end{equation*}
\begin{equation*}
\Delta\mu_{\delta}^{\textrm{ISM} - \textrm{Oph S1, VLBA}} \approx+2.7 \,\, \textrm{mas}\,\textrm{yr}^{-1} \,\,,
\end{equation*}
\begin{equation*}
\Delta v_{\textrm{t}}^{\textrm{ISM} - \textrm{Oph S1, VLBA}} \approx 1.8 \pm 1.1 \,\, \textrm{km}\,\textrm{s}^{-1} \,\,,
\end{equation*}
which does not differ significantly from the results obtained with {\it Gaia} astrometry.

\subsection{Role of stellar feedback}\label{section:5.3}

Stellar feedback (radiation, winds, supernovae) is capable of accelerating the ISM in the vicinity of the driving sources. We usually lack the information about the proper motion of the ISM, which makes the measurements presented in this work key for characterising the impact of feedback on ISM through observations. This requires a careful identification of the primary feedback sources in the studied region. Since our ISM measurement focuses on a single region and is based on only two epochs, the available data constrain the level of detail of our feedback discussion. We focus mainly on Oph~S1 and HD~147889, which are believed to be the primary local sources of feedback \citep{Howard2021}. Among the more massive USco stars in the vicinity of L1688 are $\sigma$~Sco, $i$~Sco, and $\rho$~Oph.

Let us briefly consider individual dynamical forces that can affect kinematics of the ISM, starting with the radiation fluxes from the massive stars surrounding our studied region. This requires assumptions about the spectral type of the stars. Based on the mass obtained by \citet{Ordonez2025} and the tables from \citet{Eker2018}, Oph~S1 is assumed to be a B6V star (with a sub-solar companion), although the spectral class of this star is still disputed \citep[for example, compare with][]{Mookerjea2021} and the chosen classification should be taken as the lower-mass limit. HD~147889 is noted to be a binary B2IV/V+B2IV/V star, based on its the behaviour of the observed spectral lines \citep{Haffner1995}. $i$-Sco is identified as a B2.5V star with no confirmed companion \citep{Negueruela2024}. \citet{MaizApellaniz2021} investigated several massive stars and found that $\sigma$~Sco is composed of two unresolved components of spectral types B1III and B1V, including a third lower-mass component ($\sigma$~Sco~B) that we chose to omit from our analysis. The classification of $\rho$~Oph is significantly more complex \citep[e.g.,][]{Cordiner2013}. The primary component is a binary star of a spectral type B2/3V+B2V. Two more stars can be found in the vicinity of the primary component, HD~147932 and HD~147888, with spectral types of B5V and B3V, respectively. In addition to the spectral types, we use angular distances and the assumed distance of L1688 (140~pc) to derive the projected distances $\vec{d}_{\textrm{t}}$ of the ISM from all of the mentioned stars.

\begin{figure}
 \centering
 \includegraphics[width=\columnwidth]{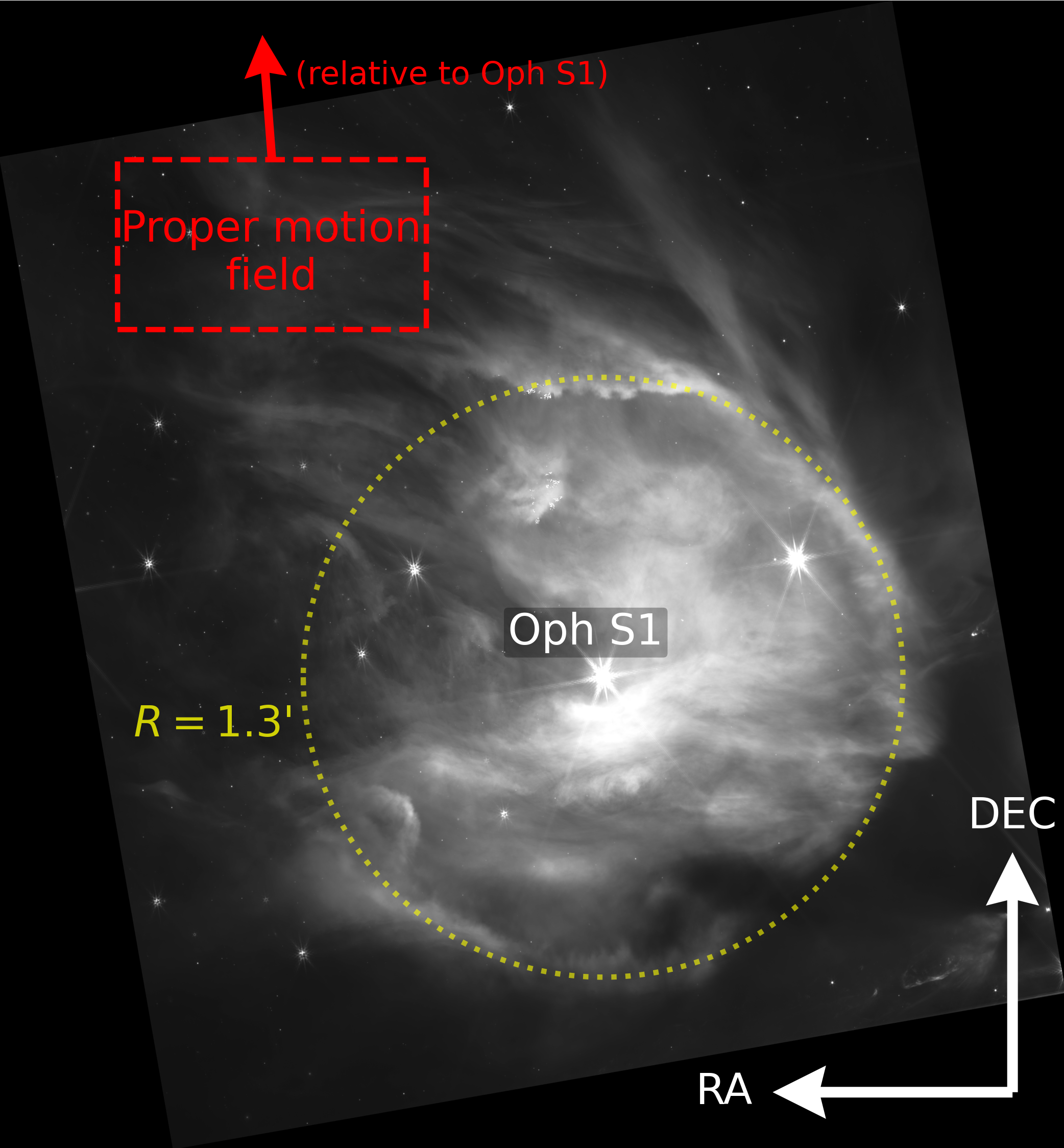}
 \caption{JWST (F444W) image of a cavity in L1688 formed by feedback from Oph~S1. The investigated proper motion field is located to the north-east within the red rectangle -- the red arrow displays the motion of this field relative to Oph~S1. The absence of an ISM wall can be noted towards east and north-east. The head of L1688 is located roughly in the south-west direction.}
 \label{fig:ophS1}
\end{figure}

We use the tables from \citet{Eker2018} to estimate stellar luminosities relative to the Sun and convert them to fluxes $(F_{\alpha^*}, F_{\delta})$ at the location of the ISM
\begin{equation*}
\vec{F} = \frac{L \vec{d}_{\textrm{t}}}{4 \pi |d_{\textrm{t}}|^3} \,\,.
\end{equation*}
By comparing the fluxes relative to Oph~S1, we find that Oph~S1 is the dominating source of radiation in the region, closely followed by HD~147889 (flux ratio of $\sim0.48$). The contribution from the remaining stars is insignificant in the studied region (combined flux ratio of $\sim0.04$ relative to Oph~S1). We note that the flux ratios may vary by approximately $0.05 |\vec{F}|_{\textrm{Oph~S1}}$, depending on the choice of the spectral classes. By summing over all of the flux vectors and normalising by the absolute flux from Oph~S1, we predict a radiatively driven motion in the direction $(F_{\alpha^*}, F_{\delta}) \approx (+1.06, +0.95) |\vec{F}|_{\textrm{Oph~S1}}$, from Oph~S1 towards the missing wall of the surrounding cavity. We note that we neglected interstellar extinction between the stellar sources and the ISM, which is an assumption that likely holds in the near vicinity of Oph~S1, but can significantly reduce the UV flux from the other stars. This further supports Oph~S1 as the primary radiation source in the studied region. If we compare the proper motion direction relative to Oph~S1 with the outwards direction from Oph~S1 towards the ISM, we find that the two vectors are misaligned in the plane of the sky ($\sim 30^{\circ}$), but the importance of this result depends on the presently unknown 3D configuration of the star and the ISM features in the studied field.

The direct radiation pressure generated by the stars is likely negligible compared to the pressure produced by the associated ionisation and photodissociation regions \citep[e.g.,][]{Krumholz2009,Matzner2015}. The environment surrounding Oph~S1 was previously studied by \citet{Mookerjea2021} using multiple tracers of ionised, photo-dissociated, and neutral medium. The ionised gas appears to extend up 0.8\arcmin away from the star with the outer parts being dominated by a photo-dissociated gas. The authors concluded that the photodissociation region surrounding the star appears to be freely expanding in the north-east direction, which coincides with our investigated ISM field and the broken cavity wall. For this reason, we highlight the need of future observational and theoretical examinations of the radiation processes generated by Oph~S1 that affect kinematics of the surrounding ISM.

In the case of the stellar winds, we note that the theoretical models presented by \citet{Krticka2014} suggest that the winds from Oph~S1 are expected to only have a second-order effect. The winds from the more massive stars, especially HD~147889, $\sigma$~Sco, and $\rho$~Oph, should be the dominating sources of this feedback process and could be quantified in future studies using numerical simulations. However, early B-type stars should have winds that are less important in momentum transport when compared with the stellar radiation \citep[e.g.,][]{Matzner2015}.

Lastly, let us consider a supernova-driven acceleration of ISM. \citet{Neuhauser2020} showed that the most recent Sco-Cen supernova ($\sim 2$~Myr ago) occurred in the vicinity of USco, with the estimated present-day coordinates of the supernova position being $(l,b)_{\textrm{SN}}=(-16\pm4, +15\pm3)^{\circ}$. By utilising the available supernova progenitor information and uncertainties together with the Monte~Carlo sampling technique, we were able to generate possible supernova locations and ISM proper motions. We estimated the alignment between ISM proper motion vectors and the directions away from the predicted supernova position towards our ISM field. Motivated by the larger-scale nature of this feedback process, we take the proper motions relative to the USco population, which was largely born prior to the considered supernova event. The alignment is defined by an angular separation of $\Delta\theta \leq 1^{\circ}$. Using this criterion, we find an alignment probability of about 88\% (71\%, if using the median absolute deviation). A similar analysis using Oph~S1 as the kinematic reference frame yields lower but still significant alignment probabilities ($>50$\%). While the exact momentum budget of a supernova-driven process depends on the ISM density distribution and other unknown parameters, it is generally believed that such events are capable of driving flows affecting shapes of molecular clouds, particularly in Sco-Cen \citep[e.g.,][]{Alves2025,Alves2026}. Nevertheless, we acknowledge that our analysis remains limited to the projected plane-of-the-sky information, and therefore cannot uniquely establish a physical association without additional constraints on the 3D configuration and kinematics of the ISM.

Within the limitations of the available data, we conclude that the pressure generated by the photodissociation region surrounding Oph~S1 together with the recent Sco-Cen supernova event are the most likely contributors in the evolution of ISM kinematics within the investigated field. Particularly, Oph~S1 shows hints of ISM clearing from its vicinity when looking at the multi-wavelength images from JWST \citep[e.g.,][]{Nakamura2025}. The star appears to have created a nearly circular cavity in its surrounding (Fig.~\ref{fig:ophS1}) with a radius of $\approx1.3\arcmin$ and the structure of the wall being interrupted in the north-east direction, towards which also lies the ISM investigated in this work. Based on the described morphological structure of this cavity, we propose that the ISM studied in this work appears to have been accelerated away from the star at some point in the past.

\begin{figure*}
 \centering
 \includegraphics[width=\textwidth]{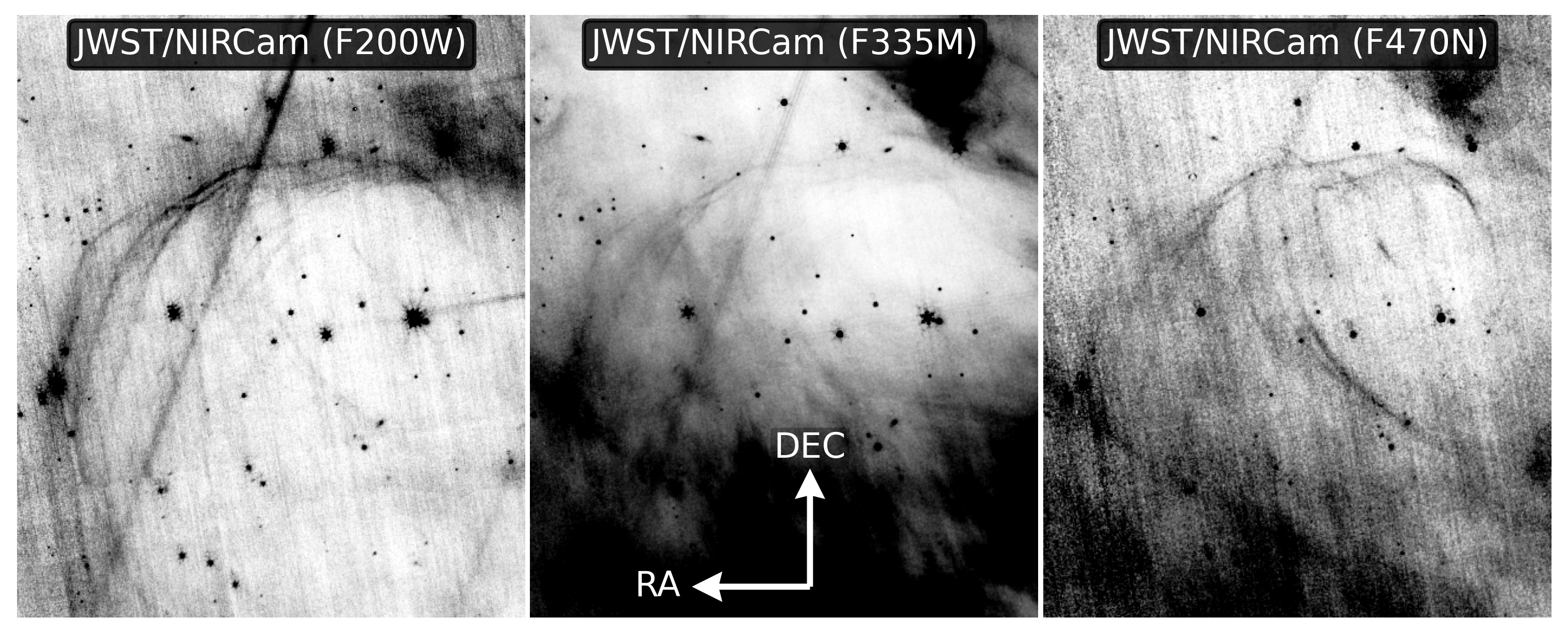}
 \caption{Negative images of the Oph~ISM~4 structure in three different JWST filters (program ID: DD~2739) reprojected using WCS from the HST image. Reprocessing mentioned in Sect.~\ref{section:2} was not applied to these images.}
 \label{fig:oph6}
\end{figure*}

\subsection{Thin shell-like structure in L1688}\label{section:5.4}

The shell-like structure of Oph~ISM~4 is displayed in Fig~\ref{fig:oph6}. This feature appears brightest relative to the surrounding ISM in the F200W filter, dimmest in the F335M filter, and shows visible differences in the intensity distribution when comparing with the F470N filter. Compared to the other features displayed in Fig.~\ref{fig:infomap}, the structure of Oph~ISM~4 is less diffuse and less intricate. The small width of the structure walls ranges from about $\approx 400$~mas to $\approx 750$~mas, corresponding to spatial scales of 55--105~AU at a distance of 140~pc. Furthermore, we note that the geometry of Oph~ISM~4 might be suggestive of an origin in the direction of Oph~S1, which needs to be verified using kinematic information.

Because the proper motion field does not directly cover Oph~ISM~4, assigning the median field motion to this structure may not provide an accurate estimate of its kinematics. As an alternative, we obtained a separate measurement for the local proper motion of this HST sub-field, defined by coordinates $X_{\textrm{min}}=230$~px, $X_{\textrm{max}}=613$~px, $Y_{\textrm{min}}=900$~px, $Y_{\textrm{max}}=1148$~px, which coincides with the corresponding rectangle displayed in Fig.~\ref{fig:infomap}. It is clear from the measured motion $(\mu_{\alpha^*},\mu_\delta)^{\textrm{Oph~ISM~4}}=(-12.3\pm0.9, -23.8\pm0.7)$ mas\,yr$^{-1}$ that it behaves as an outlier when compared with the proper motion field (specifically in $\mu_{\alpha^*}$), and it better agrees with the motion of the older (Antares or Pop~2) population of stars (see Fig.~\ref{fig:main}). This measurement suggests a motion in the west or north-west direction
\begin{equation*}
\Delta\mu_{\alpha^*}^{\textrm{Oph ISM 6} - \textrm{Oph S1}} \approx-9.2 \,\, \textrm{mas}\,\textrm{yr}^{-1} \,\,,
\end{equation*}
\begin{equation*}
\Delta\mu_{\delta}^{\textrm{Oph ISM 6} - \textrm{Oph S1}} \approx+2.9 \,\, \textrm{mas}\,\textrm{yr}^{-1} \,\,,
\end{equation*}
\begin{equation*}
\Delta v_{\textrm{t}}^{\textrm{Oph ISM 6} - \textrm{Oph S1}} \approx 6.4 \,\, \textrm{km}\,\textrm{s}^{-1} \,\,.
\end{equation*}
By visually inspecting the quality of the alignments using this motion measurement and the median motion of the field, we find that a better alignment is achieved when using an offset vector derived from the latter. This is likely a result of the presence of a prominent diffraction spike from GSS~37 that likely negatively affected the accuracy of image registration. We therefore favour the median field proper motion over the feature-targeted measurement, but note that the kinematics of Oph~ISM~4 remain poorly constrained.

Based on its low velocity relative to Oph~S1 (or the USco population) and the absence of strong molecular hydrogen emission (as supported by JWST images), the present-day nature of this object is likely not that of a Herbig-Haro object. However, given the good image alignment when using the median of the proper motion field, we find that Oph~ISM~4 might have originated from the general direction of Oph~S1. Therefore, we do not rule out a past Herbig-Haro nature of the object. We note that a jet with an associated Herbig-Haro object is capable of puncturing through cavities, such as the one seen around Oph~S1 (Fig.~\ref{fig:ophS1}), which might also have dynamically shaped the kinematic structure of the ISM studied in this work. If this is the case, the kinematic age of Oph~ISM~4 is $\lesssim70$~kyr, assuming no significant acceleration or deceleration during most of its travel-time, which can serve as an age constraint for Oph~S1, if confirmed. In comparison, the dynamical timescales of Herbig-Haro objects are of the same order \citep[up to a few 10~kyr,][]{Reipurth2001}.

As an alternative explanation, we considered the resemblance between the morphology of this structure and that of the walls of some supernova remnants, such as those seen in the optical images of Vela~SNR or Cygnus~Loop. Given the nearly circular form of the walls and its small radius of 0.01~pc, one might expect the structure to be expanding. However, the analysed images from HST and JWST show no hints of a noticeable expansion -- on the contrary, the north-western section of the walls shows signs of an inward motion, but this might be a result of systematic effects, similar to those identified in Oph~ISM~1 and 3. The lack of a measurable expansion and the absence of other ISM indicators suggest that our object is not a supernova remnant.

Another possibility is that Oph~ISM~4 is actually located in the background of L1688 and is larger, which would explain the lack of expansion. Regardless of the measurement, the proper motion of Oph~ISM~4 confirms that the structure is kinematically and, at least somewhat, spatially associated with L1688. This rules out the proposed hypothesis of Oph~ISM~4 being a significantly more distant object.

We conclude that we cannot unambiguously identify the correct proper motion of the shell-like structure Oph~ISM~4, but its morphology and the better alignment with the median of the proper motion field favour an origin roughly in the direction of Oph~S1. If confirmed, this would be suggestive of the object's Herbig-Haro nature at some point in the past. Although its origin remains uncertain, the thin shell-like structure Oph~ISM~4 might represent an important feature for our understanding of ISM structure and dynamics, with available data suggesting approximate pressure equilibrium between its interior and exterior environments.

\subsection{Notes on measurement uncertainties}\label{section:5.5}

In Sect.~\ref{section:5.1} we derived the median (robust to outliers) of the distribution together with its median absolute deviation and the standard error on the median. We proposed that we consider the latter as our uncertainty estimate for the median. This is based on the fact that aligning the images with extreme values of the distribution-allowed offsets $\textrm{median} \pm \textrm{MAD}$ results in non-optimal image alignments that manifest as clear offset residuals when comparing the images between the two epochs in the regions with the most prominent ISM features. It follows that either different parts of the analysed field display different signs of motion or that in some sub-regions SimpleITK performed poorly (presently difficult to quantify). For this reason, median absolute deviation cannot be considered as a proper uncertainty estimate of the typical local motion of the cloud. However, this assumes that a measure of the central tendency is a proper measure of the typical local ISM motion (Fig.~\ref{fig:main}). With only two-epoch images available, we are unable to verify this assumption and conclude that the whole distribution of proper motions should be considered for deriving a proper uncertainty measure (such as quantile ranges) if the provided median turns out being a non-optimal measure for the typical local motion in future studies.

The ISM features present in the analysed region are visible through starlight scattering and gas emission. The former case was previously considered in Corona Australis by \citet{CrA_motion}. The paper found a motion consistency between different parts of the cloud, which is not expected if the motion is primarily a result of the relative star-ISM motion. The impact of this relative motion on the ISM proper motion measurement is inversely proportional to the relative distance between the two objects. In our case, the primary source of radiation is most likely Oph~S1, which is in the close vicinity of the ISM, preventing us from ruling out the impact on the ISM proper motion measurement. We note that the measured transverse velocity difference (Sect.~\ref{section:5.2}) and the time baseline of 6.5~yr should result in a relative distance change of only a few AU. This suggests that the difference in the illumination of the local ISM features should be minimal, likely within the estimated velocity uncertainty. Lastly, another physical source of systematics is the presence of light echoes -- although the available data cannot rule out this effect, we note that the close match between the ISM and stellar motions point to this effect being of at most a secondary importance.

It should also be noted that the measured ISM proper motion does not need to exactly agree with the proper motion of the CO gas. Tracing ISM using dust reflection in the $H$-band should allow us to probe the outer layers of the molecular region of the cloud, somewhere between the H$_2$ layer and the outer CO layer. The still denser central parts of molecular clouds carry more inertia than these outer layers and are expected to be more resistant to acceleration induced by stellar feedback. The differences in the kinematics of the outer and inner layers of a molecular cloud will depend on the star formation time scales. We presently do not have enough information to establish the dynamical age of the molecular gas near the head of L1688. Given the lack of wide field-of-view high-quality observations in the radio, we are forced to assume a perfect kinematic match between the observed outer and the dense inner layers of the cloud. This allows us to consider our derived proper motion as a representative measurement of the typical local motion of the cloud.

\section{Conclusions}\label{section:6}

By utilising archival high-resolution images obtained by HST/JWST and working with image registration tools, we were able to measure ISM proper motion inside L1688. Not only were we able to provide an estimate of the typical local transverse motion of the cloud, we also derived the first proper motion field of dust and gas within an interstellar cloud. The presented results show how much kinematic information can be extracted from multi-epoch observations of ISM clouds. Providing additional observations for L1688 will further improve measurement precision and serve not only for verification of our findings but also allow studies of ISM turbulence.

Our main result is represented by the measured transverse motion of the investigated ISM, which slightly differs from that of the young stars associated with L1688. We conclude that the identified difference is statistically significant ($\Delta v_{\textrm{t}} \approx 3 \pm 1$~km\,s$^{-1}$), although a marginal significance should be noted when using MAD as the uncertainty estimator. If confirmed by future observations, this would suggest the possibility of non-negligible biases when using YSOs as kinematic proxies for the ISM.

We propose that the measured velocity difference is attributable to interaction with stellar feedback produced by the young stars surrounding the cloud. Similar to previous studies, our analysis suggests that Oph~S1 might be the most important source of feedback within the studied region based on the following evidence:
\begin{itemize}
    \item The morphology of the cavity surrounding Oph~S1 suggests that this star has been interacting with the surrounding ISM. An absence of reflecting or emitting material can be seen in the direction of our investigated ISM structures, suggesting that the structures are likely not shielded from the feedback generated by Oph~S1.
    \item Assuming the existence of a jet originating from the direction of Oph~S1, a possibility suggested by the presence and properties of Oph~ISM~4, we note that the cavity surrounding the star could have been punctured by a jet in the direction of the investigated ISM.
    \item Our investigation of the radiative fluxes from Oph~S1 and other massive stars in its vicinity leads us to a flux ratio that favours Oph~S1 as the dominant source of radiative feedback in the region.
\end{itemize}
Due to the limitations of the available data, we are presently unable to identify the primary process impacting the dynamics of the region. Our analysis, combined with the measured kinematic difference between the USco stellar population and the ISM, suggests that the following three processes are the most plausible candidates: protostellar feedback (in the form of a jet), an expanding photodissociation region surrounding Oph~S1, or the most recent supernova event in the USco region of Sco-Cen. We interpret the ISM morphological structure (the cavity surrounding Oph~S1) as evidence of a past interaction between Oph~S1 and the surrounding ISM, while its present-day dynamical influence remains uncertain. In contrast, the direction of the feedback associated with the supernova event that occurred near USco about 2~Myr ago is well aligned with the ISM proper motion relative to the young USco stellar population. This agreement suggests that the supernova may have contributed to the presently observed ISM kinematics. This large-scale feedback event was identified as a major factor in the evolution of the ISM throughout the Ophiuchus complex by multiple other studies.

In our analysis of the investigated images, we identify two sources of uncertainty that should be addressed by future observations. Firstly, weak position-correlated offsets between galaxy positions in the HST and JWST images suggest a possible presence of residual astrometric systematics, potentially related to distortions or the use of different filters. Secondly, several ISM structures display proper motion gradients, particularly across Oph~ISM~1, although these may partly reflect filter-dependent intensity variations rather than true kinematic substructure. A third high-resolution epoch obtained in the same or a strongly overlapping filter would allow both effects to be tested and significantly improve the robustness of the derived ISM proper motions.

We detect a relatively dim shell-like ISM feature in L1688 that we label as Oph~ISM~4. The available JWST images and the absence of measurable expansion provide no strong evidence for either a supernova remnant or a fast-moving Herbig-Haro object. However, the morphology and measured kinematics suggest that the structure is physically associated with L1688 and may have originated roughly from the direction of Oph~S1, although its exact proper motion remains uncertain. In this context, Oph~ISM~4 represents one of the key pieces of evidence supporting a possible role of jet-driven feedback affecting the studied ISM field. While its origin cannot presently be established unambiguously, it is clear that this ISM structure represent an important target for future studies of ISM dynamics within the cloud.

Follow-up studies will benefit from the presented proper motion measurements by building a 3D kinematic picture of L1688. While only bulk radial velocities are presently accessible for regions as small as the one studied in this work (through, for example, H~\textsc{i} or CO measurements), the use of the integral field spectrography and higher-resolution studies should allow us to derive radial velocities of the individual cloud sub-structures. High-spatial-resolution radial velocity and proper motion maps will lead to observationally-driven 3D kinematic studies of not only bulk ISM motions but also of the internal dynamical state of ISM clouds. Furthermore, conducting studies at different wavelengths will enable investigations of kinematics of ISM in various phases, not only scattering dust features. Finally, comparing the motions of ISM and YSOs in other star-forming clouds will help building a good statistical sample for a general validity test of the kinematic-coupling assumption, which is challenged (at least locally) by the results of this work. We conclude that ISM proper motion measurements have a strong potential for improving our understanding of ISM dynamics, an important aspect of physical processes involved in star formation and galactic dynamics.

\begin{acknowledgements}
Co-funded by the European Union (ERC, ISM-FLOW, 101055318). Views and opinions expressed are, however, those of the author(s) only and do not necessarily reflect those of the European Union or the European Research Council. Neither the European Union nor the granting authority can be held responsible for them.

This work is based on observations made with the NASA/ESA Hubble Space Telescope and the NASA/ESA/CSA James Webb Space Telescope. The HST observations are associated with program 14181, and the JWST observations with program DD~2739. The data were obtained from the Mikulski Archive for Space Telescopes (MAST) at the Space Telescope Science Institute (STScI), which is operated by AURA, Inc., under NASA contract NAS5-03127.

This work made use of data from the European Space Agency (ESA) mission {\it Gaia} (\url{https://www.cosmos.esa.int/gaia}), 
processed by the {\it Gaia} Data Processing and Analysis Consortium (DPAC, \url{https://www.cosmos.esa.int/web/gaia/dpac/consortium}). 
Funding for the DPAC has been provided by national institutions, in particular the institutions participating in the {\it Gaia} 
Multilateral Agreement.

The following Python libraries were used in this work: \texttt{numpy} \citep{numpy}, \texttt{scipy} \citep{scipy}, \texttt{astropy} \citep{astropy1,astropy2,astropy3}, \texttt{matplotlib} \citep{matplotlib}, \texttt{skimage} \citep{skimage}, \texttt{photutils} \citep{photutils}, \texttt{image\_registration} \citep{ginsburg_image_registration_2024}, and \texttt{SimpleITK} \citep{SimpleITK1,SimpleITK2,SimpleITK3}.

Generative Artificial Intelligence (AI) tools were used for language editing purposes only (GPT-5.3~mini). The authors did not use AI to generate any intellectual content or ideas presented in this work.

\end{acknowledgements}

\bibliographystyle{aa} 
\bibliography{aanda}

@ARTICLE{CrA_motion,
       author = {{Piecka}, M. and {Posch}, L. and {Meingast}, S. and {Hutschenreuter}, S. and {Rottensteiner}, A. and {Alves}, J.},
        title = "{Direct measurement of ISM proper motion with image registration}",
      journal = {\aap},
         year = 2025,
        month = sep,
       volume = {702},
          eid = {L1},
        pages = {L1},
          doi = {10.1051/0004-6361/202556622},
archivePrefix = {arXiv},
       eprint = {2509.04857},
 primaryClass = {astro-ph.GA},
       adsurl = {https://ui.adsabs.harvard.edu/abs/2025A&A...702L...1P}
}

@ARTICLE{Rottensteiner2026,
       author = {{Rottensteiner}, Alena and {Petr-Gotzens}, Monika G. and {Meingast}, Stefan and {Alves}, Jo{\~a}o and {Bertin}, Emmanuel and {Bouy}, Herv{\'e} and {Piecka}, Martin and {Ratzenb{\"o}ck}, Sebastian and {Socci}, Andrea},
        title = "{Kinematics of young stellar objects in NGC 2024 based on infrared proper motions}",
      journal = {\aap},
         year = 2026,
        month = feb,
       volume = {706},
          eid = {A210},
        pages = {A210},
          doi = {10.1051/0004-6361/202557533},
archivePrefix = {arXiv},
       eprint = {2512.02124},
 primaryClass = {astro-ph.GA},
       adsurl = {https://ui.adsabs.harvard.edu/abs/2026A&A...706A.210R}
}

@ARTICLE{Griggio2023,
       author = {{Griggio}, M. and {Nardiello}, D. and {Bedin}, L.~R.},
        title = "{Photometry and astrometry with JWST{\ensuremath{-}}II: NIRCam distortion correction}",
      journal = {Astronomische Nachrichten},
         year = 2023,
        month = mar,
       volume = {344},
       number = {3},
          eid = {e20230006},
        pages = {e20230006},
          doi = {10.1002/asna.20230006},
archivePrefix = {arXiv},
       eprint = {2212.03256},
 primaryClass = {astro-ph.IM},
       adsurl = {https://ui.adsabs.harvard.edu/abs/2023AN....34430006G}
}

@ARTICLE{Bellini2009,
       author = {{Bellini}, A. and {Bedin}, L.~R.},
        title = "{Astrometry and Photometry with HST WFC3. I. Geometric Distortion Corrections of F225W, F275W, F336W Bands of the UVIS Channel}",
      journal = {\pasp},
         year = 2009,
        month = dec,
       volume = {121},
       number = {886},
        pages = {1419},
          doi = {10.1086/649061},
archivePrefix = {arXiv},
       eprint = {0910.3250},
 primaryClass = {astro-ph.IM},
       adsurl = {https://ui.adsabs.harvard.edu/abs/2009PASP..121.1419B}
}

@ARTICLE{Alves2025,
       author = {{Alves}, Jo{\~a}o and {Lombardi}, Marco and {Lada}, Charles J.},
        title = "{HP2 Survey: V. Ophiuchus: Filament formation in a dispersing cloud complex}",
      journal = {\aap},
         year = 2025,
        month = may,
       volume = {697},
          eid = {A208},
        pages = {A208},
          doi = {10.1051/0004-6361/202452881},
archivePrefix = {arXiv},
       eprint = {2501.13931},
 primaryClass = {astro-ph.GA},
       adsurl = {https://ui.adsabs.harvard.edu/abs/2025A&A...697A.208A}
}

@INCOLLECTION{Wilking2008,
       author = {{Wilking}, B.~A. and {Gagn{\'e}}, M. and {Allen}, L.~E.},
        title = "{Star Formation in the {\ensuremath{\rho}} Ophiuchi Molecular Cloud}",
    booktitle = {Handbook of Star Forming Regions, Volume II},
         year = 2008,
       editor = {{Reipurth}, B.},
       volume = {5},
        pages = {351},
          doi = {10.48550/arXiv.0811.0005},
       adsurl = {https://ui.adsabs.harvard.edu/abs/2008hsf2.book..351W}
}

@ARTICLE{Robitaille2018,
       author = {{Robitaille}, J. -F. and {Scaife}, A.~M.~M. and {Carretti}, E. and {Haverkorn}, M. and {Crocker}, R.~M. and {Kesteven}, M.~J. and {Poppi}, S. and {Staveley-Smith}, L.},
        title = "{Interstellar magnetic cannon targeting the Galactic halo. A young bubble at the origin of the Ophiuchus and Lupus molecular complexes}",
      journal = {\aap},
         year = 2018,
        month = sep,
       volume = {617},
          eid = {A101},
        pages = {A101},
          doi = {10.1051/0004-6361/201833358},
archivePrefix = {arXiv},
       eprint = {1807.04054},
 primaryClass = {astro-ph.GA},
       adsurl = {https://ui.adsabs.harvard.edu/abs/2018A&A...617A.101R}
}

@ARTICLE{deGeus1992,
       author = {{de Geus}, E.~J.},
        title = "{Interactions of stars and interstellar matter in Scorpio Centaurus.}",
      journal = {\aap},
         year = 1992,
        month = aug,
       volume = {262},
        pages = {258-270},
       adsurl = {https://ui.adsabs.harvard.edu/abs/1992A&A...262..258D}
}

@ARTICLE{MaizApellaniz2001,
       author = {{Ma{\'\i}z-Apell{\'a}niz}, Jes{\'u}s},
        title = "{The Origin of the Local Bubble}",
      journal = {\apjl},
         year = 2001,
        month = oct,
       volume = {560},
       number = {1},
        pages = {L83-L86},
          doi = {10.1086/324016},
archivePrefix = {arXiv},
       eprint = {astro-ph/0108472},
 primaryClass = {astro-ph},
       adsurl = {https://ui.adsabs.harvard.edu/abs/2001ApJ...560L..83M}
}

@ARTICLE{Fuchs2006,
       author = {{Fuchs}, B. and {Breitschwerdt}, D. and {de Avillez}, M.~A. and {Dettbarn}, C. and {Flynn}, C.},
        title = "{The search for the origin of the Local Bubble redivivus}",
      journal = {\mnras},
         year = 2006,
        month = dec,
       volume = {373},
       number = {3},
        pages = {993-1003},
          doi = {10.1111/j.1365-2966.2006.11044.x},
archivePrefix = {arXiv},
       eprint = {astro-ph/0609227},
 primaryClass = {astro-ph},
       adsurl = {https://ui.adsabs.harvard.edu/abs/2006MNRAS.373..993F}
}

@ARTICLE{Krause2018,
       author = {{Krause}, Martin G.~H. and {Burkert}, Andreas and {Diehl}, Roland and {Fierlinger}, Katharina and {Gaczkowski}, Benjamin and {Kroell}, Daniel and {Ngoumou}, Judith and {Roccatagliata}, Veronica and {Siegert}, Thomas and {Preibisch}, Thomas},
        title = "{Surround and Squash: the impact of superbubbles on the interstellar medium in Scorpius-Centaurus OB2}",
      journal = {\aap},
         year = 2018,
        month = nov,
       volume = {619},
          eid = {A120},
        pages = {A120},
          doi = {10.1051/0004-6361/201732416},
archivePrefix = {arXiv},
       eprint = {1808.04788},
 primaryClass = {astro-ph.GA},
       adsurl = {https://ui.adsabs.harvard.edu/abs/2018A&A...619A.120K}
}

@ARTICLE{Zucker2022,
       author = {{Zucker}, Catherine and {Goodman}, Alyssa A. and {Alves}, Jo{\~a}o and {Bialy}, Shmuel and {Foley}, Michael and {Speagle}, Joshua S. and {Gro{\^I}{\texttwosuperior}schedl}, Josefa and {Finkbeiner}, Douglas P. and {Burkert}, Andreas and {Khimey}, Diana and {Swiggum}, Cameren},
        title = "{Star formation near the Sun is driven by expansion of the Local Bubble}",
      journal = {\nat},
         year = 2022,
        month = jan,
       volume = {601},
       number = {7893},
        pages = {334-337},
          doi = {10.1038/s41586-021-04286-5},
archivePrefix = {arXiv},
       eprint = {2201.05124},
 primaryClass = {astro-ph.GA},
       adsurl = {https://ui.adsabs.harvard.edu/abs/2022Natur.601..334Z}
}

@ARTICLE{Frisch1995,
       author = {{Frisch}, Priscilla C.},
        title = "{Characteristics of Nearby Interstellar Matter}",
      journal = {\ssr},
         year = 1995,
        month = may,
       volume = {72},
       number = {3-4},
        pages = {499-592},
          doi = {10.1007/BF00749006},
       adsurl = {https://ui.adsabs.harvard.edu/abs/1995SSRv...72..499F}
}

@ARTICLE{Frisch2011,
       author = {{Frisch}, Priscilla C. and {Redfield}, Seth and {Slavin}, Jonathan D.},
        title = "{The Interstellar Medium Surrounding the Sun}",
      journal = {\araa},
         year = 2011,
        month = sep,
       volume = {49},
       number = {1},
        pages = {237-279},
          doi = {10.1146/annurev-astro-081710-102613},
       adsurl = {https://ui.adsabs.harvard.edu/abs/2011ARA&A..49..237F}
}

@ARTICLE{Zucker2025,
       author = {{Zucker}, Catherine and {Redfield}, Seth and {Starecheski}, Sara and {Konietzka}, Ralf and {Linsky}, Jeffrey L.},
        title = "{The Origin of the Cluster of Local Interstellar Clouds}",
      journal = {\apj},
         year = 2025,
        month = jun,
       volume = {986},
       number = {1},
          eid = {58},
        pages = {58},
          doi = {10.3847/1538-4357/adc920},
archivePrefix = {arXiv},
       eprint = {2504.00093},
 primaryClass = {astro-ph.GA},
       adsurl = {https://ui.adsabs.harvard.edu/abs/2025ApJ...986...58Z}
}

@ARTICLE{Ellis1995,
       author = {{Ellis}, John and {Schramm}, David N.},
        title = "{Could a Nearby Supernova Explosion have Caused a Mass Extinction?}",
      journal = {Proceedings of the National Academy of Science},
         year = 1995,
        month = jan,
       volume = {92},
       number = {1},
        pages = {235-238},
          doi = {10.1073/pnas.92.1.235},
archivePrefix = {arXiv},
       eprint = {hep-ph/9303206},
 primaryClass = {hep-ph},
       adsurl = {https://ui.adsabs.harvard.edu/abs/1995PNAS...92..235E}
}

@ARTICLE{Benitez2002,
       author = {{Ben{\'\i}tez}, Narciso and {Ma{\'\i}z-Apell{\'a}niz}, Jes{\'u}s and {Canelles}, Matilde},
        title = "{Evidence for Nearby Supernova Explosions}",
      journal = {\prl},
         year = 2002,
        month = feb,
       volume = {88},
       number = {8},
        pages = {081101},
          doi = {10.1103/PhysRevLett.88.081101},
archivePrefix = {arXiv},
       eprint = {astro-ph/0201018},
 primaryClass = {astro-ph},
       adsurl = {https://ui.adsabs.harvard.edu/abs/2002PhRvL..88h1101B}
}

@ARTICLE{Breitschwerdt2016,
       author = {{Breitschwerdt}, D. and {Feige}, J. and {Schulreich}, M.~M. and {Avillez}, M.~A. De. and {Dettbarn}, C. and {Fuchs}, B.},
        title = "{The locations of recent supernovae near the Sun from modelling $^{60}$Fe transport}",
      journal = {\nat},
         year = 2016,
        month = apr,
       volume = {532},
       number = {7597},
        pages = {73-76},
          doi = {10.1038/nature17424},
       adsurl = {https://ui.adsabs.harvard.edu/abs/2016Natur.532...73B}
}

@ARTICLE{Schulreich2017,
       author = {{Schulreich}, M.~M. and {Breitschwerdt}, D. and {Feige}, J. and {Dettbarn}, C.},
        title = "{Numerical studies on the link between radioisotopic signatures on Earth and the formation of the Local Bubble. I. $^{60}$Fe transport to the solar system by turbulent mixing of ejecta from nearby supernovae into a locally homogeneous interstellar medium}",
      journal = {\aap},
         year = 2017,
        month = aug,
       volume = {604},
          eid = {A81},
        pages = {A81},
          doi = {10.1051/0004-6361/201629837},
archivePrefix = {arXiv},
       eprint = {1704.08221},
 primaryClass = {astro-ph.HE},
       adsurl = {https://ui.adsabs.harvard.edu/abs/2017A&A...604A..81S}
}

@ARTICLE{Zucker2020,
       author = {{Zucker}, Catherine and {Speagle}, Joshua S. and {Schlafly}, Edward F. and {Green}, Gregory M. and {Finkbeiner}, Douglas P. and {Goodman}, Alyssa and {Alves}, Jo{\~a}o},
        title = "{A compendium of distances to molecular clouds in the Star Formation Handbook}",
      journal = {\aap},
         year = 2020,
        month = jan,
       volume = {633},
          eid = {A51},
        pages = {A51},
          doi = {10.1051/0004-6361/201936145},
archivePrefix = {arXiv},
       eprint = {2001.00591},
 primaryClass = {astro-ph.GA},
       adsurl = {https://ui.adsabs.harvard.edu/abs/2020A&A...633A..51Z}
}

@ARTICLE{Montmerle1983,
       author = {{Montmerle}, T. and {Koch-Miramond}, L. and {Falgarone}, E. and {Grindlay}, J.~E.},
        title = "{Einstein observations of the Rho Ophiuchi dark cloud : an X-ray christmas tree.}",
      journal = {\apj},
         year = 1983,
        month = jun,
       volume = {269},
        pages = {182-201},
          doi = {10.1086/161029},
       adsurl = {https://ui.adsabs.harvard.edu/abs/1983ApJ...269..182M}
}

@ARTICLE{Lomax2014,
       author = {{Lomax}, O. and {Whitworth}, A.~P. and {Hubber}, D.~A. and {Stamatellos}, D. and {Walch}, S.},
        title = "{Simulating star formation in Ophiuchus}",
      journal = {\mnras},
         year = 2014,
        month = apr,
       volume = {439},
       number = {3},
        pages = {3039-3050},
          doi = {10.1093/mnras/stu177},
archivePrefix = {arXiv},
       eprint = {1401.7237},
 primaryClass = {astro-ph.SR},
       adsurl = {https://ui.adsabs.harvard.edu/abs/2014MNRAS.439.3039L}
}

@ARTICLE{Tachihara2000,
       author = {{Tachihara}, Kengo and {Mizuno}, Akira and {Fukui}, Yasuo},
        title = "{C$^{18}$O Observations of the Dense Cloud Cores and Star Formation in Ophiuchus}",
      journal = {\apj},
         year = 2000,
        month = jan,
       volume = {528},
       number = {2},
        pages = {817-840},
          doi = {10.1086/308189},
       adsurl = {https://ui.adsabs.harvard.edu/abs/2000ApJ...528..817T}
}

@ARTICLE{Williams2019,
       author = {{Williams}, Jonathan P. and {Cieza}, Lucas and {Hales}, Antonio and {Ansdell}, Megan and {Ruiz-Rodriguez}, Dary and {Casassus}, Simon and {Perez}, Sebastian and {Zurlo}, Alice},
        title = "{The Ophiuchus DIsk Survey Employing ALMA (ODISEA): Disk Dust Mass Distributions across Protostellar Evolutionary Classes}",
      journal = {\apjl},
         year = 2019,
        month = apr,
       volume = {875},
       number = {2},
          eid = {L9},
        pages = {L9},
          doi = {10.3847/2041-8213/ab1338},
archivePrefix = {arXiv},
       eprint = {1904.06471},
 primaryClass = {astro-ph.SR},
       adsurl = {https://ui.adsabs.harvard.edu/abs/2019ApJ...875L...9W}
}

@ARTICLE{Testi2022,
       author = {{Testi}, L. and {Natta}, A. and {Manara}, C.~F. and {de Gregorio Monsalvo}, I. and {Lodato}, G. and {Lopez}, C. and {Muzic}, K. and {Pascucci}, I. and {Sanchis}, E. and {Miranda}, A. Santamaria and {Scholz}, A. and {De Simone}, M. and {Williams}, J.~P.},
        title = "{The protoplanetary disk population in the {\ensuremath{\rho}}-Ophiuchi region L1688 and the time evolution of Class II YSOs}",
      journal = {\aap},
         year = 2022,
        month = jul,
       volume = {663},
          eid = {A98},
        pages = {A98},
          doi = {10.1051/0004-6361/202141380},
archivePrefix = {arXiv},
       eprint = {2201.04079},
 primaryClass = {astro-ph.SR},
       adsurl = {https://ui.adsabs.harvard.edu/abs/2022A&A...663A..98T}
}

@ARTICLE{Ratzenbock2023,
       author = {{Ratzenb{\"o}ck}, Sebastian and {Gro{\ss}schedl}, Josefa E. and {Alves}, Jo{\~a}o and {Miret-Roig}, N{\'u}ria and {Bomze}, Immanuel and {Forbes}, John and {Goodman}, Alyssa and {Hacar}, {\'A}lvaro and {Lin}, Doug and {Meingast}, Stefan and {M{\"o}ller}, Torsten and {Piecka}, Martin and {Posch}, Laura and {Rottensteiner}, Alena and {Swiggum}, Cameren and {Zucker}, Catherine},
        title = "{The star formation history of the Sco-Cen association. Coherent star formation patterns in space and time}",
      journal = {\aap},
         year = 2023,
        month = oct,
       volume = {678},
          eid = {A71},
        pages = {A71},
          doi = {10.1051/0004-6361/202346901},
archivePrefix = {arXiv},
       eprint = {2302.07853},
 primaryClass = {astro-ph.SR},
       adsurl = {https://ui.adsabs.harvard.edu/abs/2023A&A...678A..71R}
}

@ARTICLE{Grasser2021,
       author = {{Grasser}, Natalie and {Ratzenb{\"o}ck}, Sebastian and {Alves}, Jo{\~a}o and {Gro{\ss}schedl}, Josefa and {Meingast}, Stefan and {Zucker}, Catherine and {Hacar}, Alvaro and {Lada}, Charles and {Goodman}, Alyssa and {Lombardi}, Marco and {Forbes}, John C. and {Bomze}, Immanuel M. and {M{\"o}ller}, Torsten},
        title = "{The {\ensuremath{\rho}} Ophiuchi region revisited with Gaia EDR3. Two young populations, new members, and old impostors}",
      journal = {\aap},
         year = 2021,
        month = aug,
       volume = {652},
          eid = {A2},
        pages = {A2},
          doi = {10.1051/0004-6361/202140438},
archivePrefix = {arXiv},
       eprint = {2101.12200},
 primaryClass = {astro-ph.SR},
       adsurl = {https://ui.adsabs.harvard.edu/abs/2021A&A...652A...2G}
}

@ARTICLE{SigMA,
       author = {{Ratzenb{\"o}ck}, Sebastian and {Gro{\ss}schedl}, Josefa E. and {M{\"o}ller}, Torsten and {Alves}, Jo{\~a}o and {Bomze}, Immanuel and {Meingast}, Stefan},
        title = "{Significance mode analysis (SigMA) for hierarchical structures. An application to the Sco-Cen OB association}",
      journal = {\aap},
         year = 2023,
        month = sep,
       volume = {677},
          eid = {A59},
        pages = {A59},
          doi = {10.1051/0004-6361/202243690},
archivePrefix = {arXiv},
       eprint = {2211.14225},
 primaryClass = {astro-ph.GA},
       adsurl = {https://ui.adsabs.harvard.edu/abs/2023A&A...677A..59R}
}

@ARTICLE{Hutschenreuter2026,
       author = {{Hutschenreuter}, S. and {Alves}, J. and {Posch}, L. and {Gro{\ss}schedl}, J. and {Piecka}, M. and {Miret-Roig}, N. and {Ratzenb{\"o}ck}, S. and {Swiggum}, C.},
        title = "{The velocity field of the Scorpius-Centaurus OB association: I. Method and general properties}",
      journal = {\aap},
         year = 2026,
        month = jan,
       volume = {705},
          eid = {A108},
        pages = {A108},
          doi = {10.1051/0004-6361/202557273},
       adsurl = {https://ui.adsabs.harvard.edu/abs/2026A&A...705A.108H}
}

@ARTICLE{MiretRoig2022,
       author = {{Miret-Roig}, N. and {Galli}, P.~A.~B. and {Olivares}, J. and {Bouy}, H. and {Alves}, J. and {Barrado}, D.},
        title = "{The star formation history of Upper Scorpius and Ophiuchus. A 7D picture: positions, kinematics, and dynamical traceback ages}",
      journal = {\aap},
         year = 2022,
        month = nov,
       volume = {667},
          eid = {A163},
        pages = {A163},
          doi = {10.1051/0004-6361/202244709},
archivePrefix = {arXiv},
       eprint = {2209.12938},
 primaryClass = {astro-ph.GA},
       adsurl = {https://ui.adsabs.harvard.edu/abs/2022A&A...667A.163M}
}

@ARTICLE{Posch2025,
       author = {{Posch}, Laura and {Alves}, Jo{\~a}o and {Miret-Roig}, N{\'u}ria and {Ratzenb{\"o}ck}, Sebastian and {Gro{\ss}schedl}, Josefa and {Meingast}, Stefan and {Swiggum}, Cameren and {Konietzka}, Ralf},
        title = "{The physical properties of cluster chains}",
      journal = {\aap},
         year = 2025,
        month = jan,
       volume = {693},
          eid = {A175},
        pages = {A175},
          doi = {10.1051/0004-6361/202451312},
archivePrefix = {arXiv},
       eprint = {2410.18080},
 primaryClass = {astro-ph.GA},
       adsurl = {https://ui.adsabs.harvard.edu/abs/2025A&A...693A.175P}
}

@ARTICLE{Simpson2008,
       author = {{Simpson}, R.~J. and {Nutter}, D. and {Ward-Thompson}, D.},
        title = "{The initial conditions of star formation - VIII. An observational study of the Ophiuchus cloud L1688 and implications for the pre-stellar core mass function}",
      journal = {\mnras},
         year = 2008,
        month = nov,
       volume = {391},
       number = {1},
        pages = {205-214},
          doi = {10.1111/j.1365-2966.2008.13750.x},
archivePrefix = {arXiv},
       eprint = {0807.4382},
 primaryClass = {astro-ph},
       adsurl = {https://ui.adsabs.harvard.edu/abs/2008MNRAS.391..205S}
}

@ARTICLE{Loren1989,
       author = {{Loren}, Robert B.},
        title = "{The Cobwebs of Ophiuchus. I. Strands of 13CO: The Mass Distribution}",
      journal = {\apj},
         year = 1989,
        month = mar,
       volume = {338},
        pages = {902},
          doi = {10.1086/167244},
       adsurl = {https://ui.adsabs.harvard.edu/abs/1989ApJ...338..902L}
}

@ARTICLE{Gomez2003,
       author = {{G{\'o}mez}, M. and {Stark}, D.~P. and {Whitney}, B.~A. and {Churchwell}, E.},
        title = "{Jets and Herbig-Haro Objects in the {\ensuremath{\rho}} Ophiuchi Embedded Cluster}",
      journal = {\aj},
         year = 2003,
        month = aug,
       volume = {126},
       number = {2},
        pages = {863-886},
          doi = {10.1086/376741},
       adsurl = {https://ui.adsabs.harvard.edu/abs/2003AJ....126..863G}
}

@ARTICLE{Chen2019,
       author = {{Chen}, Hope How-Huan and {Pineda}, Jaime E. and {Goodman}, Alyssa A. and {Burkert}, Andreas and {Offner}, Stella S.~R. and {Friesen}, Rachel K. and {Myers}, Philip C. and {Alves}, Felipe and {Arce}, H{\'e}ctor G. and {Caselli}, Paola and {Chac{\'o}n-Tanarro}, Ana and {Chen}, Michael Chun-Yuan and {Di Francesco}, James and {Ginsburg}, Adam and {Keown}, Jared and {Kirk}, Helen and {Martin}, Peter G. and {Matzner}, Christopher and {Punanova}, Anna and {Redaelli}, Elena and {Rosolowsky}, Erik and {Scibelli}, Samantha and {Seo}, Youngmin and {Shirley}, Yancy and {Singh}, Ayushi and {GAS Collaboration}},
        title = "{Droplets. I. Pressure-dominated Coherent Structures in L1688 and B18}",
      journal = {\apj},
         year = 2019,
        month = jun,
       volume = {877},
       number = {2},
          eid = {93},
        pages = {93},
          doi = {10.3847/1538-4357/ab1a40},
archivePrefix = {arXiv},
       eprint = {1809.10223},
 primaryClass = {astro-ph.GA},
       adsurl = {https://ui.adsabs.harvard.edu/abs/2019ApJ...877...93C}
}

@ARTICLE{Nakamura2025,
       author = {{Nakamura}, Fumitaka and {Kawabe}, Ryohei and {Huang}, Shuo and {Saigo}, Kazuya and {Hirano}, Naomi and {Takakuwa}, Shigehisa and {Kamazaki}, Takeshi and {Tamura}, Motohide and {Di Francesco}, James and {Friesen}, Rachel and {Iwasaki}, Kazunari and {Hara}, Chihomi},
        title = "{Unveiling Stellar Feedback and Cloud Structure in the {\ensuremath{\rho}} Ophiuchi-A Region Using ALMA and JWST: Discovery of Substellar Cores, C$^{18}$O Striations, and Protostellar Outflows}",
      journal = {\apj},
         year = 2025,
        month = dec,
       volume = {994},
       number = {2},
          eid = {225},
        pages = {225},
          doi = {10.3847/1538-4357/adfddd},
archivePrefix = {arXiv},
       eprint = {2509.01122},
 primaryClass = {astro-ph.GA},
       adsurl = {https://ui.adsabs.harvard.edu/abs/2025ApJ...994..225N}
}

@ARTICLE{Howard2021,
       author = {{Howard}, A.~D.~P. and {Whitworth}, A.~P. and {Griffin}, M.~J. and {Marsh}, K.~A. and {Smith}, M.~W.~L.},
        title = "{A PPMAP analysis of the filamentary structures in Ophiuchus L1688 and L1689}",
      journal = {\mnras},
         year = 2021,
        month = jul,
       volume = {504},
       number = {4},
        pages = {6157-6178},
          doi = {10.1093/mnras/stab1166},
archivePrefix = {arXiv},
       eprint = {2104.04007},
 primaryClass = {astro-ph.GA},
       adsurl = {https://ui.adsabs.harvard.edu/abs/2021MNRAS.504.6157H}
}

@ARTICLE{Ordonez2025,
       author = {{Ord{\'o}{\~n}ez-Toro}, Jazm{\'\i}n and {Dzib}, Sergio A. and {Loinard}, Laurent and {Ortiz-Le{\'o}n}, Gisela and {Kounkel}, Marina A. and {Galli}, Phillip A.~B. and {Masqu{\'e}}, Josep M. and {Dupuy}, Trent J. and {Quiroga-Nu{\~n}ez}, Luis H. and {Rodr{\'\i}guez}, Luis F.},
        title = "{VLBA detections in the Oph-S1 binary system near periastron confirmation of its orbital elements and mass}",
      journal = {\mnras},
         year = 2025,
        month = apr,
       volume = {538},
       number = {3},
        pages = {1784-1788},
          doi = {10.1093/mnras/staf396},
archivePrefix = {arXiv},
       eprint = {2503.04594},
 primaryClass = {astro-ph.SR},
       adsurl = {https://ui.adsabs.harvard.edu/abs/2025MNRAS.538.1784O}
}

@ARTICLE{Eker2018,
       author = {{Eker}, Z. and {Bak{\i}{\textcommabelow s}}, V. and {Bilir}, S. and {Soydugan}, F. and {Steer}, I. and {Soydugan}, E. and {Bak{\i}{\textcommabelow s}}, H. and {Ali{\c{c}}avu{\textcommabelow s}}, F. and {Aslan}, G. and {Alpsoy}, M.},
        title = "{Interrelated main-sequence mass-luminosity, mass-radius, and mass-effective temperature relations}",
      journal = {\mnras},
         year = 2018,
        month = oct,
       volume = {479},
       number = {4},
        pages = {5491-5511},
          doi = {10.1093/mnras/sty1834},
archivePrefix = {arXiv},
       eprint = {1807.02568},
 primaryClass = {astro-ph.SR},
       adsurl = {https://ui.adsabs.harvard.edu/abs/2018MNRAS.479.5491E}
}

@ARTICLE{Cordiner2013,
       author = {{Cordiner}, M.~A. and {Fossey}, S.~J. and {Smith}, A.~M. and {Sarre}, P.~J.},
        title = "{Small-scale Structure of the Interstellar Medium toward {\ensuremath{\rho}} Oph Stars: Diffuse Band Observations}",
      journal = {\apjl},
         year = 2013,
        month = feb,
       volume = {764},
       number = {1},
          eid = {L10},
        pages = {L10},
          doi = {10.1088/2041-8205/764/1/L10},
archivePrefix = {arXiv},
       eprint = {1301.6167},
 primaryClass = {astro-ph.GA},
       adsurl = {https://ui.adsabs.harvard.edu/abs/2013ApJ...764L..10C}
}

@ARTICLE{Wilking1983,
       author = {{Wilking}, B.~A. and {Lada}, C.~J.},
        title = "{The discovery of new embedded sources in the centrally condensed coreof the rho Ophiuchi dark cloud : the formation of a bound cluster ?}",
      journal = {\apj},
         year = 1983,
        month = nov,
       volume = {274},
        pages = {698-716},
          doi = {10.1086/161482},
       adsurl = {https://ui.adsabs.harvard.edu/abs/1983ApJ...274..698W}
}

@ARTICLE{Andre2007,
       author = {{Andr{\'e}}, Ph. and {Belloche}, A. and {Motte}, F. and {Peretto}, N.},
        title = "{The initial conditions of star formation in the Ophiuchus main cloud: Kinematics of the protocluster condensations}",
      journal = {\aap},
         year = 2007,
        month = sep,
       volume = {472},
       number = {2},
        pages = {519-535},
          doi = {10.1051/0004-6361:20077422},
archivePrefix = {arXiv},
       eprint = {0706.1535},
 primaryClass = {astro-ph},
       adsurl = {https://ui.adsabs.harvard.edu/abs/2007A&A...472..519A}
}

@ARTICLE{Rigliaco2016,
       author = {{Rigliaco}, E. and {Wilking}, B. and {Meyer}, M.~R. and {Jeffries}, R.~D. and {Cottaar}, M. and {Frasca}, A. and {Wright}, N.~J. and {Bayo}, A. and {Bonito}, R. and {Damiani}, F. and {Jackson}, R.~J. and {Jim{\'e}nez-Esteban}, F. and {Kalari}, V.~M. and {Klutsch}, A. and {Lanzafame}, A.~C. and {Sacco}, G. and {Gilmore}, G. and {Randich}, S. and {Alfaro}, E.~J. and {Bragaglia}, A. and {Costado}, M.~T. and {Franciosini}, E. and {Lardo}, C. and {Monaco}, L. and {Morbidelli}, L. and {Prisinzano}, L. and {Sousa}, S.~G. and {Zaggia}, S.},
        title = "{The Gaia-ESO Survey: Dynamical analysis of the L1688 region in Ophiuchus}",
      journal = {\aap},
         year = 2016,
        month = apr,
       volume = {588},
          eid = {A123},
        pages = {A123},
          doi = {10.1051/0004-6361/201527253},
archivePrefix = {arXiv},
       eprint = {1601.05209},
 primaryClass = {astro-ph.SR},
       adsurl = {https://ui.adsabs.harvard.edu/abs/2016A&A...588A.123R}
}

@ARTICLE{Yun2021,
       author = {{Yun}, Hyeong-Sik and {Lee}, Jeong-Eun and {Evans}, Neal J. and {Offner}, Stella S.~R. and {Heyer}, Mark H. and {Cho}, Jungyeon and {Gaches}, Brandt A.~L. and {Yang}, Yao-Lun and {Chen}, How-Huan and {Choi}, Yunhee and {Lee}, Yong-Hee and {Baek}, Giseon and {Choi}, Minho and {Kim}, Jongsoo and {Kang}, Hyunwoo and {Lee}, Seokho and {Tatematsu}, Ken'ichi},
        title = "{Turbulent Properties in Star-forming Molecular Clouds Down to the Sonic Scale. II. Investigating the Relation between Turbulence and Star-forming Environments in Molecular Clouds}",
      journal = {\apj},
         year = 2021,
        month = nov,
       volume = {921},
       number = {1},
          eid = {31},
        pages = {31},
          doi = {10.3847/1538-4357/ac193e},
       adsurl = {https://ui.adsabs.harvard.edu/abs/2021ApJ...921...31Y}
}

@ARTICLE{Le2024,
       author = {{L{\^e}}, Ng{\^a}n and {Tram}, Le Ngoc and {Karska}, Agata and {Hoang}, Thiem and {Diep}, Pham Ngoc and {Hanasz}, Micha{\l} and {Ngoc}, Nguyen Bich and {Phuong}, Nguyen Thi and {Menten}, Karl M. and {Wyrowski}, Friedrich and {Nguyen}, Dieu D. and {Hoang}, Thuong Duc and {Khang}, Nguyen Minh},
        title = "{Mapping and characterizing magnetic fields in the Rho Ophiuchus-A molecular cloud with SOFIA/HAWC+}",
      journal = {\aap},
         year = 2024,
        month = oct,
       volume = {690},
          eid = {A191},
        pages = {A191},
          doi = {10.1051/0004-6361/202348008},
archivePrefix = {arXiv},
       eprint = {2408.17122},
 primaryClass = {astro-ph.GA},
       adsurl = {https://ui.adsabs.harvard.edu/abs/2024A&A...690A.191L}
}

@ARTICLE{Bontemps2001,
       author = {{Bontemps}, S. and {Andr{\'e}}, P. and {Kaas}, A.~A. and {Nordh}, L. and {Olofsson}, G. and {Huldtgren}, M. and {Abergel}, A. and {Blommaert}, J. and {Boulanger}, F. and {Burgdorf}, M. and {Cesarsky}, C.~J. and {Cesarsky}, D. and {Copet}, E. and {Davies}, J. and {Falgarone}, E. and {Lagache}, G. and {Montmerle}, T. and {P{\'e}rault}, M. and {Persi}, P. and {Prusti}, T. and {Puget}, J.~L. and {Sibille}, F.},
        title = "{ISOCAM observations of the rho Ophiuchi cloud: Luminosity and mass functions of the pre-main sequence embedded cluster}",
      journal = {\aap},
         year = 2001,
        month = jun,
       volume = {372},
        pages = {173-194},
          doi = {10.1051/0004-6361:20010474},
archivePrefix = {arXiv},
       eprint = {astro-ph/0103373},
 primaryClass = {astro-ph},
       adsurl = {https://ui.adsabs.harvard.edu/abs/2001A&A...372..173B}
}

@ARTICLE{Wilking1987,
       author = {{Wilking}, B.~A. and {Claussen}, M.~J.},
        title = "{Water Masers Associated with Low-Mass Stars: A Survey of the Rho Ophiuchi Infrared Cluster}",
      journal = {\apjl},
         year = 1987,
        month = sep,
       volume = {320},
        pages = {L133},
          doi = {10.1086/184989},
       adsurl = {https://ui.adsabs.harvard.edu/abs/1987ApJ...320L.133W}
}

@ARTICLE{Covey2006,
       author = {{Covey}, Kevin R. and {Greene}, Thomas P. and {Doppmann}, Greg W. and {Lada}, Charles J.},
        title = "{The Radial Velocity Distribution of Class I and Flat-Spectrum Protostars}",
      journal = {\aj},
         year = 2006,
        month = jan,
       volume = {131},
       number = {1},
        pages = {512-519},
          doi = {10.1086/498064},
archivePrefix = {arXiv},
       eprint = {astro-ph/0509264},
 primaryClass = {astro-ph},
       adsurl = {https://ui.adsabs.harvard.edu/abs/2006AJ....131..512C}
}

@ARTICLE{VISIONS1,
       author = {{Meingast}, Stefan and {Alves}, Jo{\~a}o and {Bouy}, Herv{\'e} and {Petr-Gotzens}, Monika G. and {F{\"u}rnkranz}, Verena and {Gro{\ss}schedl}, Josefa E. and {Hernandez}, David and {Rottensteiner}, Alena and {Arnaboldi}, Magda and {Ascenso}, Joana and {Bayo}, Amelia and {Br{\"a}ndli}, Erik and {Brown}, Anthony G.~A. and {Forbrich}, Jan and {Goodman}, Alyssa and {Hacar}, Alvaro and {Hasenberger}, Birgit and {K{\"o}hler}, Rainer and {Kubiak}, Karolina and {Kuhn}, Michael and {Lada}, Charles and {Leschinski}, Kieran and {Lombardi}, Marco and {Mardones}, Diego and {Mascetti}, Laura and {Miret-Roig}, N{\'u}ria and {Moitinho}, Andr{\'e} and {Mu{\v{z}}i{\'c}}, Koraljka and {Piecka}, Martin and {Posch}, Laura and {Prusti}, Timo and {Pe{\~n}a Ram{\'\i}rez}, Karla and {Ramlau}, Ronny and {Ratzenb{\"o}ck}, Sebastian and {Sacco}, Germano and {Swiggum}, Cameren and {Teixeira}, Paula Stella and {Urban}, Vanessa and {Zari}, Eleonora and {Zucker}, Catherine},
        title = "{VISIONS: the VISTA Star Formation Atlas. I. Survey overview}",
      journal = {\aap},
         year = 2023,
        month = may,
       volume = {673},
          eid = {A58},
        pages = {A58},
          doi = {10.1051/0004-6361/202245771},
archivePrefix = {arXiv},
       eprint = {2303.08831},
 primaryClass = {astro-ph.GA},
       adsurl = {https://ui.adsabs.harvard.edu/abs/2023A&A...673A..58M}
}

@ARTICLE{VISIONS2,
       author = {{Meingast}, Stefan and {Bouy}, Herv{\'e} and {F{\"u}rnkranz}, Verena and {Hernandez}, David and {Rottensteiner}, Alena and {Br{\"a}ndli}, Erik},
        title = "{VISIONS: the VISTA Star Formation Atlas. II. The data processing pipeline}",
      journal = {\aap},
         year = 2023,
        month = may,
       volume = {673},
          eid = {A59},
        pages = {A59},
          doi = {10.1051/0004-6361/202245772},
archivePrefix = {arXiv},
       eprint = {2303.08840},
 primaryClass = {astro-ph.IM},
       adsurl = {https://ui.adsabs.harvard.edu/abs/2023A&A...673A..59M}
}

@book{ITK,
  author       = {Ibanez, Luis and Schroeder, Will and Ng, Lydia and Cates, Josh},
  title        = {The ITK Software Guide},
  year         = {2005},
  publisher    = {Kitware, Inc.},
  edition      = {Second Edition},
  address      = {Clifton Park, NY},
  url          = {https://itk.org/ItkSoftwareGuide.pdf}
}

@ARTICLE{SimpleITK3,
  author       = {Beare, Richard and Lowekamp, Bradley and Yaniv, Ziv},
  title        = {Image Segmentation, Registration and Characterization in R with SimpleITK},
  journal      = {Journal of Statistical Software},
  volume       = {86},
  number       = {8},
  pages        = {1--35},
  year         = {2018},
  month        = {September},
  doi          = {10.18637/jss.v086.i08},
  url          = {https://doi.org/10.18637/jss.v086.i08}
}

@article{SimpleITK2,
  author       = {Yaniv, Ziv and Lowekamp, Bradley C. and Johnson, Hans J. and Beare, Richard},
  title        = {SimpleITK Image-Analysis Notebooks: a Collaborative Environment for Education and Reproducible Research},
  journal      = {Journal of Digital Imaging},
  volume       = {31},
  pages        = {290--303},
  year         = {2018},
  doi          = {10.1007/s10278-017-0037-8},
  url          = {https://doi.org/10.1007/s10278-017-0037-8},
  publisher    = {Springer}
}

@article{SimpleITK1,
  author       = {Lowekamp, Bradley C. and Chen, David T. and Ibanez, Luis and Blezek, Daniel},
  title        = {The Design of SimpleITK},
  journal      = {Frontiers in Neuroinformatics},
  volume       = {7},
  year         = {2013},
  pages        = {45},
  doi          = {10.3389/fninf.2013.00045},
  url          = {https://www.frontiersin.org/articles/10.3389/fninf.2013.00045},
  publisher    = {Frontiers}
}

@ARTICLE{GaiaDR2,
       author = {{Gaia Collaboration} and {Brown}, A.~G.~A. and {Vallenari}, A. and {Prusti}, T. and {de Bruijne}, J.~H.~J. and {Babusiaux}, C. and {Bailer-Jones}, C.~A.~L. and {Biermann}, M. and {Evans}, D.~W. and {Eyer}, L. and {Jansen}, F. and {Jordi}, C. and {Klioner}, S.~A. and {Lammers}, U. and {Lindegren}, L. and {Luri}, X. and {Mignard}, F. and {Panem}, C. and {Pourbaix}, D. and {Randich}, S. and {Sartoretti}, P. and {Siddiqui}, H.~I. and {Soubiran}, C. and {van Leeuwen}, F. and {Walton}, N.~A. and {Arenou}, F. and {Bastian}, U. and {Cropper}, M. and {Drimmel}, R. and {Katz}, D. and {Lattanzi}, M.~G. and {Bakker}, J. and {Cacciari}, C. and {Casta{\~n}eda}, J. and {Chaoul}, L. and {Cheek}, N. and {De Angeli}, F. and {Fabricius}, C. and {Guerra}, R. and {Holl}, B. and {Masana}, E. and {Messineo}, R. and {Mowlavi}, N. and {Nienartowicz}, K. and {Panuzzo}, P. and {Portell}, J. and {Riello}, M. and {Seabroke}, G.~M. and {Tanga}, P. and {Th{\'e}venin}, F. and {Gracia-Abril}, G. and {Comoretto}, G. and {Garcia-Reinaldos}, M. and {Teyssier}, D. and {Altmann}, M. and {Andrae}, R. and {Audard}, M. and {Bellas-Velidis}, I. and {Benson}, K. and {Berthier}, J. and {Blomme}, R. and {Burgess}, P. and {Busso}, G. and {Carry}, B. and {Cellino}, A. and {Clementini}, G. and {Clotet}, M. and {Creevey}, O. and {Davidson}, M. and {De Ridder}, J. and {Delchambre}, L. and {Dell'Oro}, A. and {Ducourant}, C. and {Fern{\'a}ndez-Hern{\'a}ndez}, J. and {Fouesneau}, M. and {Fr{\'e}mat}, Y. and {Galluccio}, L. and {Garc{\'\i}a-Torres}, M. and {Gonz{\'a}lez-N{\'u}{\~n}ez}, J. and {Gonz{\'a}lez-Vidal}, J.~J. and {Gosset}, E. and {Guy}, L.~P. and {Halbwachs}, J. -L. and {Hambly}, N.~C. and {Harrison}, D.~L. and {Hern{\'a}ndez}, J. and {Hestroffer}, D. and {Hodgkin}, S.~T. and {Hutton}, A. and {Jasniewicz}, G. and {Jean-Antoine-Piccolo}, A. and {Jordan}, S. and {Korn}, A.~J. and {Krone-Martins}, A. and {Lanzafame}, A.~C. and {Lebzelter}, T. and {L{\"o}ffler}, W. and {Manteiga}, M. and {Marrese}, P.~M. and {Mart{\'\i}n-Fleitas}, J.~M. and {Moitinho}, A. and {Mora}, A. and {Muinonen}, K. and {Osinde}, J. and {Pancino}, E. and {Pauwels}, T. and {Petit}, J. -M. and {Recio-Blanco}, A. and {Richards}, P.~J. and {Rimoldini}, L. and {Robin}, A.~C. and {Sarro}, L.~M. and {Siopis}, C. and {Smith}, M. and {Sozzetti}, A. and {S{\"u}veges}, M. and {Torra}, J. and {van Reeven}, W. and {Abbas}, U. and {Abreu Aramburu}, A. and {Accart}, S. and {Aerts}, C. and {Altavilla}, G. and {{\'A}lvarez}, M.~A. and {Alvarez}, R. and {Alves}, J. and {Anderson}, R.~I. and {Andrei}, A.~H. and {Anglada Varela}, E. and {Antiche}, E. and {Antoja}, T. and {Arcay}, B. and {Astraatmadja}, T.~L. and {Bach}, N. and {Baker}, S.~G. and {Balaguer-N{\'u}{\~n}ez}, L. and {Balm}, P. and {Barache}, C. and {Barata}, C. and {Barbato}, D. and {Barblan}, F. and {Barklem}, P.~S. and {Barrado}, D. and {Barros}, M. and {Barstow}, M.~A. and {Bartholom{\'e} Mu{\~n}oz}, S. and {Bassilana}, J. -L. and {Becciani}, U. and {Bellazzini}, M. and {Berihuete}, A. and {Bertone}, S. and {Bianchi}, L. and {Bienaym{\'e}}, O. and {Blanco-Cuaresma}, S. and {Boch}, T. and {Boeche}, C. and {Bombrun}, A. and {Borrachero}, R. and {Bossini}, D. and {Bouquillon}, S. and {Bourda}, G. and {Bragaglia}, A. and {Bramante}, L. and {Breddels}, M.~A. and {Bressan}, A. and {Brouillet}, N. and {Br{\"u}semeister}, T. and {Brugaletta}, E. and {Bucciarelli}, B. and {Burlacu}, A. and {Busonero}, D. and {Butkevich}, A.~G. and {Buzzi}, R. and {Caffau}, E. and {Cancelliere}, R. and {Cannizzaro}, G. and {Cantat-Gaudin}, T. and {Carballo}, R. and {Carlucci}, T. and {Carrasco}, J.~M. and {Casamiquela}, L. and {Castellani}, M. and {Castro-Ginard}, A. and {Charlot}, P. and {Chemin}, L. and {Chiavassa}, A. and {Cocozza}, G. and {Costigan}, G. and {Cowell}, S. and {Crifo}, F. and {Crosta}, M. and {Crowley}, C. and {Cuypers}, J. and {Dafonte}, C. and {Damerdji}, Y. and {Dapergolas}, A. and {David}, P. and {David}, M. and {de Laverny}, P. and {De Luise}, F.},
        title = "{Gaia Data Release 2. Summary of the contents and survey properties}",
      journal = {\aap},
         year = 2018,
        month = aug,
       volume = {616},
          eid = {A1},
        pages = {A1},
          doi = {10.1051/0004-6361/201833051},
archivePrefix = {arXiv},
       eprint = {1804.09365},
 primaryClass = {astro-ph.GA},
       adsurl = {https://ui.adsabs.harvard.edu/abs/2018A&A...616A...1G}
}

@ARTICLE{GaiaDR3,
       author = {{Gaia Collaboration} and {Vallenari}, A. and {Brown}, A.~G.~A. and {Prusti}, T. and {de Bruijne}, J.~H.~J. and {Arenou}, F. and {Babusiaux}, C. and {Biermann}, M. and {Creevey}, O.~L. and {Ducourant}, C. and {Evans}, D.~W. and {Eyer}, L. and {Guerra}, R. and {Hutton}, A. and {Jordi}, C. and {Klioner}, S.~A. and {Lammers}, U.~L. and {Lindegren}, L. and {Luri}, X. and {Mignard}, F. and {Panem}, C. and {Pourbaix}, D. and {Randich}, S. and {Sartoretti}, P. and {Soubiran}, C. and {Tanga}, P. and {Walton}, N.~A. and {Bailer-Jones}, C.~A.~L. and {Bastian}, U. and {Drimmel}, R. and {Jansen}, F. and {Katz}, D. and {Lattanzi}, M.~G. and {van Leeuwen}, F. and {Bakker}, J. and {Cacciari}, C. and {Casta{\~n}eda}, J. and {De Angeli}, F. and {Fabricius}, C. and {Fouesneau}, M. and {Fr{\'e}mat}, Y. and {Galluccio}, L. and {Guerrier}, A. and {Heiter}, U. and {Masana}, E. and {Messineo}, R. and {Mowlavi}, N. and {Nicolas}, C. and {Nienartowicz}, K. and {Pailler}, F. and {Panuzzo}, P. and {Riclet}, F. and {Roux}, W. and {Seabroke}, G.~M. and {Sordo}, R. and {Th{\'e}venin}, F. and {Gracia-Abril}, G. and {Portell}, J. and {Teyssier}, D. and {Altmann}, M. and {Andrae}, R. and {Audard}, M. and {Bellas-Velidis}, I. and {Benson}, K. and {Berthier}, J. and {Blomme}, R. and {Burgess}, P.~W. and {Busonero}, D. and {Busso}, G. and {C{\'a}novas}, H. and {Carry}, B. and {Cellino}, A. and {Cheek}, N. and {Clementini}, G. and {Damerdji}, Y. and {Davidson}, M. and {de Teodoro}, P. and {Nu{\~n}ez Campos}, M. and {Delchambre}, L. and {Dell'Oro}, A. and {Esquej}, P. and {Fern{\'a}ndez-Hern{\'a}ndez}, J. and {Fraile}, E. and {Garabato}, D. and {Garc{\'\i}a-Lario}, P. and {Gosset}, E. and {Haigron}, R. and {Halbwachs}, J. -L. and {Hambly}, N.~C. and {Harrison}, D.~L. and {Hern{\'a}ndez}, J. and {Hestroffer}, D. and {Hodgkin}, S.~T. and {Holl}, B. and {Jan{\ss}en}, K. and {Jevardat de Fombelle}, G. and {Jordan}, S. and {Krone-Martins}, A. and {Lanzafame}, A.~C. and {L{\"o}ffler}, W. and {Marchal}, O. and {Marrese}, P.~M. and {Moitinho}, A. and {Muinonen}, K. and {Osborne}, P. and {Pancino}, E. and {Pauwels}, T. and {Recio-Blanco}, A. and {Reyl{\'e}}, C. and {Riello}, M. and {Rimoldini}, L. and {Roegiers}, T. and {Rybizki}, J. and {Sarro}, L.~M. and {Siopis}, C. and {Smith}, M. and {Sozzetti}, A. and {Utrilla}, E. and {van Leeuwen}, M. and {Abbas}, U. and {{\'A}brah{\'a}m}, P. and {Abreu Aramburu}, A. and {Aerts}, C. and {Aguado}, J.~J. and {Ajaj}, M. and {Aldea-Montero}, F. and {Altavilla}, G. and {{\'A}lvarez}, M.~A. and {Alves}, J. and {Anders}, F. and {Anderson}, R.~I. and {Anglada Varela}, E. and {Antoja}, T. and {Baines}, D. and {Baker}, S.~G. and {Balaguer-N{\'u}{\~n}ez}, L. and {Balbinot}, E. and {Balog}, Z. and {Barache}, C. and {Barbato}, D. and {Barros}, M. and {Barstow}, M.~A. and {Bartolom{\'e}}, S. and {Bassilana}, J. -L. and {Bauchet}, N. and {Becciani}, U. and {Bellazzini}, M. and {Berihuete}, A. and {Bernet}, M. and {Bertone}, S. and {Bianchi}, L. and {Binnenfeld}, A. and {Blanco-Cuaresma}, S. and {Blazere}, A. and {Boch}, T. and {Bombrun}, A. and {Bossini}, D. and {Bouquillon}, S. and {Bragaglia}, A. and {Bramante}, L. and {Breedt}, E. and {Bressan}, A. and {Brouillet}, N. and {Brugaletta}, E. and {Bucciarelli}, B. and {Burlacu}, A. and {Butkevich}, A.~G. and {Buzzi}, R. and {Caffau}, E. and {Cancelliere}, R. and {Cantat-Gaudin}, T. and {Carballo}, R. and {Carlucci}, T. and {Carnerero}, M.~I. and {Carrasco}, J.~M. and {Casamiquela}, L. and {Castellani}, M. and {Castro-Ginard}, A. and {Chaoul}, L. and {Charlot}, P. and {Chemin}, L. and {Chiaramida}, V. and {Chiavassa}, A. and {Chornay}, N. and {Comoretto}, G. and {Contursi}, G. and {Cooper}, W.~J. and {Cornez}, T. and {Cowell}, S. and {Crifo}, F. and {Cropper}, M. and {Crosta}, M. and {Crowley}, C. and {Dafonte}, C. and {Dapergolas}, A. and {David}, M. and {David}, P. and {de Laverny}, P. and {De Luise}, F. and {De March}, R. and {De Ridder}, J. and {de Souza}, R. and {de Torres}, A. and {del Peloso}, E.~F. and {del Pozo}, E. and {Delbo}, M. and {Delgado}, A. and {Delisle}, J. -B. and {Demouchy}, C. and {Dharmawardena}, T.~E. and {Di Matteo}, P. and {Diakite}, S. and {Diener}, C. and {Distefano}, E. and {Dolding}, C. and {Edvardsson}, B. and {Enke}, H. and {Fabre}, C. and {Fabrizio}, M. and {Faigler}, S. and {Fedorets}, G. and {Fernique}, P. and {Fienga}, A. and {Figueras}, F. and {Fournier}, Y. and {Fouron}, C. and {Fragkoudi}, F. and {Gai}, M. and {Garcia-Gutierrez}, A. and {Garcia-Reinaldos}, M. and {Garc{\'\i}a-Torres}, M. and {Garofalo}, A. and {Gavel}, A. and {Gavras}, P. and {Gerlach}, E. and {Geyer}, R. and {Giacobbe}, P. and {Gilmore}, G. and {Girona}, S. and {Giuffrida}, G. and {Gomel}, R. and {Gomez}, A. and {Gonz{\'a}lez-N{\'u}{\~n}ez}, J. and {Gonz{\'a}lez-Santamar{\'\i}a}, I. and {Gonz{\'a}lez-Vidal}, J.~J. and {Granvik}, M. and {Guillout}, P. and {Guiraud}, J. and {Guti{\'e}rrez-S{\'a}nchez}, R. and {Guy}, L.~P. and {Hatzidimitriou}, D. and {Hauser}, M. and {Haywood}, M. and {Helmer}, A. and {Helmi}, A. and {Sarmiento}, M.~H. and {Hidalgo}, S.~L. and {Hilger}, T. and {H{\l}adczuk}, N. and {Hobbs}, D. and {Holland}, G. and {Huckle}, H.~E. and {Jardine}, K. and {Jasniewicz}, G. and {Jean-Antoine Piccolo}, A. and {Jim{\'e}nez-Arranz}, {\'O}. and {Jorissen}, A. and {Juaristi Campillo}, J. and {Julbe}, F. and {Karbevska}, L. and {Kervella}, P. and {Khanna}, S. and {Kontizas}, M. and {Kordopatis}, G. and {Korn}, A.~J. and {K{\'o}sp{\'a}l}, {\'A}. and {Kostrzewa-Rutkowska}, Z. and {Kruszy{\'n}ska}, K. and {Kun}, M. and {Laizeau}, P. and {Lambert}, S. and {Lanza}, A.~F. and {Lasne}, Y. and {Le Campion}, J. -F. and {Lebreton}, Y. and {Lebzelter}, T. and {Leccia}, S. and {Leclerc}, N. and {Lecoeur-Taibi}, I. and {Liao}, S. and {Licata}, E.~L. and {Lindstr{\o}m}, H.~E.~P. and {Lister}, T.~A. and {Livanou}, E. and {Lobel}, A. and {Lorca}, A. and {Loup}, C. and {Madrero Pardo}, P. and {Magdaleno Romeo}, A. and {Managau}, S. and {Mann}, R.~G. and {Manteiga}, M. and {Marchant}, J.~M. and {Marconi}, M. and {Marcos}, J. and {Marcos Santos}, M.~M.~S. and {Mar{\'\i}n Pina}, D. and {Marinoni}, S. and {Marocco}, F. and {Marshall}, D.~J. and {Martin Polo}, L. and {Mart{\'\i}n-Fleitas}, J.~M. and {Marton}, G. and {Mary}, N. and {Masip}, A. and {Massari}, D. and {Mastrobuono-Battisti}, A. and {Mazeh}, T. and {McMillan}, P.~J. and {Messina}, S. and {Michalik}, D. and {Millar}, N.~R. and {Mints}, A. and {Molina}, D. and {Molinaro}, R. and {Moln{\'a}r}, L. and {Monari}, G. and {Mongui{\'o}}, M. and {Montegriffo}, P. and {Montero}, A. and {Mor}, R. and {Mora}, A. and {Morbidelli}, R. and {Morel}, T. and {Morris}, D. and {Muraveva}, T. and {Murphy}, C.~P. and {Musella}, I. and {Nagy}, Z. and {Noval}, L. and {Oca{\~n}a}, F. and {Ogden}, A. and {Ordenovic}, C. and {Osinde}, J.~O. and {Pagani}, C. and {Pagano}, I. and {Palaversa}, L. and {Palicio}, P.~A. and {Pallas-Quintela}, L. and {Panahi}, A. and {Payne-Wardenaar}, S. and {Pe{\~n}alosa Esteller}, X. and {Penttil{\"a}}, A. and {Pichon}, B. and {Piersimoni}, A.~M. and {Pineau}, F. -X. and {Plachy}, E. and {Plum}, G. and {Poggio}, E. and {Pr{\v{s}}a}, A. and {Pulone}, L. and {Racero}, E. and {Ragaini}, S. and {Rainer}, M. and {Raiteri}, C.~M. and {Rambaux}, N. and {Ramos}, P. and {Ramos-Lerate}, M. and {Re Fiorentin}, P. and {Regibo}, S. and {Richards}, P.~J. and {Rios Diaz}, C. and {Ripepi}, V. and {Riva}, A. and {Rix}, H. -W. and {Rixon}, G. and {Robichon}, N. and {Robin}, A.~C. and {Robin}, C. and {Roelens}, M. and {Rogues}, H.~R.~O. and {Rohrbasser}, L. and {Romero-G{\'o}mez}, M. and {Rowell}, N. and {Royer}, F. and {Ruz Mieres}, D. and {Rybicki}, K.~A. and {Sadowski}, G. and {S{\'a}ez N{\'u}{\~n}ez}, A. and {Sagrist{\`a} Sell{\'e}s}, A. and {Sahlmann}, J. and {Salguero}, E. and {Samaras}, N. and {Sanchez Gimenez}, V. and {Sanna}, N. and {Santove{\~n}a}, R. and {Sarasso}, M. and {Schultheis}, M. and {Sciacca}, E. and {Segol}, M. and {Segovia}, J.~C. and {S{\'e}gransan}, D. and {Semeux}, D. and {Shahaf}, S. and {Siddiqui}, H.~I. and {Siebert}, A. and {Siltala}, L. and {Silvelo}, A. and {Slezak}, E. and {Slezak}, I. and {Smart}, R.~L. and {Snaith}, O.~N. and {Solano}, E. and {Solitro}, F. and {Souami}, D. and {Souchay}, J. and {Spagna}, A. and {Spina}, L. and {Spoto}, F. and {Steele}, I.~A. and {Steidelm{\"u}ller}, H. and {Stephenson}, C.~A. and {S{\"u}veges}, M. and {Surdej}, J. and {Szabados}, L. and {Szegedi-Elek}, E. and {Taris}, F. and {Taylor}, M.~B. and {Teixeira}, R. and {Tolomei}, L. and {Tonello}, N. and {Torra}, F. and {Torra}, J. and {Torralba Elipe}, G. and {Trabucchi}, M. and {Tsounis}, A.~T. and {Turon}, C. and {Ulla}, A. and {Unger}, N. and {Vaillant}, M.~V. and {van Dillen}, E. and {van Reeven}, W. and {Vanel}, O. and {Vecchiato}, A. and {Viala}, Y. and {Vicente}, D. and {Voutsinas}, S. and {Weiler}, M. and {Wevers}, T. and {Wyrzykowski}, {\L}. and {Yoldas}, A. and {Yvard}, P. and {Zhao}, H. and {Zorec}, J. and {Zucker}, S. and {Zwitter}, T.},
        title = "{Gaia Data Release 3. Summary of the content and survey properties}",
      journal = {\aap},
         year = 2023,
        month = jun,
       volume = {674},
          eid = {A1},
        pages = {A1},
          doi = {10.1051/0004-6361/202243940},
archivePrefix = {arXiv},
       eprint = {2208.00211},
 primaryClass = {astro-ph.GA},
       adsurl = {https://ui.adsabs.harvard.edu/abs/2023A&A...674A...1G}
}

@ARTICLE{Piecka2024,
       author = {{Piecka}, M. and {Hutschenreuter}, S. and {Alves}, J.},
        title = "{Towards a complete picture of the Sco-Cen outflow}",
      journal = {\aap},
         year = 2024,
        month = sep,
       volume = {689},
          eid = {A84},
        pages = {A84},
          doi = {10.1051/0004-6361/202450936},
archivePrefix = {arXiv},
       eprint = {2407.13226},
 primaryClass = {astro-ph.GA},
       adsurl = {https://ui.adsabs.harvard.edu/abs/2024A&A...689A..84P}
}

@ARTICLE{Ducourant2017,
       author = {{Ducourant}, C. and {Teixeira}, R. and {Krone-Martins}, A. and {Bontemps}, S. and {Despois}, D. and {Galli}, P.~A.~B. and {Bouy}, H. and {Le Campion}, J.~F. and {Rapaport}, M. and {Cuillandre}, J.~C.},
        title = "{Proper motion survey and kinematic analysis of the {\ensuremath{\rho}} Ophiuchi embedded cluster}",
      journal = {\aap},
         year = 2017,
        month = jan,
       volume = {597},
          eid = {A90},
        pages = {A90},
          doi = {10.1051/0004-6361/201527574},
archivePrefix = {arXiv},
       eprint = {1609.04963},
 primaryClass = {astro-ph.SR},
       adsurl = {https://ui.adsabs.harvard.edu/abs/2017A&A...597A..90D}
}

@ARTICLE{Megeath2022,
       author = {{Megeath}, S.~T. and {Gutermuth}, R.~A. and {Kounkel}, M.~A.},
        title = "{Low Mass Stars as Tracers of Star and Cluster Formation}",
      journal = {\pasp},
         year = 2022,
        month = apr,
       volume = {134},
       number = {1034},
          eid = {042001},
        pages = {042001},
          doi = {10.1088/1538-3873/ac4c9c},
archivePrefix = {arXiv},
       eprint = {2203.03655},
 primaryClass = {astro-ph.GA},
       adsurl = {https://ui.adsabs.harvard.edu/abs/2022PASP..134d2001M}
}

@ARTICLE{Grossschedl2021,
       author = {{Gro{\ss}schedl}, Josefa E. and {Alves}, Jo{\~a}o and {Meingast}, Stefan and {Herbst-Kiss}, Gabor},
        title = "{3D dynamics of the Orion cloud complex. Discovery of coherent radial gas motions at the 100-pc scale}",
      journal = {\aap},
         year = 2021,
        month = mar,
       volume = {647},
          eid = {A91},
        pages = {A91},
          doi = {10.1051/0004-6361/202038913},
archivePrefix = {arXiv},
       eprint = {2007.07254},
 primaryClass = {astro-ph.SR},
       adsurl = {https://ui.adsabs.harvard.edu/abs/2021A&A...647A..91G}
}

@ARTICLE{Zhang2023,
       author = {{Zhang}, Miaomiao},
        title = "{Distances to Nearby Molecular Clouds Traced by Young Stars}",
      journal = {\apjs},
         year = 2023,
        month = apr,
       volume = {265},
       number = {2},
          eid = {59},
        pages = {59},
          doi = {10.3847/1538-4365/acc1e8},
archivePrefix = {arXiv},
       eprint = {2303.01053},
 primaryClass = {astro-ph.GA},
       adsurl = {https://ui.adsabs.harvard.edu/abs/2023ApJS..265...59Z}
}

@software{SWarp,
       author = {{Bertin}, Emmanuel},
        title = "{SWarp: Resampling and Co-adding FITS Images Together}",
 howpublished = {Astrophysics Source Code Library, record ascl:1010.068},
         year = 2010,
        month = oct,
          eid = {ascl:1010.068},
archivePrefix = {ascl},
       eprint = {1010.068},
       adsurl = {https://ui.adsabs.harvard.edu/abs/2010ascl.soft10068B}
}

@ARTICLE{Hipparcos2007,
       author = {{van Leeuwen}, F.},
        title = "{Validation of the new Hipparcos reduction}",
      journal = {\aap},
         year = 2007,
        month = nov,
       volume = {474},
       number = {2},
        pages = {653-664},
          doi = {10.1051/0004-6361:20078357},
archivePrefix = {arXiv},
       eprint = {0708.1752},
 primaryClass = {astro-ph},
       adsurl = {https://ui.adsabs.harvard.edu/abs/2007A&A...474..653V}
}

@ARTICLE{Hodapp2026,
       author = {{Hodapp}, Klaus W. and {Boogert}, Adwin and {Johnstone}, Doug and {Le Gouellec}, Valentin J.~M. and {Tsiakaliari}, Eleni and {Fraser}, Helen J. and {Chu}, Laurie L. and {Greene}, Thomas and {Rieke}, Marcia J.},
        title = "{The Outflow of the B335 Protostar II: After the Outburst}",
      journal = {arXiv e-prints},
         year = 2026,
        month = feb,
          eid = {arXiv:2602.12060},
        pages = {arXiv:2602.12060},
          doi = {10.48550/arXiv.2602.12060},
archivePrefix = {arXiv},
       eprint = {2602.12060},
 primaryClass = {astro-ph.SR},
       adsurl = {https://ui.adsabs.harvard.edu/abs/2026arXiv260212060H}
}

@ARTICLE{Lopez2022,
       author = {{L{\'o}pez}, Rosario and {Estalella}, Robert and {Beltr{\'a}n}, Maria T. and {Massi}, Fabrizio and {Acosta-Pulido}, Jos{\'e} A. and {Girart}, Josep M.},
        title = "{Collision of protostellar jets in the star-forming region IC 1396N. Analysis of knot proper motions}",
      journal = {\aap},
         year = 2022,
        month = may,
       volume = {661},
          eid = {A106},
        pages = {A106},
          doi = {10.1051/0004-6361/202243380},
archivePrefix = {arXiv},
       eprint = {2203.13853},
 primaryClass = {astro-ph.SR},
       adsurl = {https://ui.adsabs.harvard.edu/abs/2022A&A...661A.106L}
}

@ARTICLE{Barsony2012,
       author = {{Barsony}, Mary and {Haisch}, Karl E. and {Marsh}, Kenneth A. and {McCarthy}, Chris},
        title = "{A Significant Population of Candidate New Members of the {\ensuremath{\rho}} Ophiuchi Cluster}",
      journal = {\apj},
         year = 2012,
        month = may,
       volume = {751},
       number = {1},
          eid = {22},
        pages = {22},
          doi = {10.1088/0004-637X/751/1/22},
archivePrefix = {arXiv},
       eprint = {1206.4552},
 primaryClass = {astro-ph.GA},
       adsurl = {https://ui.adsabs.harvard.edu/abs/2012ApJ...751...22B}
}

@ARTICLE{astropy1,
       author = {{Astropy Collaboration} and {Robitaille}, Thomas P. and {Tollerud}, Erik J. and {Greenfield}, Perry and {Droettboom}, Michael and {Bray}, Erik and {Aldcroft}, Tom and {Davis}, Matt and {Ginsburg}, Adam and {Price-Whelan}, Adrian M. and {Kerzendorf}, Wolfgang E. and {Conley}, Alexander and {Crighton}, Neil and {Barbary}, Kyle and {Muna}, Demitri and {Ferguson}, Henry and {Grollier}, Fr{\'e}d{\'e}ric and {Parikh}, Madhura M. and {Nair}, Prasanth H. and {Unther}, Hans M. and {Deil}, Christoph and {Woillez}, Julien and {Conseil}, Simon and {Kramer}, Roban and {Turner}, James E.~H. and {Singer}, Leo and {Fox}, Ryan and {Weaver}, Benjamin A. and {Zabalza}, Victor and {Edwards}, Zachary I. and {Azalee Bostroem}, K. and {Burke}, D.~J. and {Casey}, Andrew R. and {Crawford}, Steven M. and {Dencheva}, Nadia and {Ely}, Justin and {Jenness}, Tim and {Labrie}, Kathleen and {Lim}, Pey Lian and {Pierfederici}, Francesco and {Pontzen}, Andrew and {Ptak}, Andy and {Refsdal}, Brian and {Servillat}, Mathieu and {Streicher}, Ole},
        title = "{Astropy: A community Python package for astronomy}",
      journal = {\aap},
         year = 2013,
        month = oct,
       volume = {558},
          eid = {A33},
        pages = {A33},
          doi = {10.1051/0004-6361/201322068},
archivePrefix = {arXiv},
       eprint = {1307.6212},
 primaryClass = {astro-ph.IM},
       adsurl = {https://ui.adsabs.harvard.edu/abs/2013A&A...558A..33A}
}

@ARTICLE{astropy2,
       author = {{Astropy Collaboration} and {Price-Whelan}, A.~M. and {Sip{\H{o}}cz}, B.~M. and {G{\"u}nther}, H.~M. and {Lim}, P.~L. and {Crawford}, S.~M. and {Conseil}, S. and {Shupe}, D.~L. and {Craig}, M.~W. and {Dencheva}, N. and {Ginsburg}, A. and {VanderPlas}, J.~T. and {Bradley}, L.~D. and {P{\'e}rez-Su{\'a}rez}, D. and {de Val-Borro}, M. and {Aldcroft}, T.~L. and {Cruz}, K.~L. and {Robitaille}, T.~P. and {Tollerud}, E.~J. and {Ardelean}, C. and {Babej}, T. and {Bach}, Y.~P. and {Bachetti}, M. and {Bakanov}, A.~V. and {Bamford}, S.~P. and {Barentsen}, G. and {Barmby}, P. and {Baumbach}, A. and {Berry}, K.~L. and {Biscani}, F. and {Boquien}, M. and {Bostroem}, K.~A. and {Bouma}, L.~G. and {Brammer}, G.~B. and {Bray}, E.~M. and {Breytenbach}, H. and {Buddelmeijer}, H. and {Burke}, D.~J. and {Calderone}, G. and {Cano Rodr{\'\i}guez}, J.~L. and {Cara}, M. and {Cardoso}, J.~V.~M. and {Cheedella}, S. and {Copin}, Y. and {Corrales}, L. and {Crichton}, D. and {D'Avella}, D. and {Deil}, C. and {Depagne}, {\'E}. and {Dietrich}, J.~P. and {Donath}, A. and {Droettboom}, M. and {Earl}, N. and {Erben}, T. and {Fabbro}, S. and {Ferreira}, L.~A. and {Finethy}, T. and {Fox}, R.~T. and {Garrison}, L.~H. and {Gibbons}, S.~L.~J. and {Goldstein}, D.~A. and {Gommers}, R. and {Greco}, J.~P. and {Greenfield}, P. and {Groener}, A.~M. and {Grollier}, F. and {Hagen}, A. and {Hirst}, P. and {Homeier}, D. and {Horton}, A.~J. and {Hosseinzadeh}, G. and {Hu}, L. and {Hunkeler}, J.~S. and {Ivezi{\'c}}, {\v{Z}}. and {Jain}, A. and {Jenness}, T. and {Kanarek}, G. and {Kendrew}, S. and {Kern}, N.~S. and {Kerzendorf}, W.~E. and {Khvalko}, A. and {King}, J. and {Kirkby}, D. and {Kulkarni}, A.~M. and {Kumar}, A. and {Lee}, A. and {Lenz}, D. and {Littlefair}, S.~P. and {Ma}, Z. and {Macleod}, D.~M. and {Mastropietro}, M. and {McCully}, C. and {Montagnac}, S. and {Morris}, B.~M. and {Mueller}, M. and {Mumford}, S.~J. and {Muna}, D. and {Murphy}, N.~A. and {Nelson}, S. and {Nguyen}, G.~H. and {Ninan}, J.~P. and {N{\"o}the}, M. and {Ogaz}, S. and {Oh}, S. and {Parejko}, J.~K. and {Parley}, N. and {Pascual}, S. and {Patil}, R. and {Patil}, A.~A. and {Plunkett}, A.~L. and {Prochaska}, J.~X. and {Rastogi}, T. and {Reddy Janga}, V. and {Sabater}, J. and {Sakurikar}, P. and {Seifert}, M. and {Sherbert}, L.~E. and {Sherwood-Taylor}, H. and {Shih}, A.~Y. and {Sick}, J. and {Silbiger}, M.~T. and {Singanamalla}, S. and {Singer}, L.~P. and {Sladen}, P.~H. and {Sooley}, K.~A. and {Sornarajah}, S. and {Streicher}, O. and {Teuben}, P. and {Thomas}, S.~W. and {Tremblay}, G.~R. and {Turner}, J.~E.~H. and {Terr{\'o}n}, V. and {van Kerkwijk}, M.~H. and {de la Vega}, A. and {Watkins}, L.~L. and {Weaver}, B.~A. and {Whitmore}, J.~B. and {Woillez}, J. and {Zabalza}, V. and {Astropy Contributors}},
        title = "{The Astropy Project: Building an Open-science Project and Status of the v2.0 Core Package}",
      journal = {\aj},
         year = 2018,
        month = sep,
       volume = {156},
       number = {3},
          eid = {123},
        pages = {123},
          doi = {10.3847/1538-3881/aabc4f},
archivePrefix = {arXiv},
       eprint = {1801.02634},
 primaryClass = {astro-ph.IM},
       adsurl = {https://ui.adsabs.harvard.edu/abs/2018AJ....156..123A}
}

@ARTICLE{astropy3,
       author = {{Astropy Collaboration} and {Price-Whelan}, Adrian M. and {Lim}, Pey Lian and {Earl}, Nicholas and {Starkman}, Nathaniel and {Bradley}, Larry and {Shupe}, David L. and {Patil}, Aarya A. and {Corrales}, Lia and {Brasseur}, C.~E. and {N{\"o}the}, Maximilian and {Donath}, Axel and {Tollerud}, Erik and {Morris}, Brett M. and {Ginsburg}, Adam and {Vaher}, Eero and {Weaver}, Benjamin A. and {Tocknell}, James and {Jamieson}, William and {van Kerkwijk}, Marten H. and {Robitaille}, Thomas P. and {Merry}, Bruce and {Bachetti}, Matteo and {G{\"u}nther}, H. Moritz and {Aldcroft}, Thomas L. and {Alvarado-Montes}, Jaime A. and {Archibald}, Anne M. and {B{\'o}di}, Attila and {Bapat}, Shreyas and {Barentsen}, Geert and {Baz{\'a}n}, Juanjo and {Biswas}, Manish and {Boquien}, M{\'e}d{\'e}ric and {Burke}, D.~J. and {Cara}, Daria and {Cara}, Mihai and {Conroy}, Kyle E. and {Conseil}, Simon and {Craig}, Matthew W. and {Cross}, Robert M. and {Cruz}, Kelle L. and {D'Eugenio}, Francesco and {Dencheva}, Nadia and {Devillepoix}, Hadrien A.~R. and {Dietrich}, J{\"o}rg P. and {Eigenbrot}, Arthur Davis and {Erben}, Thomas and {Ferreira}, Leonardo and {Foreman-Mackey}, Daniel and {Fox}, Ryan and {Freij}, Nabil and {Garg}, Suyog and {Geda}, Robel and {Glattly}, Lauren and {Gondhalekar}, Yash and {Gordon}, Karl D. and {Grant}, David and {Greenfield}, Perry and {Groener}, Austen M. and {Guest}, Steve and {Gurovich}, Sebastian and {Handberg}, Rasmus and {Hart}, Akeem and {Hatfield-Dodds}, Zac and {Homeier}, Derek and {Hosseinzadeh}, Griffin and {Jenness}, Tim and {Jones}, Craig K. and {Joseph}, Prajwel and {Kalmbach}, J. Bryce and {Karamehmetoglu}, Emir and {Ka{\l}uszy{\'n}ski}, Miko{\l}aj and {Kelley}, Michael S.~P. and {Kern}, Nicholas and {Kerzendorf}, Wolfgang E. and {Koch}, Eric W. and {Kulumani}, Shankar and {Lee}, Antony and {Ly}, Chun and {Ma}, Zhiyuan and {MacBride}, Conor and {Maljaars}, Jakob M. and {Muna}, Demitri and {Murphy}, N.~A. and {Norman}, Henrik and {O'Steen}, Richard and {Oman}, Kyle A. and {Pacifici}, Camilla and {Pascual}, Sergio and {Pascual-Granado}, J. and {Patil}, Rohit R. and {Perren}, Gabriel I. and {Pickering}, Timothy E. and {Rastogi}, Tanuj and {Roulston}, Benjamin R. and {Ryan}, Daniel F. and {Rykoff}, Eli S. and {Sabater}, Jose and {Sakurikar}, Parikshit and {Salgado}, Jes{\'u}s and {Sanghi}, Aniket and {Saunders}, Nicholas and {Savchenko}, Volodymyr and {Schwardt}, Ludwig and {Seifert-Eckert}, Michael and {Shih}, Albert Y. and {Jain}, Anany Shrey and {Shukla}, Gyanendra and {Sick}, Jonathan and {Simpson}, Chris and {Singanamalla}, Sudheesh and {Singer}, Leo P. and {Singhal}, Jaladh and {Sinha}, Manodeep and {Sip{\H{o}}cz}, Brigitta M. and {Spitler}, Lee R. and {Stansby}, David and {Streicher}, Ole and {{\v{S}}umak}, Jani and {Swinbank}, John D. and {Taranu}, Dan S. and {Tewary}, Nikita and {Tremblay}, Grant R. and {de Val-Borro}, Miguel and {Van Kooten}, Samuel J. and {Vasovi{\'c}}, Zlatan and {Verma}, Shresth and {de Miranda Cardoso}, Jos{\'e} Vin{\'\i}cius and {Williams}, Peter K.~G. and {Wilson}, Tom J. and {Winkel}, Benjamin and {Wood-Vasey}, W.~M. and {Xue}, Rui and {Yoachim}, Peter and {Zhang}, Chen and {Zonca}, Andrea and {Astropy Project Contributors}},
        title = "{The Astropy Project: Sustaining and Growing a Community-oriented Open-source Project and the Latest Major Release (v5.0) of the Core Package}",
      journal = {\apj},
         year = 2022,
        month = aug,
       volume = {935},
       number = {2},
          eid = {167},
        pages = {167},
          doi = {10.3847/1538-4357/ac7c74},
archivePrefix = {arXiv},
       eprint = {2206.14220},
 primaryClass = {astro-ph.IM},
       adsurl = {https://ui.adsabs.harvard.edu/abs/2022ApJ...935..167A}
}

@ARTICLE{scipy,
  author  = {Virtanen, Pauli and Gommers, Ralf and Oliphant, Travis E. and
            Haberland, Matt and Reddy, Tyler and Cournapeau, David and
            Burovski, Evgeni and Peterson, Pearu and Weckesser, Warren and
            Bright, Jonathan and {van der Walt}, St{\'e}fan J. and
            Brett, Matthew and Wilson, Joshua and Millman, K. Jarrod and
            Mayorov, Nikolay and Nelson, Andrew R. J. and Jones, Eric and
            Kern, Robert and Larson, Eric and Carey, C J and
            Polat, {\.I}lhan and Feng, Yu and Moore, Eric W. and
            {VanderPlas}, Jake and Laxalde, Denis and Perktold, Josef and
            Cimrman, Robert and Henriksen, Ian and Quintero, E. A. and
            Harris, Charles R. and Archibald, Anne M. and
            Ribeiro, Ant{\^o}nio H. and Pedregosa, Fabian and
            {van Mulbregt}, Paul and {SciPy 1.0 Contributors}},
  title   = {{{SciPy} 1.0: Fundamental Algorithms for Scientific
            Computing in Python}},
  journal = {Nature Methods},
  year    = {2020},
  volume  = {17},
  pages   = {261--272},
  adsurl  = {https://rdcu.be/b08Wh},
  doi     = {10.1038/s41592-019-0686-2},
}

@Article{         numpy,
 title         = {Array programming with {NumPy}},
 author        = {Charles R. Harris and K. Jarrod Millman and St{\'{e}}fan J.
                 van der Walt and Ralf Gommers and Pauli Virtanen and David
                 Cournapeau and Eric Wieser and Julian Taylor and Sebastian
                 Berg and Nathaniel J. Smith and Robert Kern and Matti Picus
                 and Stephan Hoyer and Marten H. van Kerkwijk and Matthew
                 Brett and Allan Haldane and Jaime Fern{\'{a}}ndez del
                 R{\'{i}}o and Mark Wiebe and Pearu Peterson and Pierre
                 G{\'{e}}rard-Marchant and Kevin Sheppard and Tyler Reddy and
                 Warren Weckesser and Hameer Abbasi and Christoph Gohlke and
                 Travis E. Oliphant},
 year          = {2020},
 month         = sep,
 journal       = {Nature},
 volume        = {585},
 number        = {7825},
 pages         = {357--362},
 doi           = {10.1038/s41586-020-2649-2},
 publisher     = {Springer Science and Business Media {LLC}},
 url           = {https://doi.org/10.1038/s41586-020-2649-2}
}

@Article{matplotlib,
  Author    = {Hunter, J. D.},
  Title     = {Matplotlib: A 2D graphics environment},
  Journal   = {Computing in Science \& Engineering},
  Volume    = {9},
  Number    = {3},
  Pages     = {90--95},
  publisher = {IEEE COMPUTER SOC},
  doi       = {10.1109/MCSE.2007.55},
  year      = 2007
}

@software{photutils,
  author       = {Larry Bradley and
                  Brigitta Sip{\H o}cz and
                  Thomas Robitaille and
                  Erik Tollerud and
                  Z\`e Vin{\'{\i}}cius and
                  Christoph Deil and
                  Kyle Barbary and
                  Tom J Wilson and
                  Ivo Busko and
                  Axel Donath and
                  Hans Moritz G{\"u}nther and
                  Mihai Cara and
                  P. L. Lim and
                  Sebastian Me{\ss}linger and
                  Zach Burnett and
                  Simon Conseil and
                  Michael Droettboom and
                  Azalee Bostroem and
                  E. M. Bray and
                  Lars Andersen Bratholm and
                  William Jamieson and
                  Adam Ginsburg and
                  Geert Barentsen and
                  Matt Craig and
                  Sergio Pascual and
                  Shivangee Rathi and
                  Marshall Perrin and
                  Brett M. Morris},
  title        = {astropy/photutils: 2.2.0},
  month        = feb,
  year         = 2025,
  publisher    = {Zenodo},
  version      = {2.2.0},
  doi          = {10.5281/zenodo.14889440},
  url          = {https://doi.org/10.5281/zenodo.14889440},
  swhid        = {swh:1:dir:11159107f27a28985192ed1118b1f2055709d093
                   ;origin=https://doi.org/10.5281/zenodo.596036;visi
                   t=swh:1:snp:ae8c4a55d349d43e53cfe9ce92e678fcfe840f
                   3b;anchor=swh:1:rel:0117f67e8888adcdfc85308287dd9c
                   854b466389;path=astropy-photutils-ffb96c5
                  },
}

@article{skimage,
  author       = {van der Walt, St{\'e}fan and Sch{\"o}nberger, Johannes L. and Nunez-Iglesias, Juan and Boulogne, Fran{\c{c}}ois and Warner, Joshua D. and Yager, Neil and Gouillart, Emmanuelle and Yu, Tony},
  title        = {scikit-image: image processing in Python},
  journal      = {PeerJ},
  year         = {2014},
  volume       = {2},
  pages        = {e453},
  doi          = {10.7717/peerj.453},
  archivePrefix= {arXiv},
  eprint       = {},
}

@misc{ginsburg_image_registration_2024,
  author       = {Ginsburg, Adam},
  title        = {image\_registration: Image Registration Methods for Astronomy, v0.2.9},
  year         = {2024},
  howpublished = {\url{https://pypi.org/project/image-registration/}},
}

@ARTICLE{Krticka2014,
       author = {{Krti{\v{c}}ka}, Ji{\v{r}}{\'\i}},
        title = "{Mass loss in main-sequence B stars}",
      journal = {\aap},
         year = 2014,
        month = apr,
       volume = {564},
          eid = {A70},
        pages = {A70},
          doi = {10.1051/0004-6361/201321980},
archivePrefix = {arXiv},
       eprint = {1401.5511},
 primaryClass = {astro-ph.SR},
       adsurl = {https://ui.adsabs.harvard.edu/abs/2014A&A...564A..70K}
}

@ARTICLE{Reipurth2001,
       author = {{Reipurth}, Bo and {Bally}, John},
        title = "{Herbig-Haro Flows: Probes of Early Stellar Evolution}",
      journal = {\araa},
         year = 2001,
        month = jan,
       volume = {39},
        pages = {403-455},
          doi = {10.1146/annurev.astro.39.1.403},
       adsurl = {https://ui.adsabs.harvard.edu/abs/2001ARA&A..39..403R}
}

@ARTICLE{MaizApellaniz2021,
       author = {{Ma{\'\i}z Apell{\'a}niz}, J. and {Barb{\'a}}, R.~H. and {Fari{\~n}a}, C. and {Sota}, A. and {Pantaleoni Gonz{\'a}lez}, M. and {Holgado}, G. and {Negueruela}, I. and {Sim{\'o}n-D{\'\i}az}, S.},
        title = "{Lucky spectroscopy, an equivalent technique to lucky imaging. II. Spatially resolved intermediate-resolution blue-violet spectroscopy of 19 close massive binaries using the William Herschel Telescope}",
      journal = {\aap},
         year = 2021,
        month = feb,
       volume = {646},
          eid = {A11},
        pages = {A11},
          doi = {10.1051/0004-6361/202039479},
archivePrefix = {arXiv},
       eprint = {2011.12250},
 primaryClass = {astro-ph.SR},
       adsurl = {https://ui.adsabs.harvard.edu/abs/2021A&A...646A..11M}
}

@ARTICLE{Haffner1995,
       author = {{Haffner}, L. Matthew and {Meyer}, David M.},
        title = "{A Search for Interstellar C 3 in the Translucent Cloud toward HD 147889}",
      journal = {\apj},
         year = 1995,
        month = nov,
       volume = {453},
        pages = {450},
          doi = {10.1086/176406},
       adsurl = {https://ui.adsabs.harvard.edu/abs/1995ApJ...453..450H}
}

@ARTICLE{Negueruela2024,
       author = {{Negueruela}, I. and {Sim{\'o}n-D{\'\i}az}, S. and {de Burgos}, A. and {Casasbuenas}, A. and {Beck}, P.~G.},
        title = "{The IACOB project: XII. New grid of northern standards for the spectral classification of B-type stars}",
      journal = {\aap},
         year = 2024,
        month = oct,
       volume = {690},
          eid = {A176},
        pages = {A176},
          doi = {10.1051/0004-6361/202449298},
archivePrefix = {arXiv},
       eprint = {2407.04163},
 primaryClass = {astro-ph.SR},
       adsurl = {https://ui.adsabs.harvard.edu/abs/2024A&A...690A.176N}
}

@ARTICLE{Hacar2016,
       author = {{Hacar}, A. and {Alves}, J. and {Forbrich}, J. and {Meingast}, S. and {Kubiak}, K. and {Gro{\ss}schedl}, J.},
        title = "{APOGEE strings: A fossil record of the gas kinematic structure}",
      journal = {\aap},
         year = 2016,
        month = may,
       volume = {589},
          eid = {A80},
        pages = {A80},
          doi = {10.1051/0004-6361/201527805},
archivePrefix = {arXiv},
       eprint = {1602.01854},
 primaryClass = {astro-ph.GA},
       adsurl = {https://ui.adsabs.harvard.edu/abs/2016A&A...589A..80H}
}

@ARTICLE{Neuhauser2020,
       author = {{Neuh{\"a}user}, R. and {Gie{\ss}ler}, F. and {Hambaryan}, V.~V.},
        title = "{A nearby recent supernova that ejected the runaway star {\ensuremath{\zeta}} Oph, the pulsar PSR B1706-16, and $^{60}$Fe found on Earth}",
      journal = {\mnras},
         year = 2020,
        month = oct,
       volume = {498},
       number = {1},
        pages = {899-917},
          doi = {10.1093/mnras/stz2629},
archivePrefix = {arXiv},
       eprint = {1909.06850},
 primaryClass = {astro-ph.HE},
       adsurl = {https://ui.adsabs.harvard.edu/abs/2020MNRAS.498..899N}
}

@ARTICLE{Matzner2015,
       author = {{Matzner}, Christopher D. and {Jumper}, Peter H.},
        title = "{Star Cluster Formation with Stellar Feedback and Large-scale Inflow}",
      journal = {\apj},
         year = 2015,
        month = dec,
       volume = {815},
       number = {1},
          eid = {68},
        pages = {68},
          doi = {10.1088/0004-637X/815/1/68},
archivePrefix = {arXiv},
       eprint = {1511.03269},
 primaryClass = {astro-ph.GA},
       adsurl = {https://ui.adsabs.harvard.edu/abs/2015ApJ...815...68M}
}

@ARTICLE{Mookerjea2021,
       author = {{Mookerjea}, B. and {Sandell}, G. and {Veena}, V.~S. and {G{\"u}sten}, R. and {Riquelme}, D. and {Wiesemeyer}, H. and {Wyrowski}, F. and {Mertens}, M.},
        title = "{Distribution of ionized, atomic, and PDR gas around S 1 in {\ensuremath{\rho}} Ophiuchus}",
      journal = {\aap},
         year = 2021,
        month = apr,
       volume = {648},
          eid = {A40},
        pages = {A40},
          doi = {10.1051/0004-6361/202040217},
archivePrefix = {arXiv},
       eprint = {2102.10720},
 primaryClass = {astro-ph.GA},
       adsurl = {https://ui.adsabs.harvard.edu/abs/2021A&A...648A..40M}
}

@ARTICLE{Krumholz2009,
       author = {{Krumholz}, Mark R. and {Matzner}, Christopher D.},
        title = "{The Dynamics of Radiation-pressure-dominated H II Regions}",
      journal = {\apj},
         year = 2009,
        month = oct,
       volume = {703},
       number = {2},
        pages = {1352-1362},
          doi = {10.1088/0004-637X/703/2/1352},
archivePrefix = {arXiv},
       eprint = {0906.4343},
 primaryClass = {astro-ph.SR},
       adsurl = {https://ui.adsabs.harvard.edu/abs/2009ApJ...703.1352K}
}

@ARTICLE{Alves2026,
       author = {{Alves}, J. and {Zucker}, C. and {Lada}, C. and {Lombardi}, M. and {Piecka}, M. and {Hutschenreuter}, S. and {Meingast}, S. and {Posch}, L. and {Hacar}, A. and {Tachihara}, K. and {Yamada}, R. and {Swiggum}, C. and {Goodman}, A. and {Wunsch}, R. and {Gro{\ss}schedl}, J. and {Burkert}, A. and {Heitsch}, F. and {En{\ss}lin}, T.},
        title = "{Accelerated gas flow along Ophiuchus B44 filament: Breaking Position-Position-Velocity degeneracy}",
      journal = {arXiv e-prints},
         year = 2026,
        month = jun,
          eid = {arXiv:2606.29508},
        pages = {arXiv:2606.29508},
          doi = {10.48550/arXiv.2606.29508},
archivePrefix = {arXiv},
       eprint = {2606.29508},
 primaryClass = {astro-ph.GA},
       adsurl = {https://ui.adsabs.harvard.edu/abs/2026arXiv260629508A}
}

\begin{appendix}

\section{Notes on image processing and astrometric calibration}\label{section:A}

\begin{figure*}
 \centering
 \includegraphics[width=0.8\textwidth]{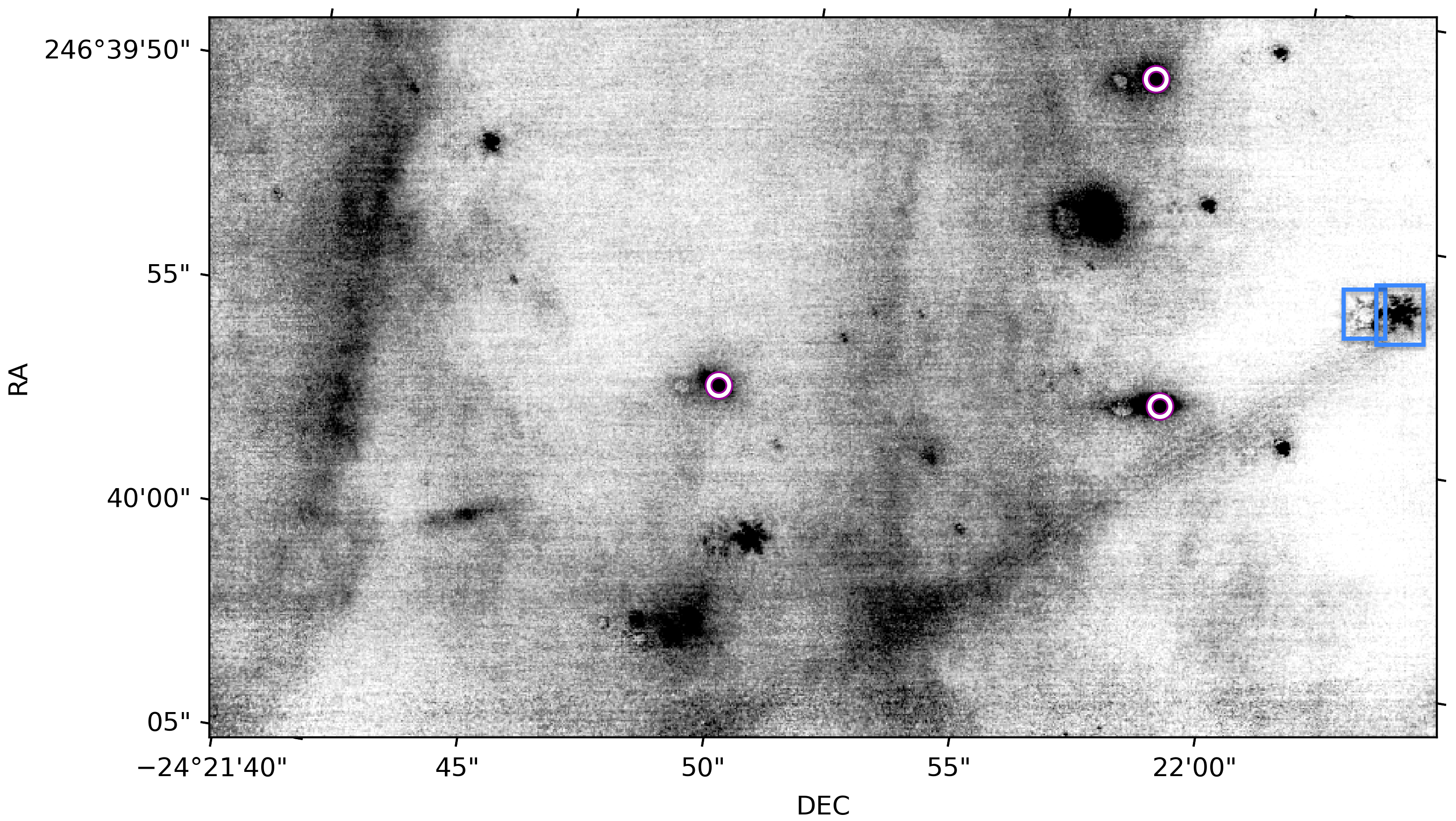}
 \caption{Example of duplication artefacts seen in the JWST (negative) images of L1688 within the Oph~ISM~3 region. These artefacts, highlighted with blue boxes for one of the sources, manifest by causing sources and ISM features to appear twice (similar to the "ghosting" effect), with a small spatial offset. Several galaxies can be seen in this figure (white circles), which were used in preparing our astrometric reference frames (Sect.~\ref{section:3}).}
 \label{fig:double}
\end{figure*}

\begin{figure*}
 \centering
 \includegraphics[width=0.8\textwidth]{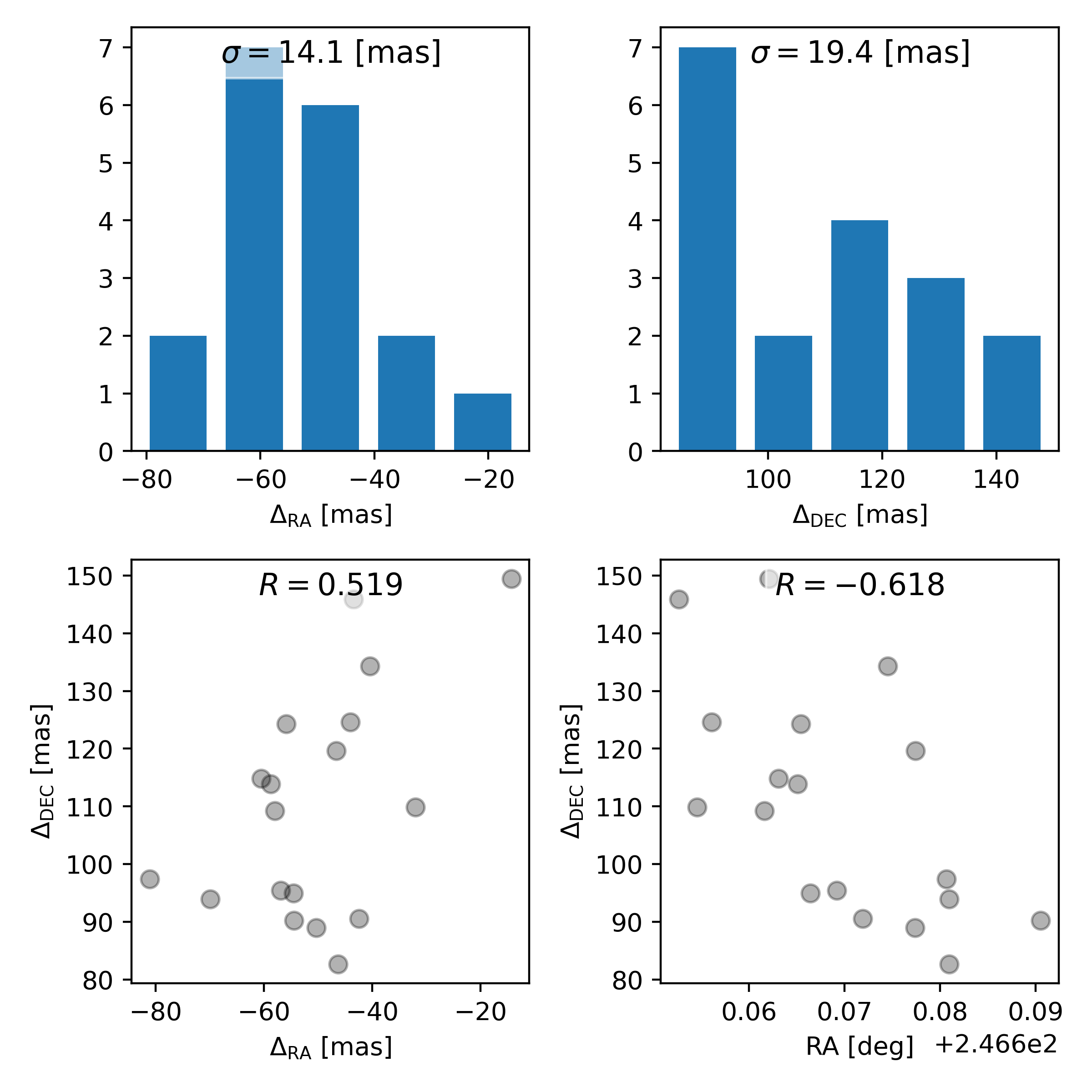}
 \caption{Histograms of positional offsets (before correction) of galaxies identified in the L1688 field (Table~\ref{table:A}) between two epochs. We note the possibly bimodal distribution in the declination offset. Two correlations that appear to be marginally significant are also displayed.}
 \label{fig:corrs}
\end{figure*}

The duplication artefacts seen in the JWST images of L1688 (Fig.~\ref{fig:double}) prompted us to reprocess these images, starting from Level 2 data. After carefully examining the individual tiles, we conclude that these artefacts resulted from a problem associated with stacking images from the visits 5~and~6 of the observation~9. The artefacts were mostly eliminated by processing the visits separately, using the JWST pipeline and then stacking the two slightly overlapping mosaics using \texttt{SWarp}. We note that the problem was not solved in the close vicinity of GSS~37. However, this region of the image was not investigated in our study and should have no impact on our analysis.

To allow proper motion measurements, images from both epochs need to be provided in the same reference frame. While the pipeline-based astrometric solutions are relatively accurate, they are not always reliable in regions lacking sources with precise astrometric information. Indeed, we noticed appreciable offsets in the positions of galaxies seen in the images. Due to the absence of a sufficient number of {\it Gaia} sources in the field (only three, compared to the required $\gtrsim100$), we decided to use these galaxies to align the images in a consistent astrometric frame of reference using the median offset vector.

We made use of \texttt{DAOStarFinder} to detect and measure positions of sources in the images. Sigma-clipping ($\sigma=3$) was used to calculate the median and the standard deviation (STD) of the fluxes in an image. The median is assumed to be a reasonable measure of the background and is subtracted from the corresponding image. In the \texttt{DAOStarFinder} setup, we chose $\texttt{FWHM}=5$~px to represent the width of a stellar profile. The $\texttt{threshold}=\textrm{STD} \cdot \textrm{SNR}_{\textrm{thresh}}$ parameter was calculated using the above mentioned standard deviation together with $\textrm{SNR}_{\textrm{thresh}}=0.2$ or $\textrm{SNR}_{\textrm{thresh}}=1.0$ for HST and JWST, respectively. The list of cross-matched galaxies identified in both images is provided in Table~\ref{table:A}. Statistical analysis of the individual offsets of galaxies (Fig.~\ref{fig:corrs}) reveals marginally significant correlations that might be hinting at image deformations. We do not address these further in our analysis but keep in mind that small systematic errors of the order of $\sim 1$~mas\,yr$^{-1}$ might affect our results.

\begin{table*}
\caption{List of extragalactic sources identified in both the HST and the JWST images.}\label{table:A}
    \begin{center}
    \begin{tabular}{r|rr|rr}
    \hline
    \hline
    Source ID & RA$_{\textrm{HST}}$ [deg] & DEC$_{\textrm{HST}}$ [deg] & RA$_{\textrm{JWST}}$ [deg] & DEC$_{\textrm{JWST}}$ [deg] \\
    \hline
    g1 & 246.68069678 & -24.36818554 & 246.68067425 & -24.36815849 \\
    g2 & 246.66512663 & -24.36618784 & 246.66511033 & -24.36615622 \\
    g3 & 246.67454274 & -24.36610615 & 246.67453152 & -24.36606885 \\
    g4 & 246.66310126 & -24.36586149 & 246.66308446 & -24.36582960 \\
    g5 & 246.65457087 & -24.36396214 & 246.65456197 & -24.36393163 \\
    g6 & 246.69052713 & -24.36367047 & 246.69051201 & -24.36364540 \\
    g7 & 246.66544954 & -24.36367642 & 246.66543401 & -24.36364190 \\
    g8 & 246.65612725 & -24.36312976 & 246.65611503 & -24.36309515 \\
    g9 & 246.66161801 & -24.36250501 & 246.66160191 & -24.36247466 \\
    g10 & 246.65268886 & -24.36215597 & 246.65267681 & -24.36211544 \\
    g11 & 246.67738874 & -24.35948889 & 246.67737476 & -24.35946417 \\
    g12 & 246.67744191 & -24.35806269 & 246.67742896 & -24.35802946 \\
    g13 & 246.66211468 & -24.35553660 & 246.66211070 & -24.35549509 \\
    g14 & 246.67193079 & -24.35526942 & 246.67191899 & -24.35524427 \\
    g15 & 246.66922502 & -24.34800747 & 246.66920922 & -24.34798097 \\
    g16 & 246.68094518 & -24.34450933 & 246.68092575 & -24.34448325 \\
    g17 & 246.66644766 & -24.34349593 & 246.66643251 & -24.34346955 \\
    g18 & 246.68095135 & -24.34248424 & 246.68093849 & -24.34246127 \\
    \hline
    \end{tabular}
    \end{center}
\end{table*}

\section{Point sources in the HST-JWST field}\label{section:B}

\begin{figure*}
 \centering
 \includegraphics[width=\textwidth]{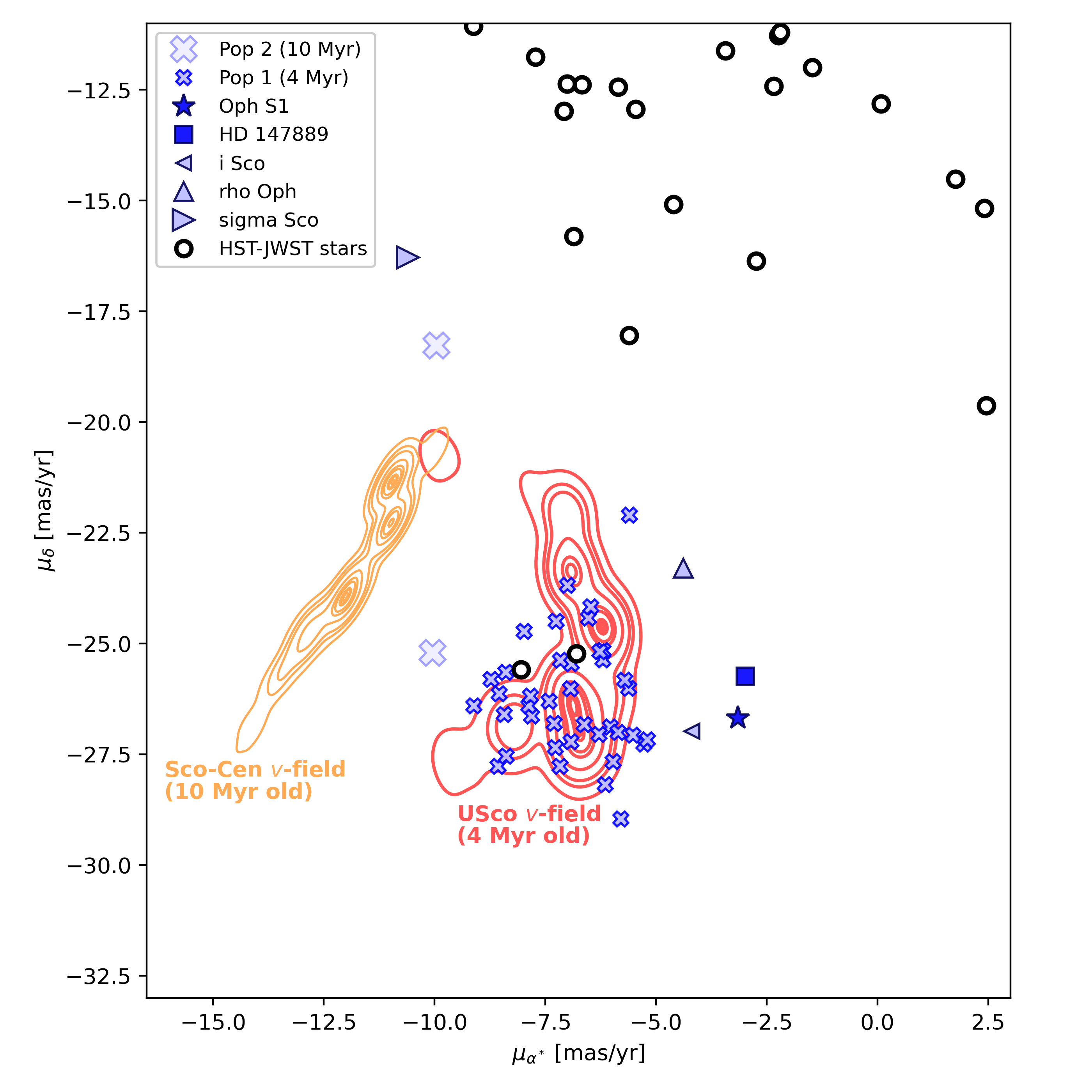}
 \caption{Similar to Fig.~\ref{fig:main}, focusing only on stars and additional stellar proper motion measurements obtained using IRAF (white points with black outlines, see Table~\ref{table:B}). Two of the new measurements align with the young USco population identified as Pop~1 by \citet{Grasser2021} and $\rho$~Oph/L1688 by \citet{Ratzenbock2023}.}
 \label{fig:mainAlt}
\end{figure*}

\begin{table*}
\caption{List of stellar sources identified in both the HST and the JWST images, with proper motions high enough to be considered in our analysis of L1688.}\label{table:B}
    \begin{center}
    \begin{tabular}{r|rr|rr|rr}
    \hline
    \hline
    Source ID & RA$_{\textrm{HST}}$ [deg] & DEC$_{\textrm{HST}}$ [deg] & RA$_{\textrm{JWST}}$ [deg] & DEC$_{\textrm{JWST}}$ [deg] & $\mu_{\alpha^*}$ [mas\,yr$^{-1}$] & $\mu_{\delta}$ [mas\,yr$^{-1}$] \\
    \hline
    1 & 246.65687374 & -24.36974620 & 246.65687722 & -24.36977235 & 1.8 & -14.5 \\
    2 & 246.66538453 & -24.36851317 & 246.66536861 & -24.36855927 & -8.0 & -25.6 \\
    3 & 246.66118808 & -24.36826265 & 246.66119294 & -24.36829800 & 2.5 & -19.6 \\
    4 & 246.67785695 & -24.36794050 & 246.67785231 & -24.36796288 & -2.3 & -12.4 \\
    5 & 246.67394530 & -24.36513471 & 246.67395008 & -24.36516204 & 2.4 & -15.2 \\
    6 & 246.65639161 & -24.36262122 & 246.65637841 & -24.36264352 & -6.7 & -12.4 \\
    7 & 246.66043126 & -24.36203012 & 246.66043142 & -24.36205319 & 0.1 & -12.8 \\
    8 & 246.68368759 & -24.35983748 & 246.68367360 & -24.35986085 & -7.1 & -13.0 \\
    9 & 246.67710799 & -24.35226015 & 246.67709641 & -24.35228255 & -5.9 & -12.4 \\
    10 & 246.67521375 & -24.35220565 & 246.67520020 & -24.35223413 & -6.9 & -15.8 \\
    11 & 246.67593687 & -24.35149174 & 246.67592579 & -24.35152424 & -5.6 & -18.0 \\
    12 & 246.66327798 & -24.35140047 & 246.66327507 & -24.35142207 & -1.5 & -12.0 \\
    13 & 246.66354446 & -24.35135181 & 246.66353769 & -24.35137273 & -3.4 & -11.6 \\
    14 & 246.68241865 & -24.35086605 & 246.68240521 & -24.35091151 & -6.8 & -25.2 \\
    15 & 246.68042834 & -24.34874443 & 246.68041923 & -24.34877161 & -4.6 & -15.1 \\
    16 & 246.68482941 & -24.34775930 & 246.68481138 & -24.34777923 & -9.1 & -11.1 \\
    17 & 246.67950206 & -24.34507138 & 246.67947748 & -24.34509331 & -12.4 & -12.2 \\
    18 & 246.66384956 & -24.34500049 & 246.66383877 & -24.34502379 & -5.5 & -12.9 \\
    19 & 246.67258515 & -24.34492744 & 246.67256990 & -24.34494861 & -7.7 & -11.8 \\
    20 & 246.67091091 & -24.34195306 & 246.67090649 & -24.34197336 & -2.2 & -11.3 \\
    21 & 246.67769584 & -24.33974339 & 246.67768199 & -24.33976566 & -7.0 & -12.4 \\
    22 & 246.66560687 & -24.33839890 & 246.66560146 & -24.33842838 & -2.7 & -16.4 \\
    23 & 246.67046242 & -24.37621174 & 246.67045808 & -24.37623191 & -2.2 & -11.2 \\
    \hline
    \end{tabular}
    \end{center}
    \tablefoot{Our conservative estimate of proper motion errors is about $2$--$3$~mas\,yr$^{-1}$ (see Appendix~\ref{section:B}).}
\end{table*}

We used the IRAF procedure described in \citet{Rottensteiner2026} to detect stellar objects and measure their proper motions. The calculated motions form a distribution centred on $(\mu_{\alpha^*},\mu_\delta)^{\textrm{stars}}_{\textrm{HST-JWST}}=(-4.2\pm0.2, -3.5\pm0.2)$ mas\,yr$^{-1}$ with standard deviations of about $4$~mas\,yr$^{-1}$. When compared with {\it Gaia} in the surrounding field of $50\arcmin \times 30\arcmin$, we find a somewhat faster median field motion of $(\mu_{\alpha^*},\mu_\delta)^{\textrm{stars}}_{\textrm{Gaia}}=(-5,-7)$~mas\,yr$^{-1}$. This difference is probably due to the infrared observations detecting a comparatively larger fraction of background targets, whose motions are closer to zero, thereby shifting the measured median towards zero.

The stars displayed in Fig.~\ref{fig:mainAlt} and their proper motions are presented in Table~\ref{table:B}. Source 11 was previously mentioned in the literature as [BHM2012b]~301 \citep{Barsony2012} and was marked as a possible member of the $\rho$~Oph cluster. Our proper motion measurement provides evidence that this star might truly be part of the Sco-Cen population of stars, although its kinematics slightly differ from those of the USco velocity field. To our knowledge, stars 2 and 14 were never previously mentioned in the literature. Interestingly, they appear to be good YSO candidates based on their kinematics matching that of the young USco population.

Since the calculated stellar proper motions are based on only two epochs, the measurement error estimates are poorly constrained. However, the standard deviations of the positional offsets of galaxies provide some insight into the proper motion errors of point sources. Using the distributions presented in Fig.~\ref{fig:corrs}, we estimate that the expected proper motion errors for point sources should be in the range of $2$--$3$~mas\,yr$^{-1}$. Due to the more complex morphology of galaxies, we note that these values should correspond to an upper estimate for the stellar proper motion errors. While not comparable to {\it Gaia}, they are sufficient for identifying additional candidates for kinematic members of the USco population of stars.

We identify three {\it Gaia} stars within our field of interest: the binary system GSS~37 and [GY92]~100. Although these sources are heavily saturated, it is still possible to roughly estimate their proper motions from the images using image registration and compare them with {\it Gaia} measurements. To mitigate the effects of saturation, we smoothed the stellar profiles by applying a $5\sigma$ Gaussian filter to the regions surrounding each source. Proper motions were then estimated using SimpleITK. We note that both sampling strategies (Sect.~\ref{section:4.2}) were disabled for this analysis, as they are not suited for such targets. By comparing our estimated motion with that obtained by {\it Gaia}:
\begin{equation*}
    (\mu_{\alpha^*},\mu_\delta)^{\textrm{GSS 37}}_{\textrm{HST-JWST}} = (-8.0,-26.8) \,\, \textrm{mas}\,\textrm{yr}^{-1} \,\,,
\end{equation*}
\begin{equation*}
    (\mu_{\alpha^*},\mu_\delta)^{\textrm{GSS 37}}_{\textrm{Gaia}} = (-4.5,-26.7) \,\, \textrm{mas}\,\textrm{yr}^{-1} \,\,,
\end{equation*}
\begin{equation*}
    (\mu_{\alpha^*},\mu_\delta)^{\textrm{[GY92]~100}}_{\textrm{HST-JWST}} = (-14.5,-25.8) \,\, \textrm{mas}\,\textrm{yr}^{-1} \,\,,
\end{equation*}
\begin{equation*}
    (\mu_{\alpha^*},\mu_\delta)^{\textrm{[GY92]~100}}_{\textrm{Gaia}} = (-15.9,-24.0) \,\, \textrm{mas}\,\textrm{yr}^{-1} \,\,,
\end{equation*}
we find that these measurements are in a good agreement, especially when taking into account our default two-epoch precision levels and considering our use of a Gaussian filter. This experiment verifies that our astrometric solution described in Sect.~\ref{section:3} can provide reliable proper motion measurements.

\section{SimpleITK in astrophysical context}\label{section:C}

\subsection{Application of SimpleITK}\label{section:C1}

The measurements of ISM proper motions are performed using image registration tools provided by the Python implementation of SimpleITK \citep{SimpleITK1,SimpleITK2,SimpleITK3}. First, sampling parameters are defined and initial image pre-processing is applied, which consists of aligning astrometric reference frames and reprojecting the images to a common grid (in our case, the one defined by HST). For the purposes of this work, we make use of no pre-processing filters (except for the bright stars mentioned in Appendix~\ref{section:B}). The regions of interest for the analysis are then selected using a sliding window approach, and the corresponding images from both epochs are loaded. A stellar mask is constructed to exclude point sources from the alignment procedure (Appendix~\ref{section:E}) and is visually inspected to ensure that stellar contamination is effectively removed.

When using SimpleITK, the moving, the reference, and the mask images are first converted into SimpleITK image objects. The registration is based on a linear translation transformation and uses a normalised correlation metric to compute similarities between images. The optimisation is carried out using a regular step gradient descent scheme with a learning rate of 1.0, a minimum step size of $10^{-5}$, and a maximum of 300 iterations. The transformation is initialised with a zero initial offset since the motion is expected to be $\lesssim5$~px. The result of the optimisation are the translation parameters that correspond to the optimal alignment vector -- we assume this to be a consequence of proper motion only.

Image alignment is performed using both flux-based (noise perturbations) and jitter-based (window-shape variations) sampling strategies ($N = 100$, see Appendix~\ref{section:C2}), resulting in a distribution of pixel-space offset vectors. These offsets are converted into world coordinates using the WCS information from the image headers. Proper motions are then computed for each sample using the known time baseline between the two epochs, with the $\cos\delta$ correction applied to the right ascension component.

The computed proper motion values correspond to the medians of these distributions, which are adopted as the motion estimates, while the associated uncertainties are given by the standard deviations. We note that these uncertainties are negligible compared to the errors on the median. As a consistency check, the alignment results are visually inspected -- a successful alignment is indicated by close morphological agreement between the aligned moving image and the reference image.

\subsection{Image sampling strategies}\label{section:C2}

By default, SimpleITK provides no options for estimating registration errors. To overcome this limitation, one needs to prepare a sampling strategy that can effectively capture the changes in the images that affect the registration results. Our sampling approach is designed to propagate both noise-induced and spatial uncertainties into the image registration procedure. Two complementary strategies are readily available: flux-based noise perturbations and spatial window jittering.

\begin{figure}
 \centering
 \includegraphics[width=\columnwidth]{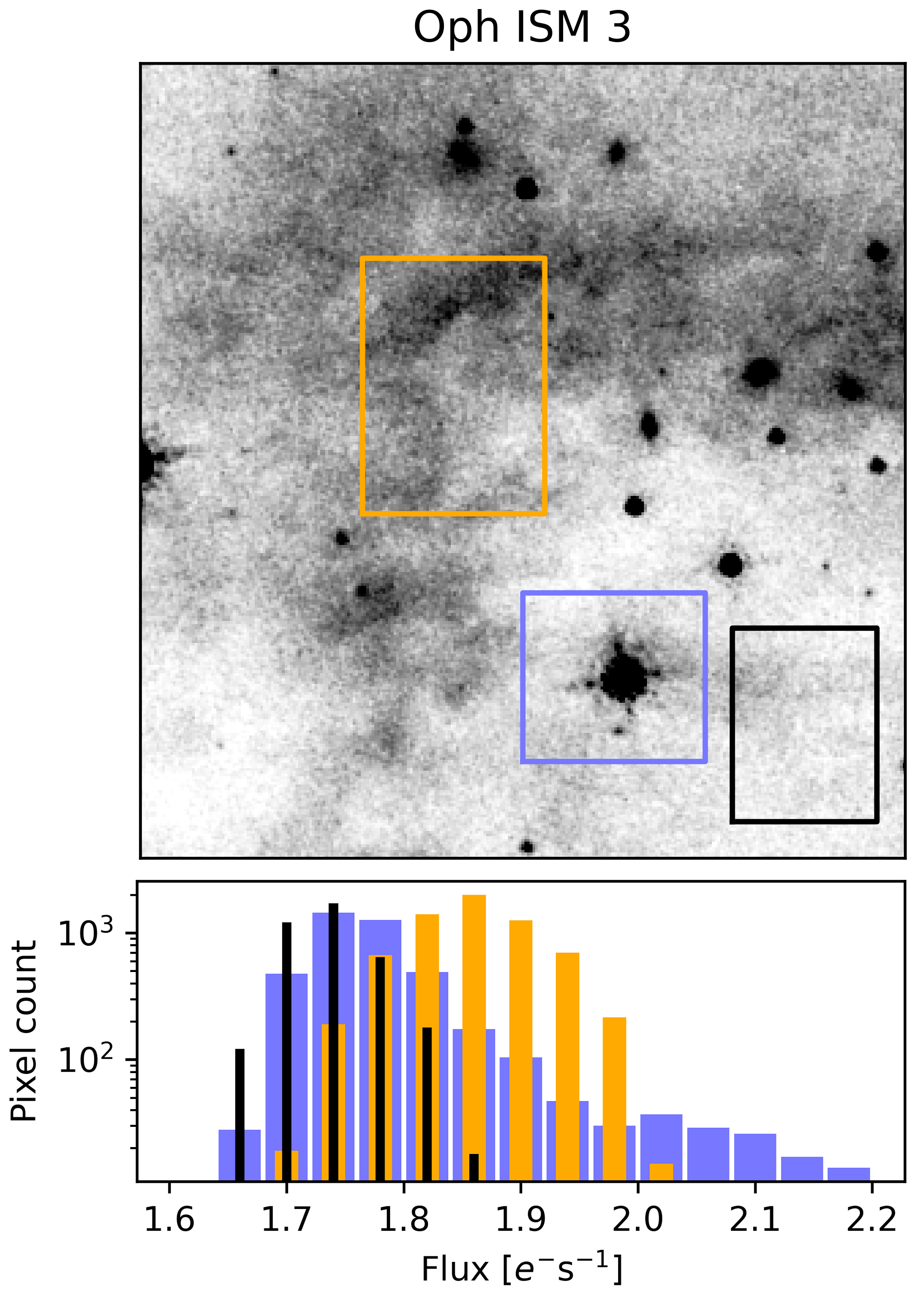}
 \caption{Flux statistics in the Oph~ISM~3 section of the HST (negative) image. The top panel shows three different regions of the image -- ISM-dominated (orange), star-dominated (blue), and noise-dominated (black). The histogram in the panel below highlights the flux distributions in these boxes, supporting our choice of a Gaussian noise model.}
 \label{fig:gauss}
\end{figure}

In the flux-based sampling, Gaussian noise is added to both the reference and the moving images in each iteration. The noise level is estimated directly from the data using mean values obtained with \texttt{skimage.restoration.estimate\_sigma}, which is applied to the analysed region in each image separately. This provides an empirical estimate of the background noise, which is used as the standard deviation of the perturbation added to each pixel. We note that while the actual noise in the images is non-Gaussian, it can be approximated as such in the case of non-stellar objects (see Fig.~\ref{fig:gauss}).

Secondly, a jitter-based sampling scheme can be employed to account for sensitivity to the choice of the spatial region. In this case, the analysis window is randomly shifted for each iteration by an integer offset drawn uniformly from a bounded interval. The maximum allowed jitter is set to $\pm 3$ pixels in both spatial directions. The resulting sub-images are then independently passed to the registration procedure. To ensure robustness, the jitter window is constrained such that the resulting sub-images remain sufficiently large for stable registration.

Together, these two sampling strategies produce a set of independent alignment solutions that capture both photometric noise and spatial selection effects. We note that depending on the image properties (presence of stellar or other ISM objects, S/N), the different sampling strategies can have a different impact on the uncertainties estimated from the samples. For example, we find that the flux-based sampling appears to be more important for capturing the uncertainties when compared with jittering in the case of Oph~ISM~4.

\subsection{Visualisation of ISM motion}\label{section:C3}

To depict the ISM proper motions corresponding to offsets of about 3--4~px in a single static image, we need to rescale the intensities of the images. We used primitive stellar masks for the HST and the JWST images by setting all pixels brighter than the ISM to \texttt{nan}, specifically all pixels with intensities $>2.8$ and $>2.6$, respectively. Then we linearly rescaled the intensity of the JWST image by using 0.05 and 0.95 as the lower and upper quantiles of both images.

\begin{figure}
 \centering
 \includegraphics[width=\columnwidth]{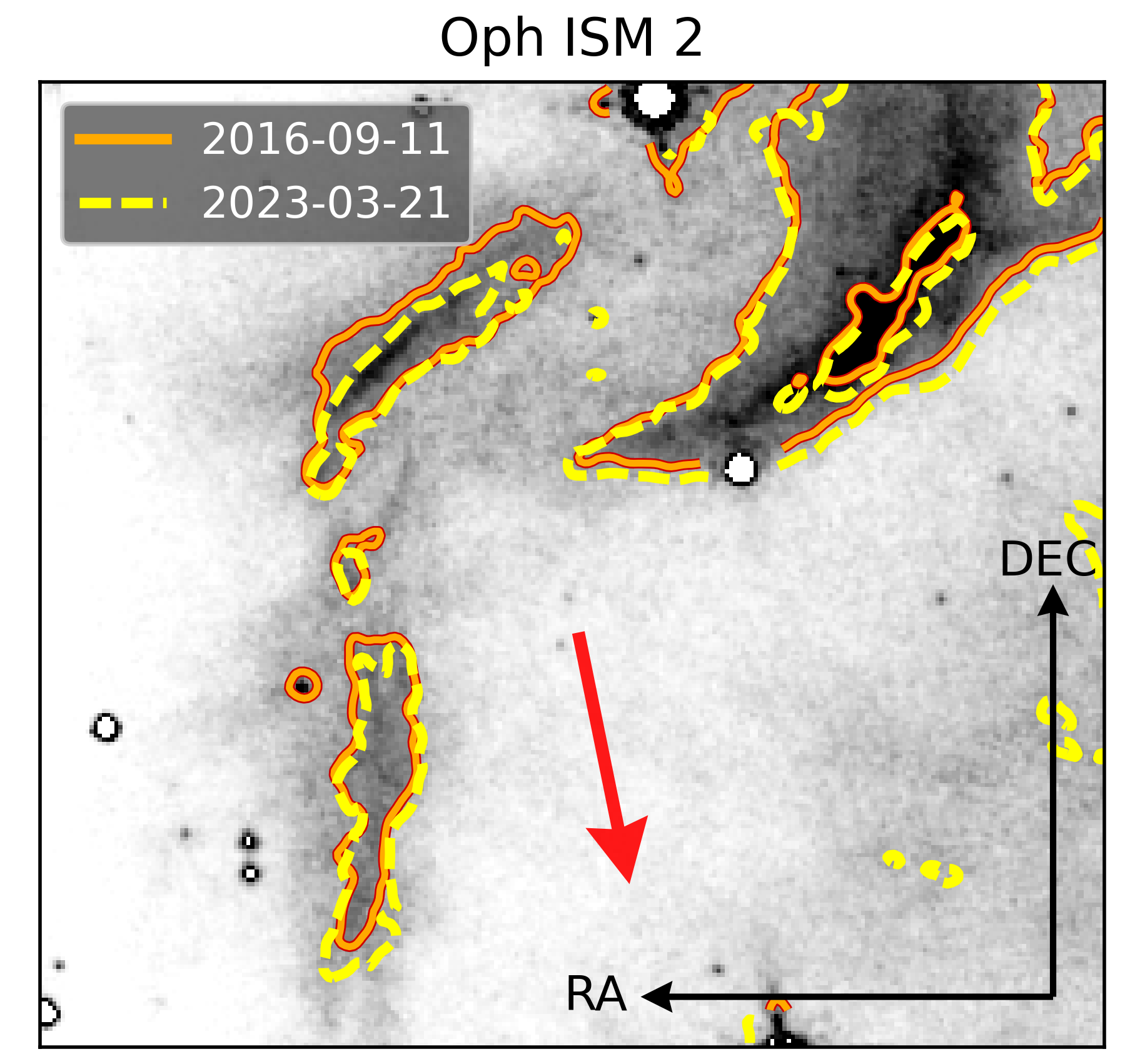}
 \caption{Visualisation of the gas proper motion for Oph~ISM~2 (uncorrected for the parallax effect). The gray-scale map displays the ISM structure as imaged by HST (negative), while the overlaid contours highlight structure positions during the 2016 (HST) and 2023 (JWST) epochs. The red arrow shows the measured direction of motion.}
 \label{fig:demo}
\end{figure}

The structure of the ISM can be visualised either by making use of skeletonisation or contours. For the ISM structures seen in Fig.~\ref{fig:infomap}, we find that contours are the better choice. First, both images were smoothed by making use of a $1.5\sigma$ Gaussian filter from \texttt{skimage} package. Our contours are calculated at the 0.86 and 0.99 quantile levels and, together with the original HST map, presented in Fig.~\ref{fig:demo}. We remind the reader that the contours are not used in proper motion calculations and serve only for visualisation purposes.

\section{Proper-motion field filter}\label{section:D}

\begin{figure}
 \centering
 \includegraphics[width=\columnwidth]{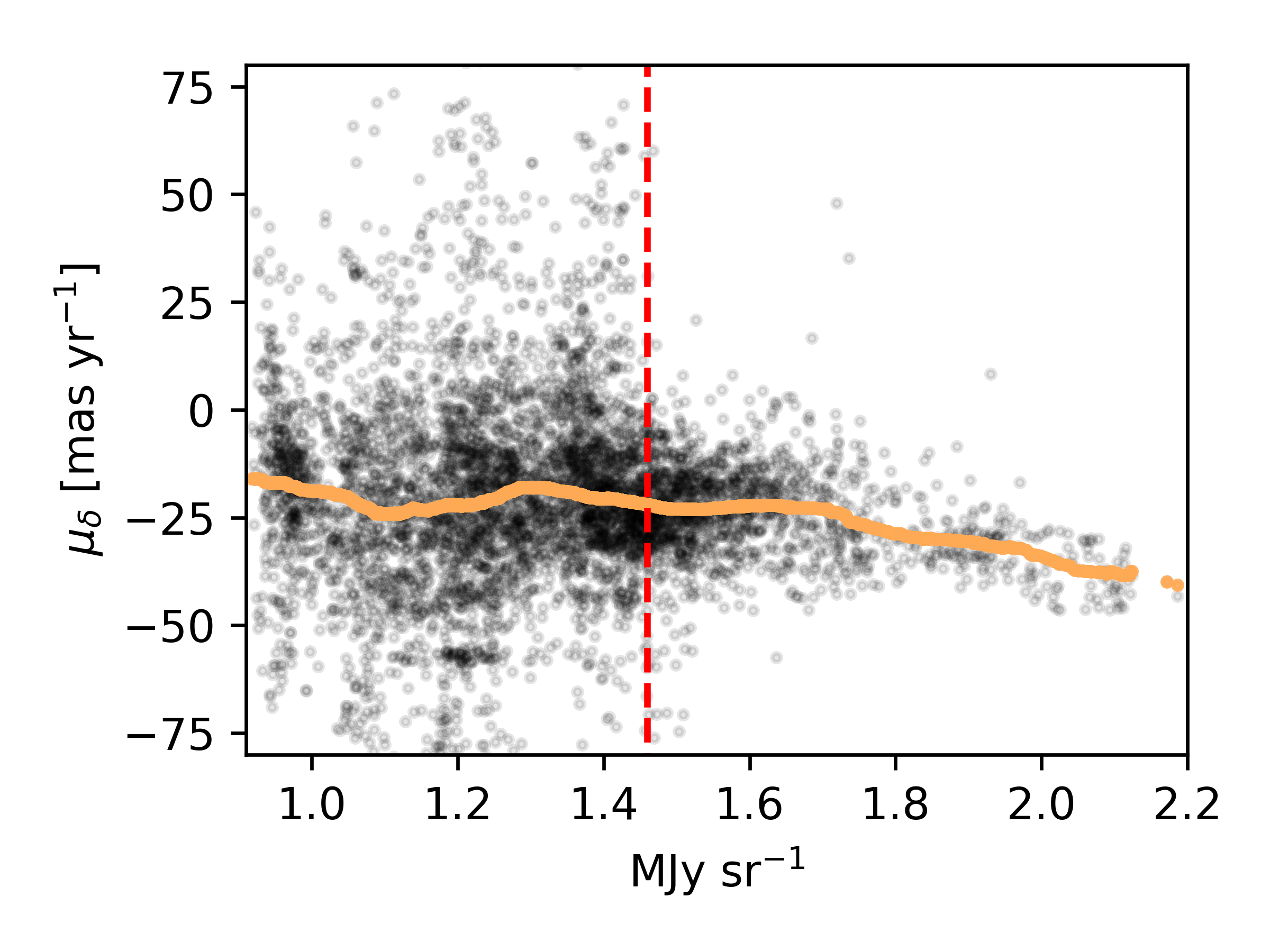}
 \caption{Visualisation of one of the measured proper motion components as a function flux in the JWST (F200W) image, uncorrected for the parallax effect. The orange line shows the running median (bin-width of 0.1 MJy\,sr$^{-1}$). Below the highlighted flux limit (75th percentile, vertical dashed line), the measurements appear more noisy, but the median of the distribution remains nearly constant down to the lowest flux levels.}
 \label{fig:fluxcut}
\end{figure}

In Fig.~\ref{fig:fluxcut} we show the declination component of the ISM proper motion from the unfiltered proper motion map as a function of flux in the JWST image. It is clear from this distribution that our choice of using only the prominently bright ISM features with the proper motion map has little-to-no impact on the results of our analysis. The median of the distribution remains unaffected down to very low fluxes, with the running median varying between $-17$~mas\,yr$^{-1}$ and $-25$~mas\,yr$^{-1}$ below the flux limit. We note the presence of a decreasing trend for this proper motion component in the bright regime. The appearance of the more negative values of this proper motion component was discussed at the end of Sect.~\ref{section:5.1} and is possibly a result of a bias associated with comparing images at higher-density regions using filters with different wavelength coverages. The few data points that skew the distribution in the bright regime do not significantly affect the proper-motion median above the chosen flux limit.

\section{Stellar-mask image for SimpleITK}\label{section:E}

We make use of a simple procedure to identify stars in images (similar to our detection of galaxies, see Appendix~\ref{section:A}) and mask them. The masking procedure begins by estimating the background statistics of the image using sigma-clipped measurements ($\sigma=2$, $\textrm{SNR}_{\textrm{thresh}}=2$), which provide robust estimates of the median and the standard deviation while mitigating the influence of bright sources. These statistics are used to define a detection threshold proportional to the background noise level. Point sources are then identified using a \texttt{DAOStarFinder}, which locates local maxima consistent with the expected point-spread function, parametrised by a full width at half maximum ($\texttt{FWHM}=3.0$). The detection is performed on a background-subtracted version of the image to enhance sensitivity to fainter sources.

For each detected source, a circular mask is constructed. The radius of this mask is scaled as a power-law function of the source flux
\begin{equation*}
R_{\textrm{mask}} = |F|^{0.3} \cdot \textrm{size}_{\textrm{mask}} \,\,,
\end{equation*}
such that brighter stars are assigned larger masking regions. The $\textrm{size}_{\textrm{mask}}$ parameters is set either to $2.5$ (near Oph~ISM~3) or $2.0$ (elsewhere), depending on the image properties of the investigated region. The power-law value and the mask size parameter were chosen based on a careful examination of images, ensuring that the point sources are sufficiently covered in both images. This strategy adequately captures both faint and bright sources without excessively masking the background. Finally, the individual masks are combined into a single binary image and \texttt{scipy.ndimage.binary\_dilation} ($\texttt{iterations}=2$) is applied to account for extended halos and residual light surrounding the sources.

\begin{figure}
 \centering
 \includegraphics[width=\columnwidth]{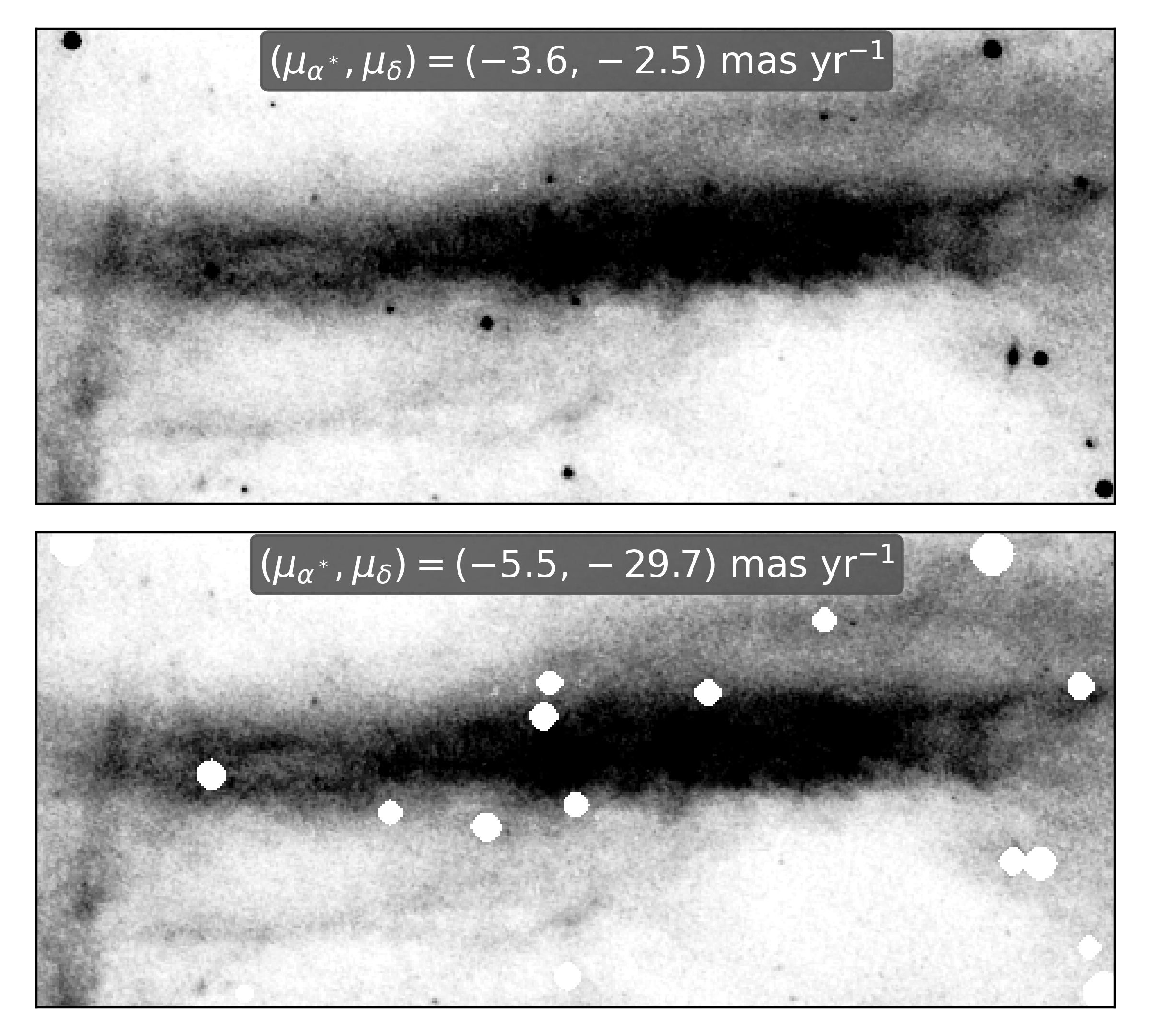}
 \caption{Demonstration of the image registration results without and with stellar masking. The displayed region (in negative) corresponds to Oph~ISM~1.}
 \label{fig:mask}
\end{figure}

Fig.~\ref{fig:mask} highlights the performance of image registration by applying SimpleITK to the whole sub-field Oph~ISM~1 (see features in Fig.~\ref{fig:infomap}), particularly the resulting alignment when including or excluding a stellar mask. For simplicity, sampling procedures were disabled for this demonstration, and the whole sub-field was used instead of a sliding window. This comparison of the image registration results makes it clear that a stellar mask is required when studying ISM proper motions using the presented method, otherwise the procedure detects mainly the motion of the background stars. Some of the fainter point sources remain unmasked in the image, but their impact on ISM proper motion measurement is minimal due to the small amount of information imprinted on the analysed section of the image. In crowded regions of the sky (e.g., near the Galactic disk), the fainter sources would also need to be carefully masked.

\end{appendix}

\end{document}